\documentclass[a4paper,preprint,showpacs,superscriptaddress]{revtex4}
\usepackage[version=3]{mhchem}
\usepackage{float,fancyhdr,amsmath,graphicx,subfig,epsfig,color,setspace,url}
\usepackage{appendix}
\newcommand{\hba}{(\frac{\beta}{2})}

\begin{document}

\title{Dissociative Electron Attachment to Polyatomic Molecules - I : Water }
\author{N. Bhargava Ram}
\email[]{nbhargavaram@tifr.res.in}
\affiliation{Tata Institute of Fundamental Research, Mumbai 400005, India}

\author{V. S. Prabhudesai}
\affiliation{Department of Physics, Weizmann Institute of Science,
Rehovot 76100, Israel}

\author{E. Krishnakumar}
\email[]{ekkumar@tifr.res.in}
\affiliation{Tata Institute of Fundamental Research,  Mumbai 400005, India}

\begin{abstract}
Using the velocity map imaging technique, we studied and characterized the process of Dissociative Electron Attachment (DEA) in polyatomic molecules like Water, Hydrogen Sulphide, Ammonia, Methane, Formic Acid and Propyl Amine. We present the details of these studies in a series of 5 articles. In the first article here, we discuss the DEA process in gas phase water (\ce{H2O} and \ce{D2O}) molecules. Electrons of 6.5 eV, 8.5 eV and 12 eV are captured by water molecules in neutral ground state to form \ce{H2O^{-*}} (\ce{D2O^{-*}}) resonant states which dissociate into an anion fragment and one or more neutrals. The details of the kinetic energy and angular distributions of the fragment anions \ce{H-} (\ce{D-}) and \ce{O-} produced from the three negative ion resonant states in the entire 2$\pi$ scattering range are discussed. Unique angular distribution patterns are observed at the 8.5 eV and 11.8 eV resonances showing dissociation dynamics beyond the axial recoil approximation. 
\end{abstract}
\pacs{34.80.Ht}

\maketitle

\section{Introduction}

Dissociative electron attachment (DEA) studies on water are of great significance for several reasons. It is the most basic molecule of life and hence, very important that we know the details of its interaction with radiation and charged particles that produce ions and radicals and consequent chemical reactions involving them. Recent studies have highlighted the contribution of the DEA process in damage to DNA and its significance in radiation induced damage to biological tissues \cite{c3boudaiffa,c3sanche1,c3sanche2}. Being the most ubiquitous molecule, water is also important as a solvent in many practical applications. Also being one of the simplest polyatomic molecules it is an ideal system to study and improve our understanding of the complex process of DEA in polyatomic molecules. Added to this, the availability of ab intio theoretical calculations on electron attachment to water \cite{c3gorfinkiel,c3h1,c3h2,c3h3,c3h4,c3h5,c3h6,c3h7} makes detailed experimental studies on the dynamics of the DEA process a timely necessity. The most important parameters in unravelling the dynamics of the DEA process are the kinetic energy and angular distribution of the products. Using the velocity map imaging technique, we measured these parameters for the \ce{H-} (\ce{D-}) and \ce{O-} ions in the complete $2\pi$ angular range with unprecedented sensitivity. In this chapter, we present these data and identify the symmetry of the resonances, the dissociation channels and the dynamics at the three resonances.

\section{Earlier work}

\begin{figure}[!htbp]
	\begin{center}
		\includegraphics[width=0.6\columnwidth]{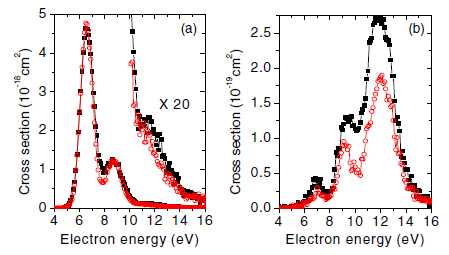}
		\caption{Cross sections for the formation of (a) \ce{H-} (\ce{D-}) and (b) \ce{O-} from DEA to \ce{H2O} (filled black squares) and \ce{D2O} (open red circles) from ref. \cite{c3rawat}}
		\label{fig3.1}
\end{center}
\end{figure}

Electron attachment to water occurs as resonances centered at 6.5, 8.5 and 11.8 eV (Figure \ref{fig3.1}) and has been widely studied to determine the exact resonance energies and widths, partial and absolute cross sections of the various anion fragments \cite{c3christo,c3rawat,c3fedor}. The electronic ground state of water is \ce{1a1^{2} 2a1^{2} 1b2^{2} 3a1^{2} 1b1^{2} -> ^{1}A1} state (\ce{C_{2v}} symmetry) and the three resonances are known to be core excited Feshbach resonances where an extra electron is attached to the singly excited states (caused by the excitation of the \ce{1b1}, \ce{3a1} and \ce{1b2} electron respectively to \ce{4a1}) of the neutral water molecule. The short lived \ce{H2O^{-*}} dissociates to give \ce{H-}, \ce{O-} or \ce{OH-} fragments through various two-body and three-body channels as listed in Table \ref{tab3.1}. Also given are the possible symmetries of the \ce{H2O^{-*}} based on Wigner-Witmer correlation rules. The cross section for \ce{H-} ion production is maximum at 6.5 eV and lowest at 11.8 eV; whereas \ce{O-} increases from 6.5 eV to 11.8 eV. The cross section for \ce{OH-} ions is lower than \ce{O-} by an order of magnitude at the above resonances \cite{c3christo}. The formation of \ce{OH-} has been a subject of controversy as some of the earlier measurements attributed its presence due to ion-molecule reaction \cite{c3compton}. Recent measurements by Fedor et al. \cite{c3fedor} confirm its presence while theoretical calculations \cite{c3h3,c3h6} rule out the possibility of its formation by direct DEA.

\begin{table}[h]
\caption{Electron attachment to \ce{H2O} and \ce{D2O} : Possible dissociation channels with threshold energy and parent molecular states based on Wigner - Witmer correlation rules.}
	\begin{center}
		\begin{tabular}{ccc}
		\hline
		\\
		From \ce{H2O^{-*}} & From \ce{D2O^{-*}} & Possible symmetry states\\
		\\
		\hline
		\\
\ce{H- + OH (X ^{2}$\Pi$)}; 4.35 eV & \ce{D- + OD (X ^{2}$\Pi$)}; 4.52 eV & \ce{A1}, \ce{A2}, \ce{B1} or \ce{B2} \\
\\
\ce{H- + OH^{*} (A ^{2}$\Sigma$)}; 8.38 eV & \ce{D- + OD^{*} (A ^{2}$\Sigma$)}; 8.38 eV & \ce{A1} or \ce{B2} \\
\\
\ce{H- + H + O}; 8.75 eV & \ce{D- + D + O}; 9.02 eV & \ce{A2}, \ce{B1} or \ce{B2} \\
\\
\ce{O- + H2}; 3.56 eV & \ce{O- + D2}; 3.68 eV & \ce{A1}, \ce{B1} or \ce{B2} \\
\\
\ce{O- + H + H}; 8.04 eV & \ce{O- + D + D}; 8.28 eV & \ce{A1}, \ce{B1} or \ce{B2} \\
\\
\ce{OH- (^{1}$\Sigma$) + H}; 3.27 eV & \ce{OD- (^{1}$\Sigma$) + D}; 3.44 eV & \ce{A1} or \ce{B2}\\
\\
\hline
\end{tabular}
\end{center}
\label{tab3.1}
\end{table}

The first measurement on angular distribution from DEA to water was reported by Trajmar and Hall \cite{c3trajmar}. They measured the angular distribution of \ce{H-} ions at 6.5 eV and identified the \ce{H2O^{-*}} resonance as \ce{^{2}B1} decaying into \ce{H- (^{1}S) + OH (X ^{2}$\Pi$)}. The angular distribution was seen to peak at $100^{\circ}$ with an asymmetry about $90^{\circ}$ and dropped off very fast at lower and higher angles. Further, they suggested using symmetry arguments that the transition matrix element for \ce{H2O^{-*}} formation would be zero for electron beam axis in the molecular plane and non zero for axis perpendicular to the molecular plane and hence, the H-OH bond dissociates when the electron beam is perpendicular to the molecular plane leading to distribution around $90^{\circ}$. They could also infer the vibrational and rotational excitation of the OH fragment from the kinetic energy distribution of \ce{H-}. 

Belic et al. \cite{c3belic} measured the energy and angular distribution of \ce{H-} and \ce{D-} ions from \ce{H2O} and \ce{D2O} respectively using conventional $127^{\circ}$ electrostatic analyzers. Based on their measurements, the three resonances were identified as having \ce{^{2}B1}, \ce{^{2}A1} and \ce{^{2}B2} symmetry respectively. The energy distribution of the \ce{H-} ions at the first and second resonance revealed the diatomic OH fragment to be in the electronic ground state but with vibrational and rotational excitation. The rotational excitation of OH is shown to be more intense at the second resonance than in the first as the energy distribution of \ce{H-} ions at 6.5 eV has better resolved vibrational structure whereas it is smoother in the second resonance at 8.5 eV. The third resonance was found to dissociate to the first excited state of \ce{OH (^{2}$\Sigma$^{+})} and ground state \ce{H-} ion. The electronically excited OH was also seen to be vibrationally and rotationally excited. 

The \ce{H-} angular distribution data at the 6.5 eV by Belic et al. \cite{c3belic} agreed with the results of Trajmar and Hall \cite{c3trajmar} reaffirming the \ce{^{2}B1} symmetry. Comparison with the electron energy loss (EEL) spectra of \ce{H2O} measured by Chutjian et al. \cite{c3chutjian} shows that the \ce{1b1^{1} 3sa1^{1} -> ^{3}B1} excited state at 7.0 eV is the parent state of the DEA resonance at 6.5 eV. An extra electron in the \ce{3sa1} results in a \ce{^{2}B1} state. The shift in the angular distribution peak to $100^{\circ}$ was explained as due to distortion of the incident electron wave from the plane wave due to direct potential scattering \cite{c3belic}. Later theoretical work by Teillet-Billy and Gauyacq \cite{c3billy} showed that the effect due to permanent dipole moment of the water molecule would explain the angular distribution exceedingly well. However, Haxton et al. \cite{c3h4} have recently shown that this backward shift is caused by a small contribution from the $d$-wave to the main contribution from the $p$-wave and the subsequent interference between them. 

The angular distribution observed by Belic et al. \cite{c3belic} at the second resonance was said to correlate with \ce{^{2}A1} state and result in \ce{H-} and OH fragments in ground state. The electronic configuration for this resonance was stated to be \ce{3a1^{-1} 3sa1^{2}} and said to match with the \ce{^{3}A1} state situated at 9.3 eV in the EEL data \cite{c3chutjian}. Continuing on same lines, the third resonance was identified as second excited state of \ce{H2O^{+} (^{2}B2)} plus two \ce{3sa1} Rydberg electrons. The \ce{H-} was accompanied with OH in the first electronic excited state (\ce{^{2}$\Sigma$^{+}}). In this way, the DEA resonances were identified as Feshbach resonances formed by the binding of two \ce{3sa1} electrons to the first three electronic states of \ce{H2O+} formed by ejecting \ce{1b1}, \ce{3a1} and \ce{1b2}. This appeared to be consistent with the known strongly antibonding nature of the \ce{3sa1} orbital making the respective potential energy surfaces repulsive leading to dissociation into \ce{H-} and OH fragments and with the calculations by Jungen et al. \cite{c3jungen}.

Curtis and Walker \cite{c3curtis} measured the kinetic energies of the \ce{D-} and \ce{O-} ions produced by DEA in \ce{D2O} over the resonance energy regions of 6.5, 8.5 and 11.5 eV using a cylindrical mirror energy analyzer and a quadrupole mass spectrometer. The kinetic energy of \ce{D-} at 6.5 eV showed the dissociation channel to be \ce{D- (^{1}S) + OD (X ^{2}$\Pi$)} with vibrational excitation of OD fragment whereas the \ce{O-} was produced with near thermal energies and accompanied by \ce{D2} in highly vibrational excited state (\ce{O- + D2}; threshold 3.68 eV). They argued that \ce{D-} at the second resonance appears via the same dissociation channel as the first resonance but from a predissociation mechanism. It is known from experiments and models of photodissociation that the potential energy surfaces of the \ce{^{2}A1} and \ce{^{1}A1} states are alike. However, the \ce{^{1}A1} state of water correlates with \ce{H + OH (A ^{2}$\Pi$)} while the ground state \ce{OH (X ^{2}$\Pi$)}, the major product in photodissociation arises from predissociation by the A \ce{^{1}B1} state. Assuming a parallel correlation for the \ce{^{2}A1} anion [\ce{D- + OD (A ^{2}$\Sigma$)}], they argued that the \ce{^{2}A1} anion state predissociates through the \ce{^{2}B1} state to give the ground state OD. According to them, \ce{O-} at the second resonance also arises from a sequential two body breakup leading to the \ce{O- + D + D} (threshold 8.28 eV) pathway. At the third resonance, two groups of \ce{D-} ions were seen; one coming from the \ce{D- (^{1}S) + OD(X ^{2}$\Pi$)} channel with kinetic energies ranging from 3 to 7 eV and the other group arising from a three body breakup (\ce{D- + D + O}) with energy partitioned in the ratio 0.4 : 0.4 : 0.2. A similar sequential two body breakup was suggested to explain the production of \ce{O-} ions with the observed kinetic energies. 

In the last few years, there have been ab initio calculations of the water anion state potential energy surfaces by Haxton et al. \cite{c3h1,c3h2,c3h3,c3h4,c3h5,c3h6,c3h7} assuming \ce{^{2}B1}, \ce{^{2}A1} and \ce{^{2}B2} symmetries. They determined the various two body dissociation channels from the topology of the potential energy surfaces and calculated the cross sections for each of these channels taking into account the initial vibrational and rotational states of the neutral water molecule target. In the case of the first \ce{^{2}B1} resonance, they also calculated the angular dependence of the electron attachment cross section and determined the \ce{H-} angular distribution which agreed perfectly with the existing measurements \cite{c3trajmar,c3belic}. They could also reproduce the $100^{\circ}$ peak and the asymmetry in the angular distribution in their calculation which they explain as mixing of partial waves. For the second and third resonances, they also take into account the Renner-Teller coupling effects (\ce{^{2}A1} and \ce{^{2}B1}) and the conical intersection (\ce{^{2}B2} and \ce{^{2}A1}) of the anion surfaces as they strongly determine the nuclear dynamics and consequently the cross sections. In one of their very recent reports \cite{c3h7}, they showed that the three body breakup is a major component of the observed cross sections in the \ce{O-} channel at the second (\ce{^{2}A1}) and third resonance (\ce{^{2}B2}) by subtracting the two-body breakup cross sections from the total cross section at the resonances. 

In the angular distribution calculations, they emphasized the role of the entrance channel amplitude in determining the electron attachment at a particular orientation \cite{c3h4}. The entrance amplitude for dissociative attachment is analogous to the dipole matrix element which controls the amplitude for photodissociation. The entrance amplitude, depends on two angles instead of one as in photodissociation. This function corresponds to the matrix element between a discrete resonance state and a background scattering wave function for an electron incident on the initial target state. The entrance amplitude depends upon the initial orientation of the molecule with respect to the incident electron beam and this dependence leads to an angular dependence in the cross section, even after it is averaged over the random orientations of the molecule with respect to the incident electron direction. Another factor is the "axial recoil" nature of the dissociation. Under, the axial recoil approximation, the orientation of the molecule in terms of the coordinates $\theta$ and $\phi$ does not change as the dissociation occurs i.e., the recoil axis which connects the atom and the diatom center of mass does not rotate during the dissociation. If this is the case, then the probability for producing dissociative fragments at a certain orientation is the same as the probability for attachment at that orientation (if the survival probability for the dissociative species is unity). If the approximation does not apply, then the dependence of the entrance amplitude upon the initial orientation of the molecule will effectively be spread over a range of final orientations, and final state- specific angular dependences are much less likely. The attachment probability is directly related to the laboratory-frame distribution when the axial recoil condition is met, requiring in the present context that the recoil axis which connects the atom and the diatom center of mass does not rotate during the dissociation. Using this procedure, the entrance amplitude has been shown to reflect the underlying shape of the \ce{1b1}, \ce{3a1} and \ce{1b2} orbitals of neutral water, from which one electron is excited into the \ce{4a1} orbital to form the \ce{^{2}B1}, \ce{^{2}A1} and \ce{^{2}B2} Feshbach resonance configurations in a very recent paper \cite{c3adaniya}. This paper also describes experiments very similar to ours and reports kinetic energy and angular distribution measurements of \ce{H-} and \ce{O-} ions especially on the second resonance. Supported with calculations of angular distribution based on axial recoil approximation and classical trajectory simulations, it was shown that the resonance at 8.5 eV undergoes structural changes due to bending mode vibrations before dissociation thus distorting the angular distributions. 

Apart from electron attachment based studies, there have been studies/ measurements of the optical absorption and electron energy loss spectra of water that strongly facilitate understanding DEA dynamics. Mota et al. \cite{c3mota} measured high resolution VUV photoabsorption spectra of water between 6 and 11 eV. Their data identifies the two absorption bands centered at 7.447 and 9.672 eV due to \ce{A1 -> B1} and \ce{A1 -> A1} transition respectively having mixed valence/Rydberg character and these excited states are known to be the parent states of DEA resonances in water at 6.5 and 8.5 eV. Electron impact measurements by Chutjian et al. \cite{c3chutjian} have revealed dipole allowed \ce{^{3,1}B1} (\ce{b1 -> 3sa1}) states centered at 7.0 and 7.4 eV (parent states of the \ce{^{2}B1} resonance anion), \ce{^{3,1}A1} (\ce{3a1 -> 3sa1}) states at 9.3 and 9.7 eV, optically forbidden transitions like \ce{^{3,1}A2} (\ce{b1 -> 3pb2}) at 8.9 and 9.1 eV and many others. A recent theoretical analysis by Rubio et al. \cite{c3rubio} on the singlet and triplet excited states of water arising from the \ce{1b1} and \ce{3a1} molecular orbital, reveals that \ce{^{3,1}B1} and \ce{^{3,1}A2} states are the lowest occurring electronic excited states with mixed valence/Rydberg character.

\section{Angular distribution of fragments under $C_{2v}$ symmetry}

Water has \ce{C_{2v}} symmetry and the various symmetry states or representations (labelled \ce{A1}, \ce{A2}, \ce{B1} and \ce{B2}) are defined by the four operations - Identity (I), rotation by $180^{\circ}$ about the symmetry axis (\ce{C2}), reflection in the molecular plane ($\sigma_{v}$) and reflection in the plane perpendicular to the molecular plane containing the symmetry axis ($\sigma_{v}^{\prime}$). The character table for \ce{C_{2v}} point group and the basis functions written in terms of spherical harmonics transforming irreducibly under this group are given in Table \ref{tab3.2}. 

\begin{table}[h]
\caption{\ce{C_{2v}} group character and basis functions}
\begin{center}
\begin{tabular}{cccccc}
\hline
\\
 & I & \ce{C2} & $\sigma_{v}$ & $\sigma_{v}^{\prime}$ & Basis Functions\\
\\
\hline
\\
\ce{A1} & 1 & 1 & 1 & 1 & $Y_{l}^{0}$ or $Y_{l}^{m} + Y_{l}^{-m}$; m even \\
\\
\ce{A2} & 1 & 1 & -1 & -1 & $Y_{l}^{m} - Y_{l}^{-m}$; m even \\
\\
\ce{B1} & 1 & -1 & -1 & 1 & $Y_{l}^{m} + Y_{l}^{-m}$; m odd \\
\\
\ce{B2} & 1 & -1 & 1 & -1 & $Y_{l}^{m} - Y_{l}^{-m}$; m odd \\
\\
\hline
\end{tabular}
\end{center}
\label{tab3.2}
\end{table}

The method to calculate the angular distribution of fragments arising from a two-body like breakup for a polyatomic molecule of a particular point group has been described by Azria et al. \cite{c3azria}. We proceed on those lines to determine the angular distribution for \ce{H-} ions produced from dissociation of water anion of various symmetries under $C_{2v}$ group for partial waves (characterized by $l$ values) whose capture is allowed. The quantity we are interested in calculating is $|<Negative ion state|Partial wave ($l$)|Initial State>|^{2}$. Here, the initial state is the electronic ground state of water molecule i.e. \ce{A1}. We make use of the $Y_{0}^{0}$ function to represent this ground state. The detailed steps leading to the angular distribution expression is illustrated for the case of \ce{A1} (neutral) to \ce{B1} (negative ion) transition for the capture of $p$-wave ($Y_{1}^{0}$). From Table \ref{tab3.2}, the basis function for a \ce{B1} state is given by $Y_{1}^{1} + Y_{1}^{-1}$. Assuming that the dissociating H-OH bond is oriented at Euler angles $(\phi,\theta,0)$ with respect to the electron beam direction; we transform the partial wave $Y_{1}^{0}$ from electron beam frame (or lab frame) along the dissociation axis. Appendix A lists the expressions for all $Y_{l}^{m}$s upto $l=3$ rotated by Euler angles $(\phi,\theta,0)$ leading to transformation from lab frame to dissociation frame. The expression arising out of the transformation of $Y_{1}^{0}$ is given by equation (A.3) in Appendix A. Similarly, we transform the basis function $Y_{1}^{1} + Y_{1}^{-1}$ defined for the molecular frame along the dissociation axis by Euler angles $(0,\beta,0)$ using the equations (A.2) and (A.4) in Appendix A. Here, $\beta$ is the angle between the molecular symmetry axis ($C_{2}$ axis) and the dissociating O-H bond. Assuming the ground state geometry of water molecule, $\beta$ is half of H-O-H bond angle i.e. $52.5^{\circ}$. Rewriting the terms appropriately and simplifying the expression, the transition amplitude $f(\theta,\phi)$ is given by: 

\begin{eqnarray}
f(\theta,\phi) & \propto & <Y_{1}^{1} + Y_{1}^{-1}| Y_{1}^{0} | Y_{0}^{0} > \mbox{all terms in dissociation frame} \nonumber \\
& \propto & \iota \sqrt{2} \sin\theta \sin\phi
\end{eqnarray}

The scattering intensity $I_{p}(\theta)$  is given by:

\begin{eqnarray}
I_{p}(\theta) & \propto & \displaystyle\int_{0}^{2\pi} |f|^{2} d\phi \nonumber \\
 & \propto & 2 \sin^{2}\theta
\end{eqnarray}

Similarly, the expression for angular distribution due to a $d$-wave ($Y_{2}^{0}$) capture under \ce{B1} state ($Y_{2}^{1} + Y_{2}^{-1}$) is given by

\begin{equation}
I_{d}(\theta) \propto \frac{3}{2} (0.64 \sin^{4}\theta + 0.36 \sin^{2}2\theta)
\end{equation}

When more than one partial wave is captured in the process, interference between the individual amplitudes can lead to forward-backward asymmetry in the angular distributions which is generally the case. We obtain a general expression taking into account the interference terms with free parameters and use this expression to fit the experimental data. In the present example of \ce{B1} symmetry, the expression for the scattering intensity $I_{p+d}(\theta)$ taking both $p$ and $d$ partial waves is given by

\begin{equation}
I_{p+d}(\theta) \propto 2 a^{2} \sin^{2}\theta + 1.5 b^{2}(0.64 \sin^{4}\theta + 0.36 \sin^{2}2\theta) + 2ab(1.05)\sin\theta \sin2\theta \cos\delta            
\end{equation}

where $a$, $b$ and $\delta$ are the fitting parameters. The origin of $\delta$ arises from the difference in the phases of the two partial waves due to potential scattering.

\begin{figure}[!ht]
\centering
  \subfloat[\ce{A1 -> A1}]{\includegraphics[width=0.4\columnwidth]{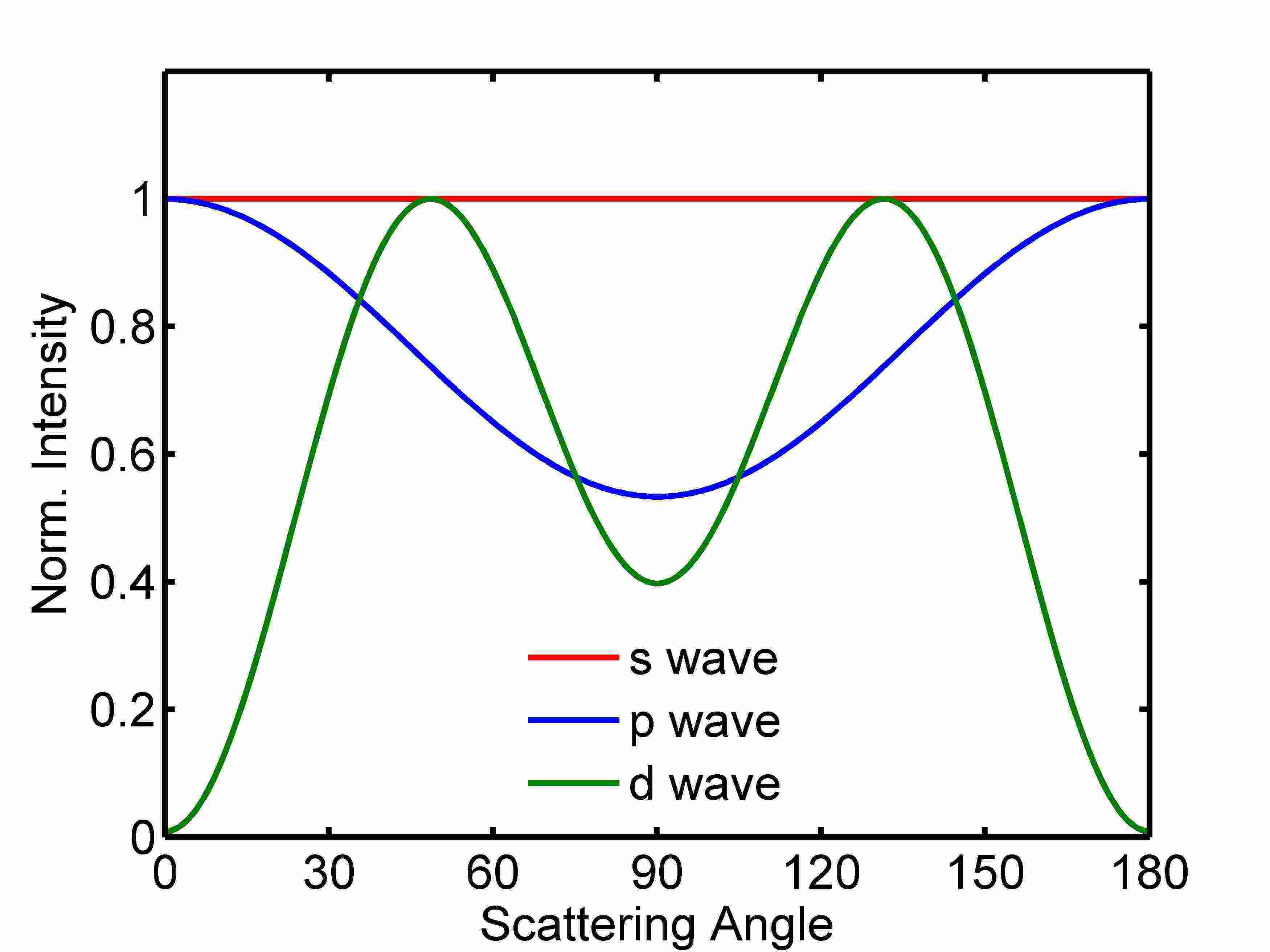}}        
  \subfloat[\ce{A1 -> A2}]{\includegraphics[width=0.4\columnwidth]{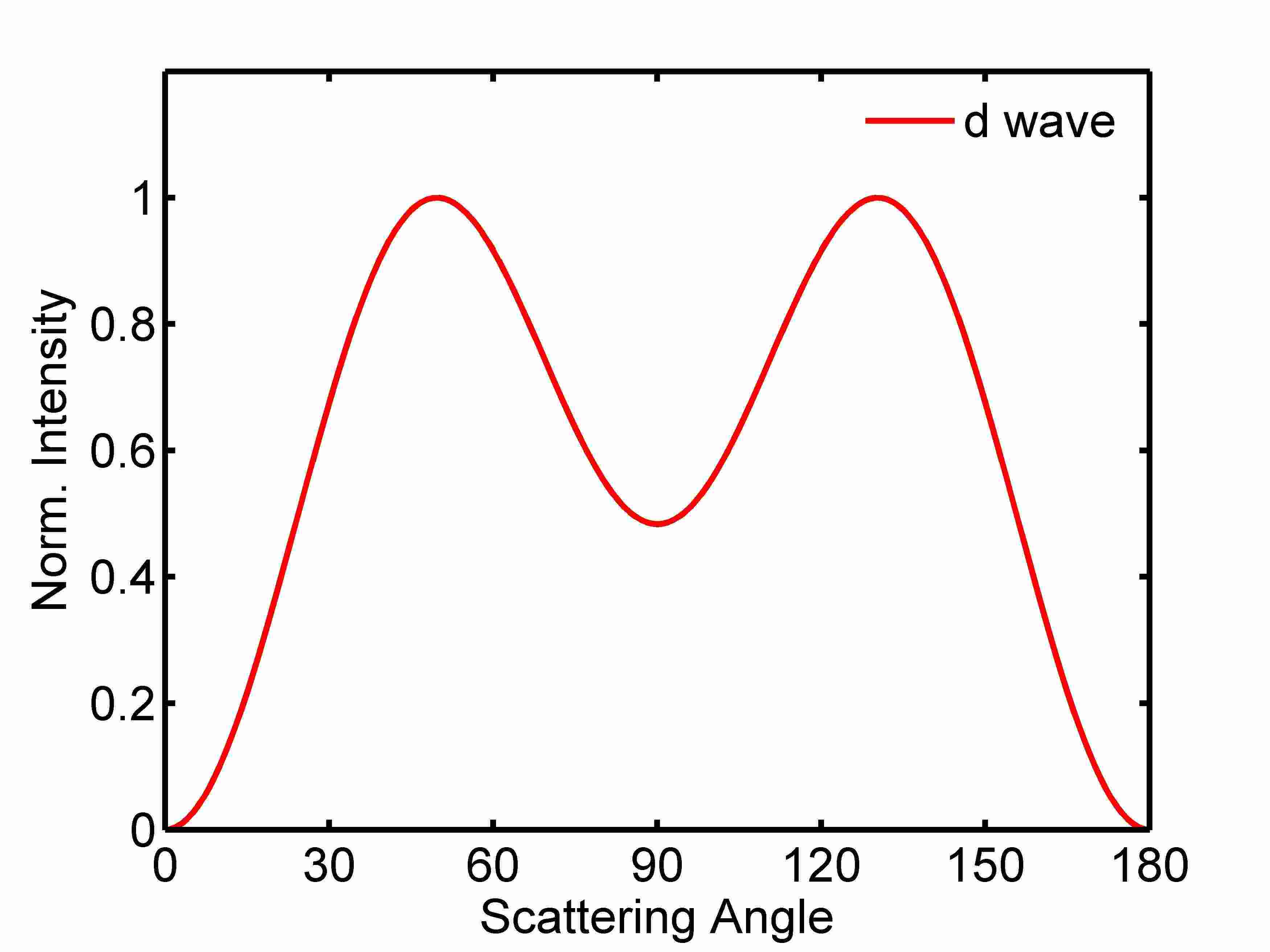}}\\
  \subfloat[\ce{A1 -> B1}]{\includegraphics[width=0.4\columnwidth]{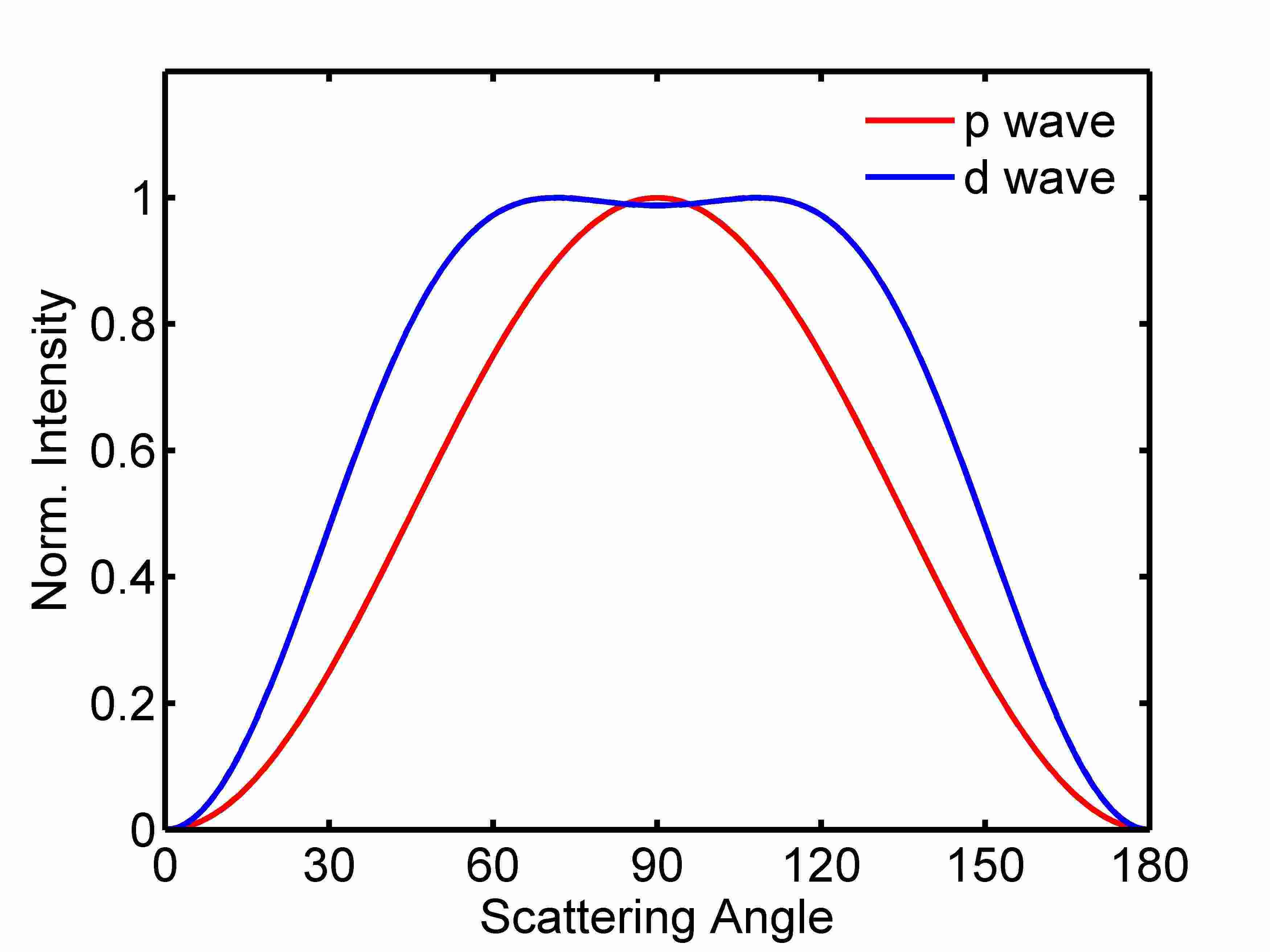}}
  \subfloat[\ce{A1 -> B2}]{\includegraphics[width=0.4\columnwidth]{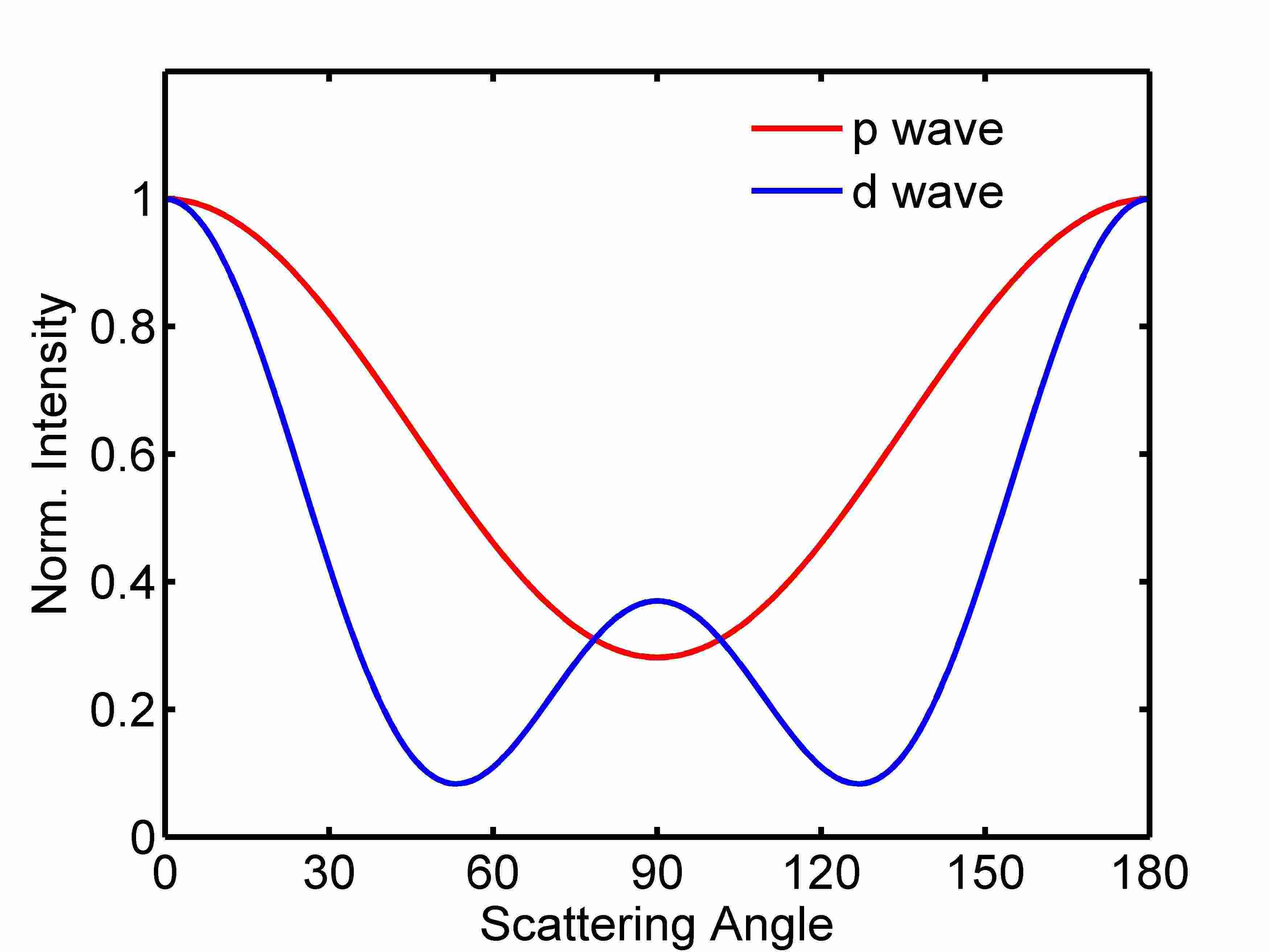}} 
  \caption{ Angular distribution curves for various symmetries under \ce{C_{2v}} point group for the lowest allowed partial waves. } 
  \label{fig3.2}
\end{figure}

The angular curves for other symmetry states - \ce{A1}, \ce{A2} and \ce{B2} are calculated similarly using appropriate basis functions listed in Table \ref{tab3.2} and partial wave expressions and following the steps in the calculation detailed for the case of \ce{B1} above. The plots for individual lowest order partial waves for the four different symmetry states of the resonance when the neutral molecule is in \ce{A1} state, like in the case of water, are given in Figure \ref{fig3.2}. All the curves are normalized to their maximum value. The expressions for scattering intensity for each symmetry and allowed partial wave are listed in Appendix B.

\section{Experimental setup}
The measurements were done using the new Velocity Map Imaging (VMI) spectrometer specially built to study low energy electron - molecule interactions, as described previously \cite{dnandirsi}. In this experiment, a magnetically collimated and pulsed electron beam interacts with an effusive molecular beam formed by a capillary array. The negative ions formed in the interaction region are extracted into a time-of-flight mass spectrometer using a suitable extraction pulse after finite delay. The extraction field and the time-of-flight tube along with a focusing electrode at the entrance of the flight tube provide the needed VMI condition. The ions are detected using a two-dimensional position sensitive detector made of three 50 mm diameter micro-channel plates in Z-stack configuration and a Wedge and Strip anode. The ions striking the detector are recorded individually for their time of arrival (t) and position (x,y) in LIST mode using a CAMAC based data acquisition system running on 'LAMPS' \cite{lamps1,lamps2}. The central slice of the Newton sphere which contains the velocity distribution of the ions with respect to the electron beam direction is obtained by selecting a narrow time window during analysis.

For imaging \ce{H-} ions we had to make small changes from the previous set up in which we imaged mostly \ce{O-} ions since for similar kinetic energies, \ce{H-} ions travel 4 times the distance as \ce{O-}. In order to take this into account, we had to limit the electron gun pulse width as well as the delay between the two pulses. Also, the operating voltages for ion extraction, focusing electrode and flight tube were increased appropriately for maintaining the VMI conditions. The increase in the bias voltage on the focusing electrode was found to affect the electron beam. This was eliminated by having a fine wire mesh on the puller electrode, which flanks the interaction region on the time-of-flight side. The addition of this wire mesh modified the VMI conditions. SIMION simulations along with actual optimization restored the imaging quality. We also found that the magnetic field used for electron beam collimation is affecting the imaging of \ce{H-} ions. Though we reduced the magnetic field strength to minimum level necessary for reasonable electron current, we find that the \ce{H-} ion images have some distortions due to it, which have not been corrected for. However, the \ce{D-} ions could be imaged without distortions. The kinetic energy of the ions is retrieved by calibrating the ion image size in terms of energy using the radius of \ce{O-} from DEA to NO or \ce{O2} at the same experimental conditions whose kinetic energy is known exactly.

We are unable to separate the \ce{O-} and \ce{OH-} masses in our spectrometer. However, as the cross sections of \ce{OH-} at the resonances are lower than \ce{O-} almost by an order of magnitude, it is safe to assume that the angular distribution and kinetic energy distribution belong to \ce{O-}.

\section{Measurements, Results and Discussion}

\begin{figure}[!htp]
\centering
  \subfloat[\ce{H-} at 5.5 eV]{\includegraphics[width=0.25\columnwidth]{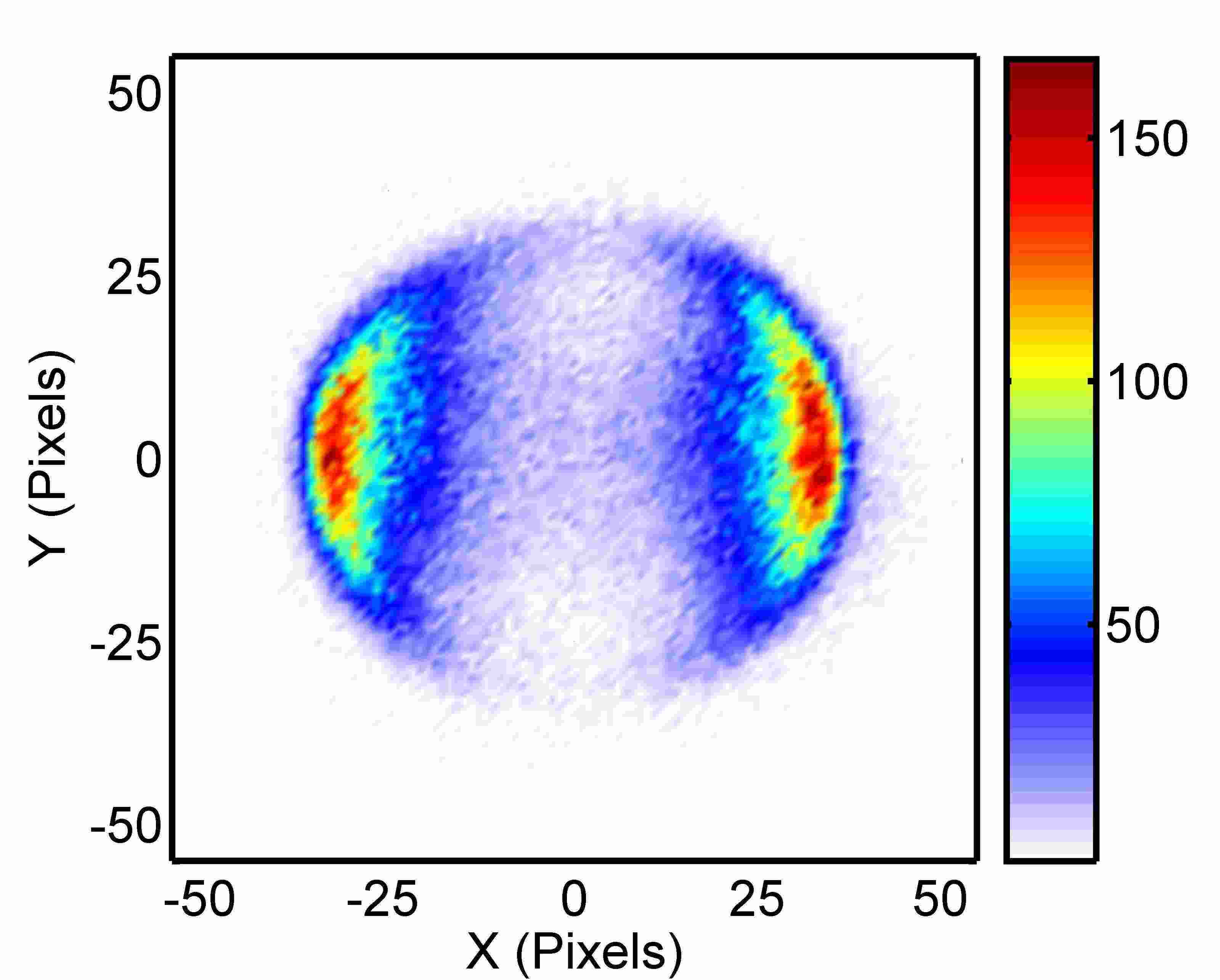}}
  \subfloat[\ce{H-} at 6.5 eV]{\includegraphics[width=0.25\columnwidth]{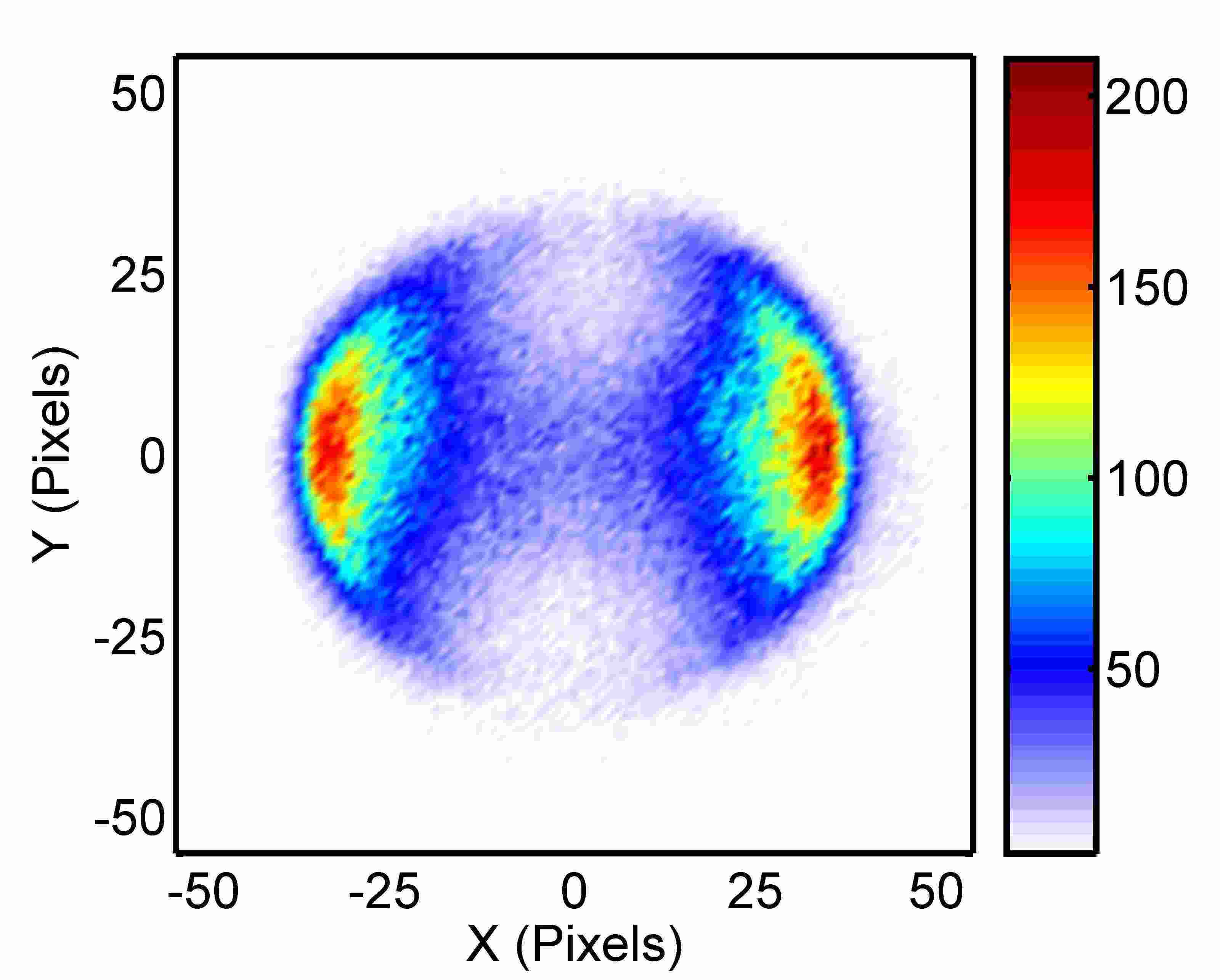}}
  \subfloat[\ce{H-} at 7.5 eV]{\includegraphics[width=0.25\columnwidth]{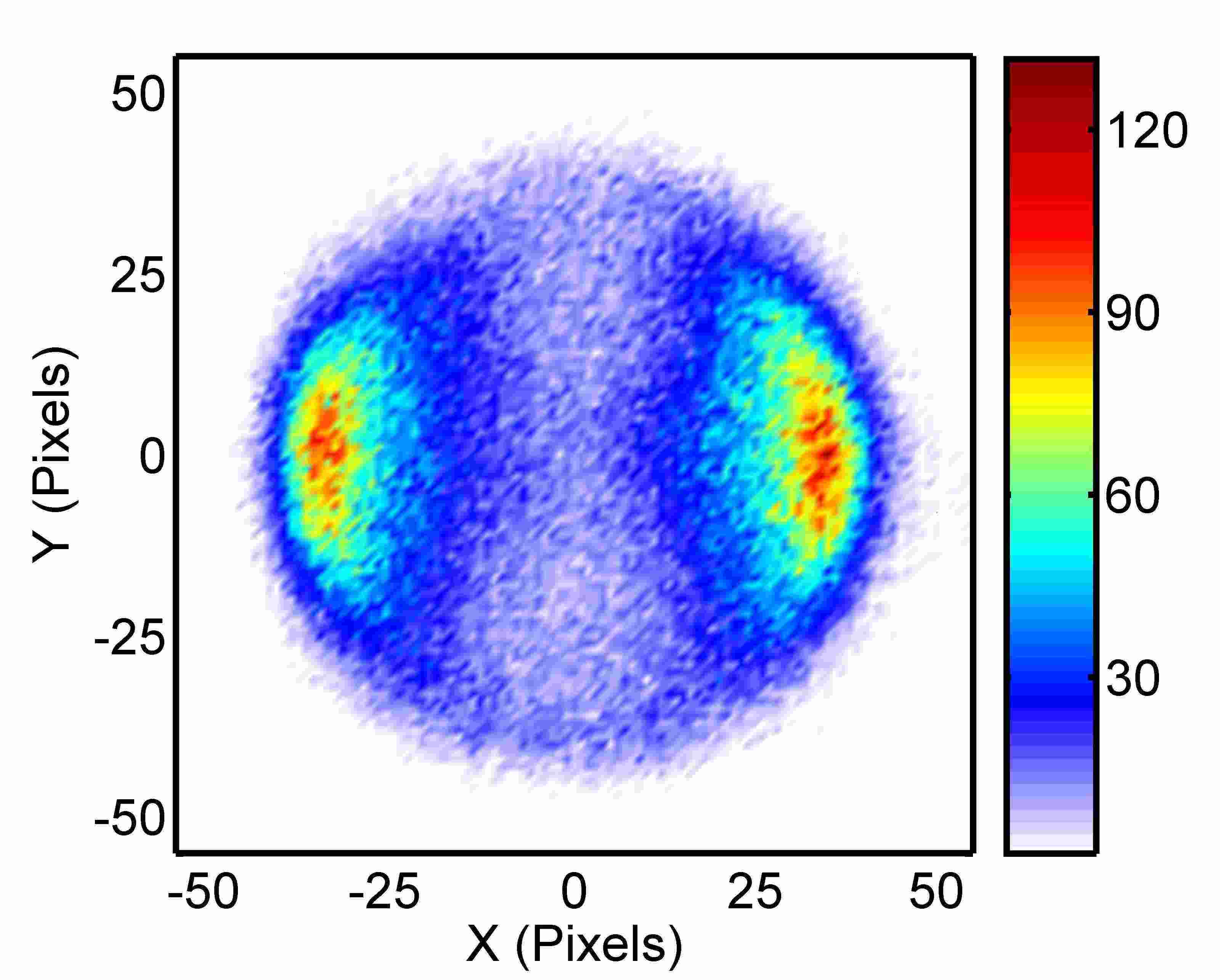}}
  \subfloat[\ce{H-} at 8.5 eV]{\includegraphics[width=0.25\columnwidth]{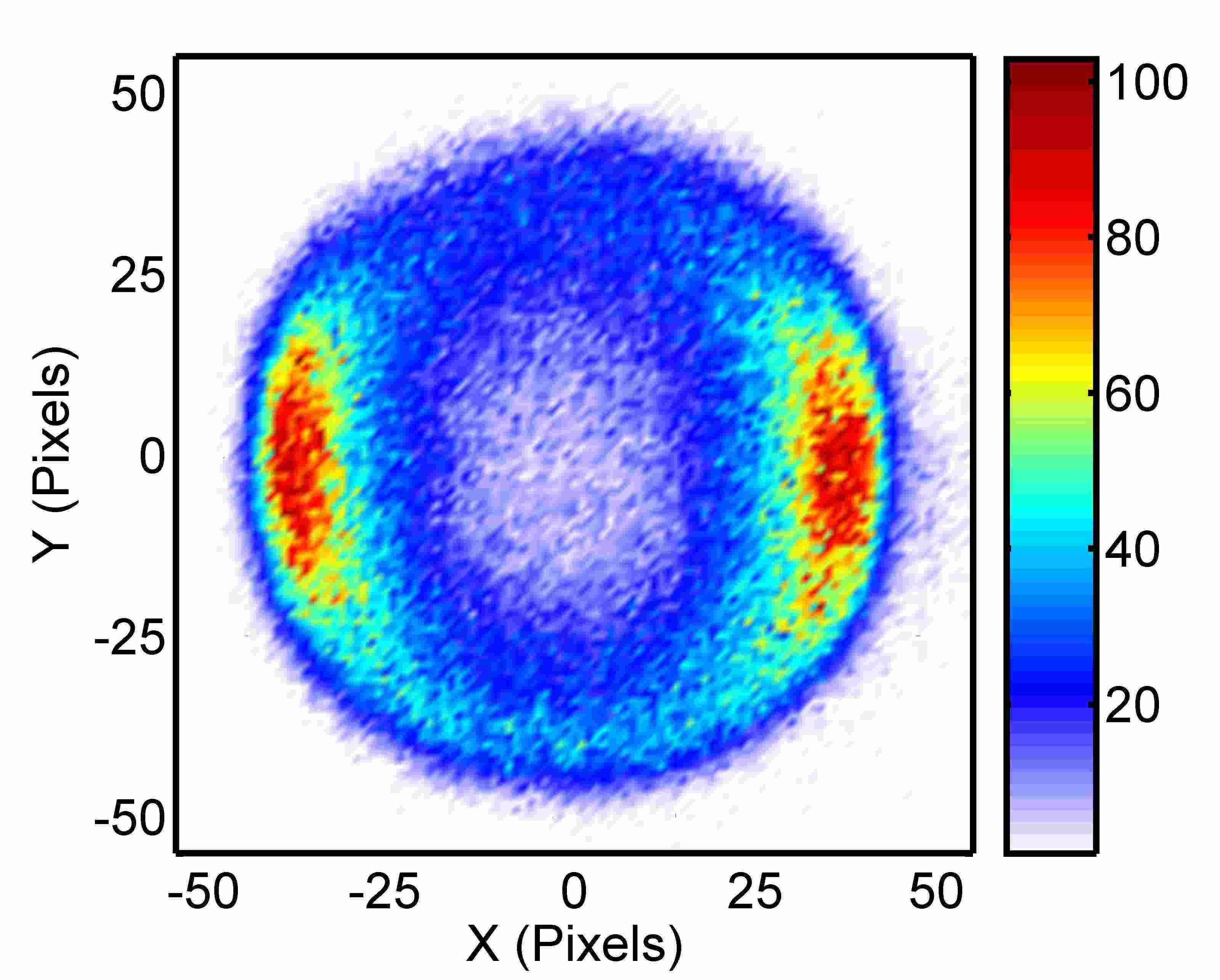}}\\
  \subfloat[\ce{H-} at 9.5 eV]{\includegraphics[width=0.25\columnwidth]{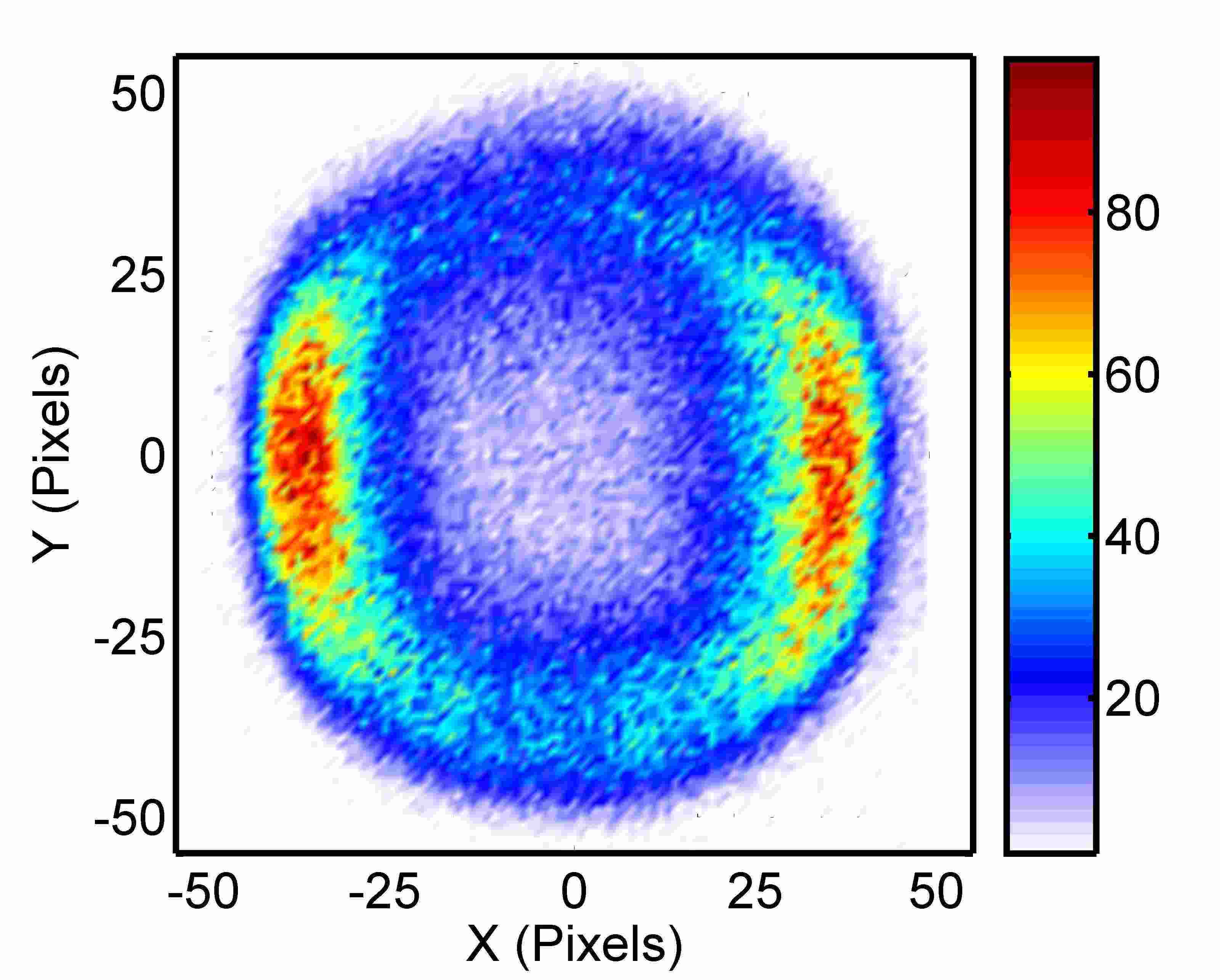}}
  \subfloat[\ce{H-} at 10.8 eV]{\includegraphics[width=0.25\columnwidth]{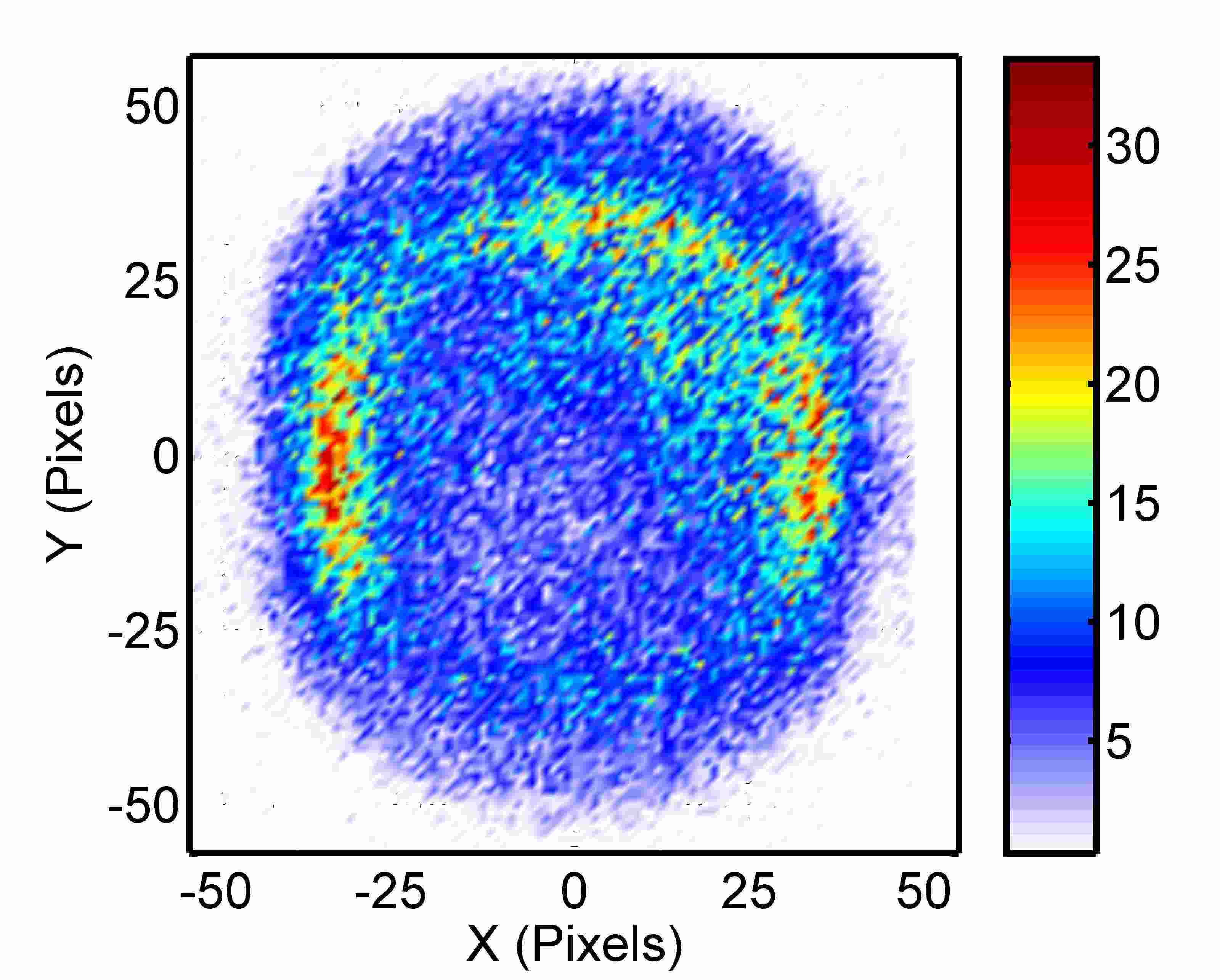}}
  \subfloat[\ce{H-} at 11.8 eV]{\includegraphics[width=0.25\columnwidth]{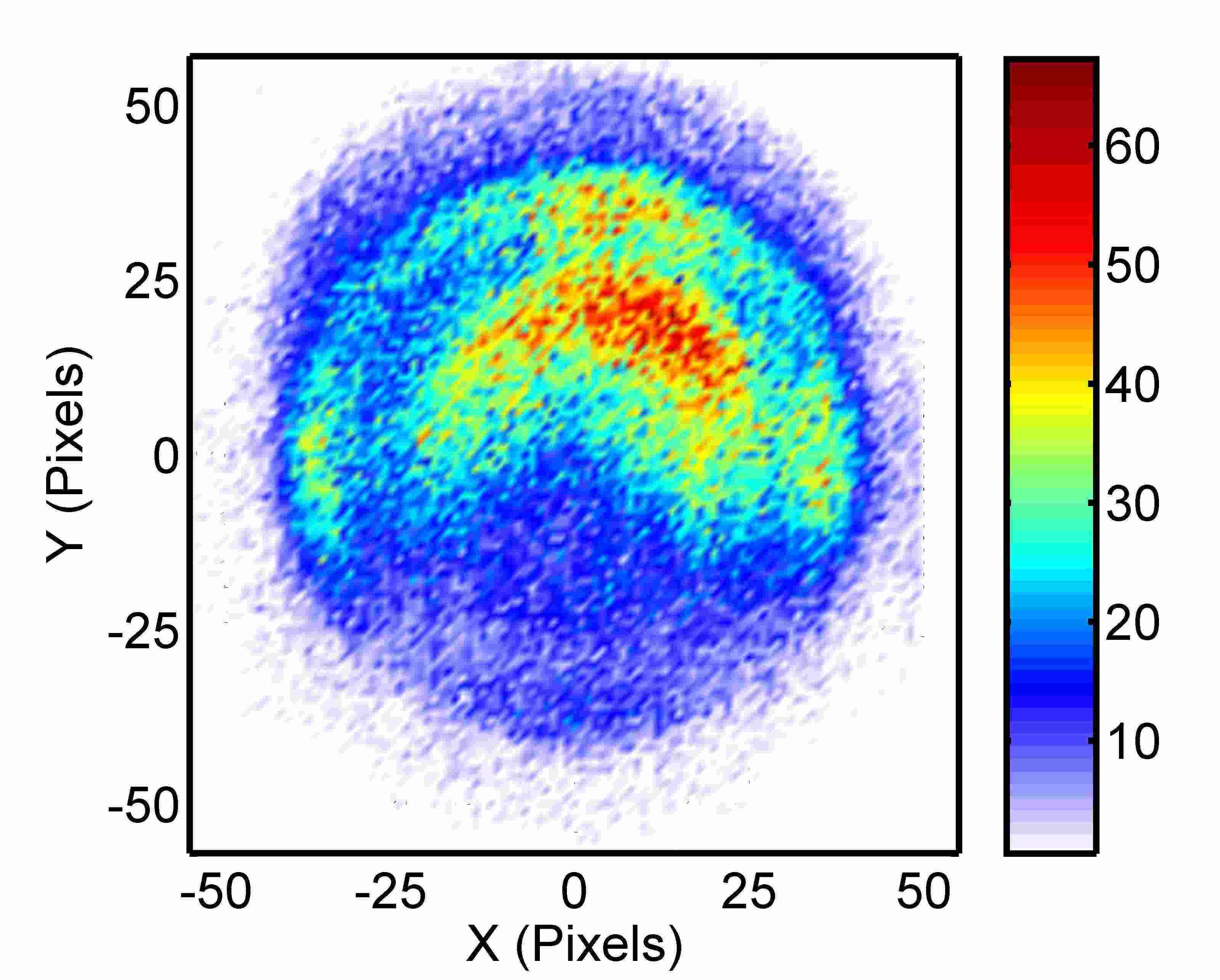}}
  \subfloat[\ce{H-} at 12.8 eV]{\includegraphics[width=0.25\columnwidth]{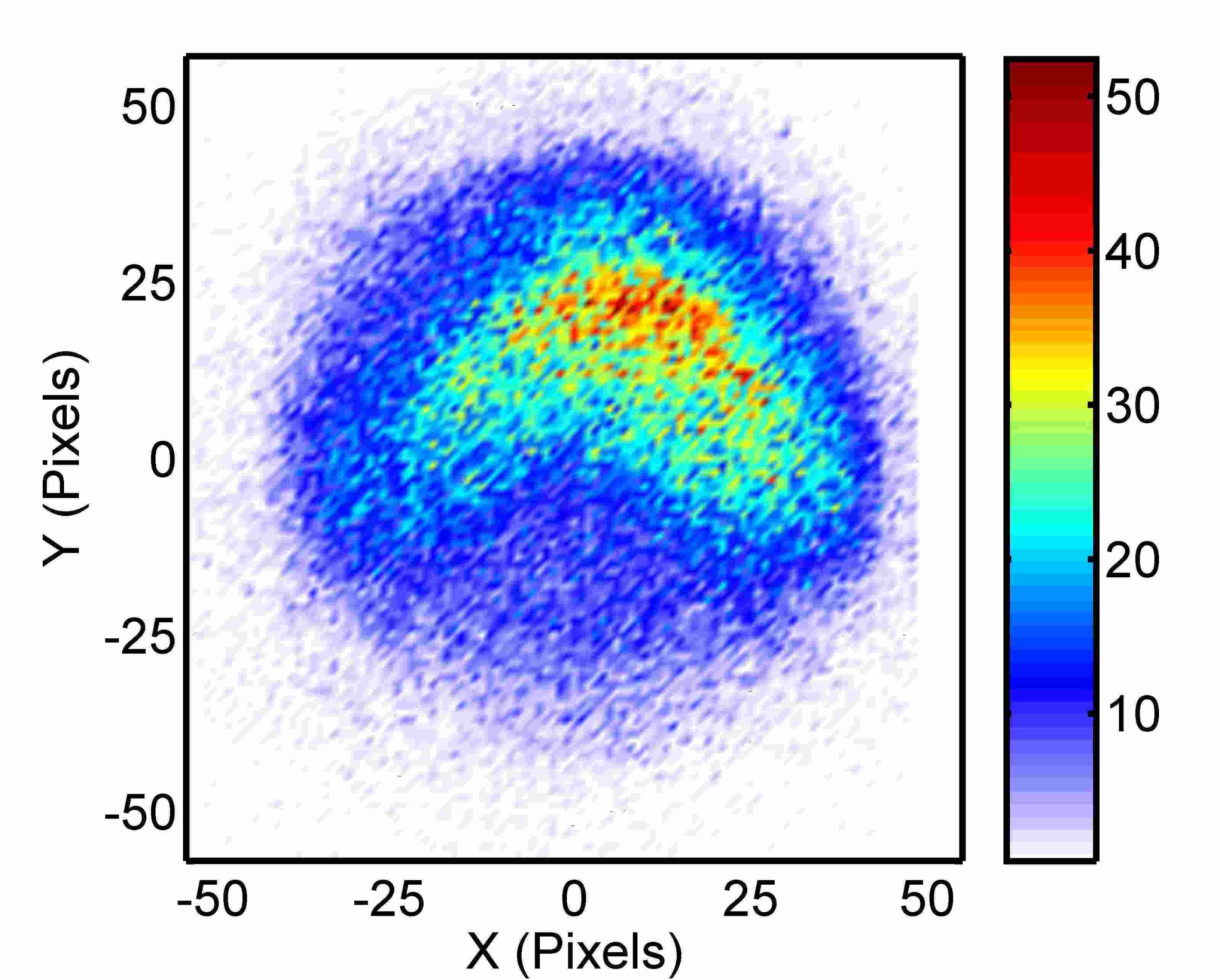}}
 \caption{Velocity images of \ce{H-} ions from DEA to \ce{H2O} at various electron energies}
 \label{fig3.3}
 \end{figure}

\begin{figure}[!h]
\centering
  \subfloat[\ce{O-} at 6 eV]{\includegraphics[width=0.25\columnwidth]{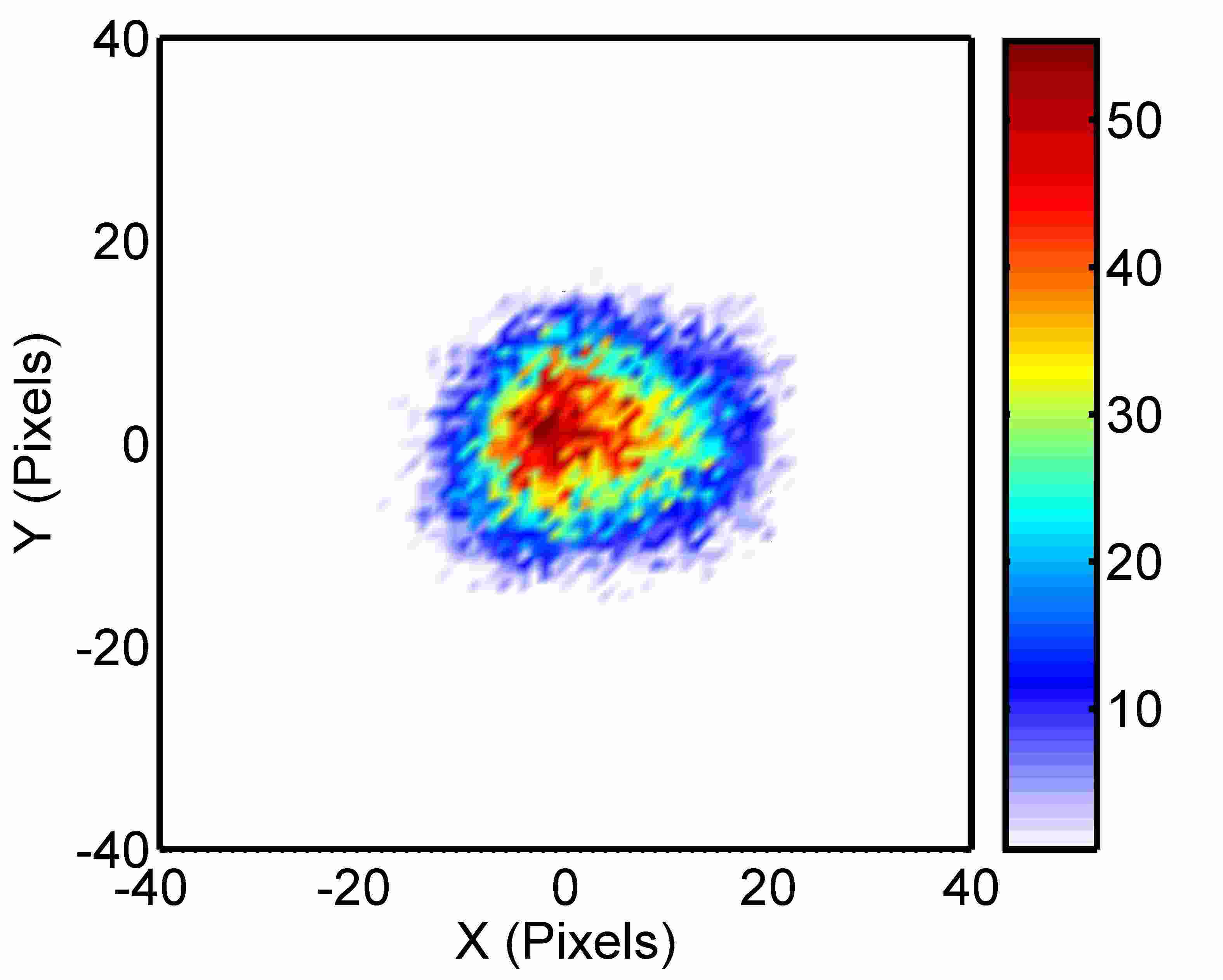}}
  \subfloat[\ce{O-} at 7 eV]{\includegraphics[width=0.25\columnwidth]{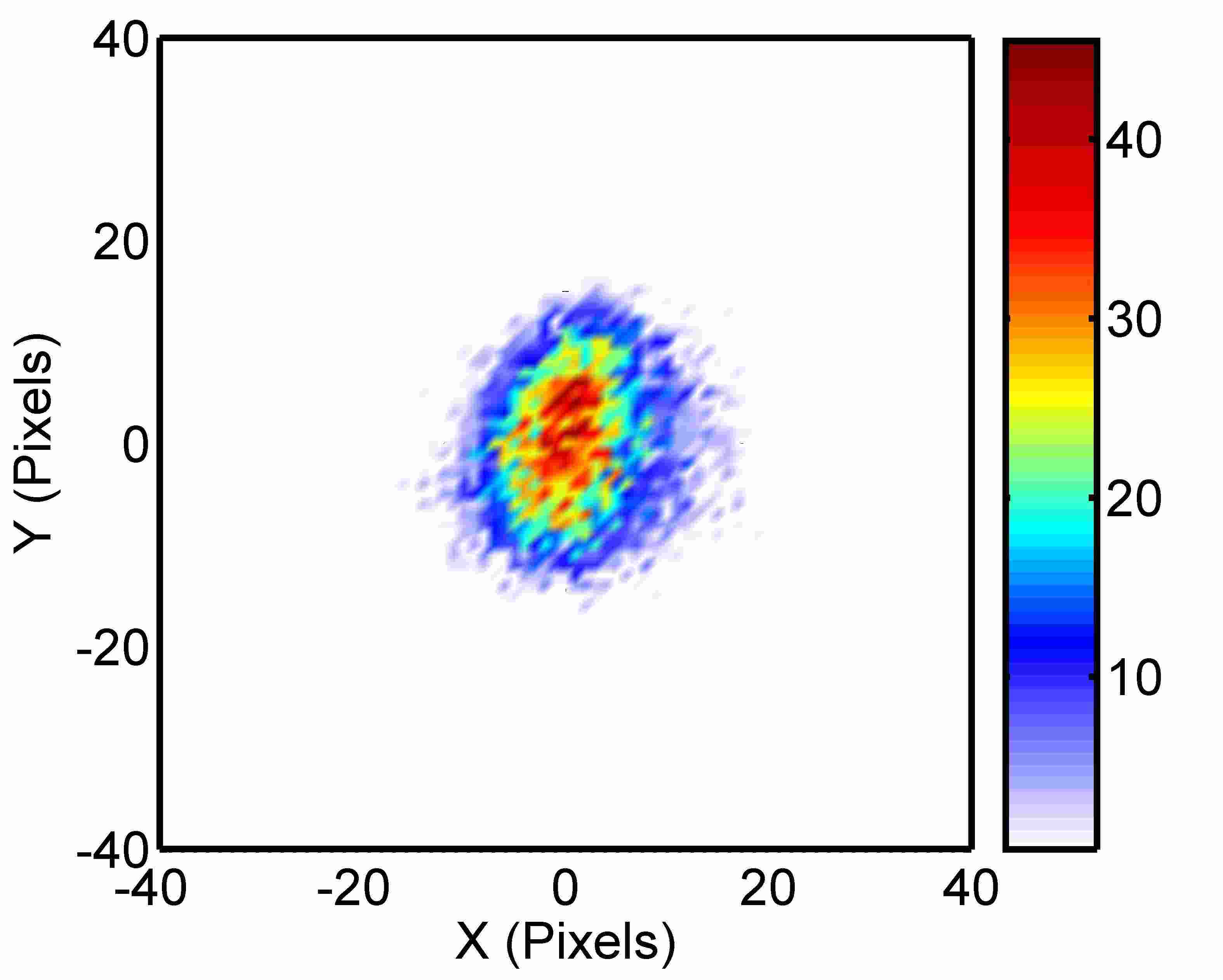}}
  \subfloat[\ce{O-} at 8 eV]{\includegraphics[width=0.25\columnwidth]{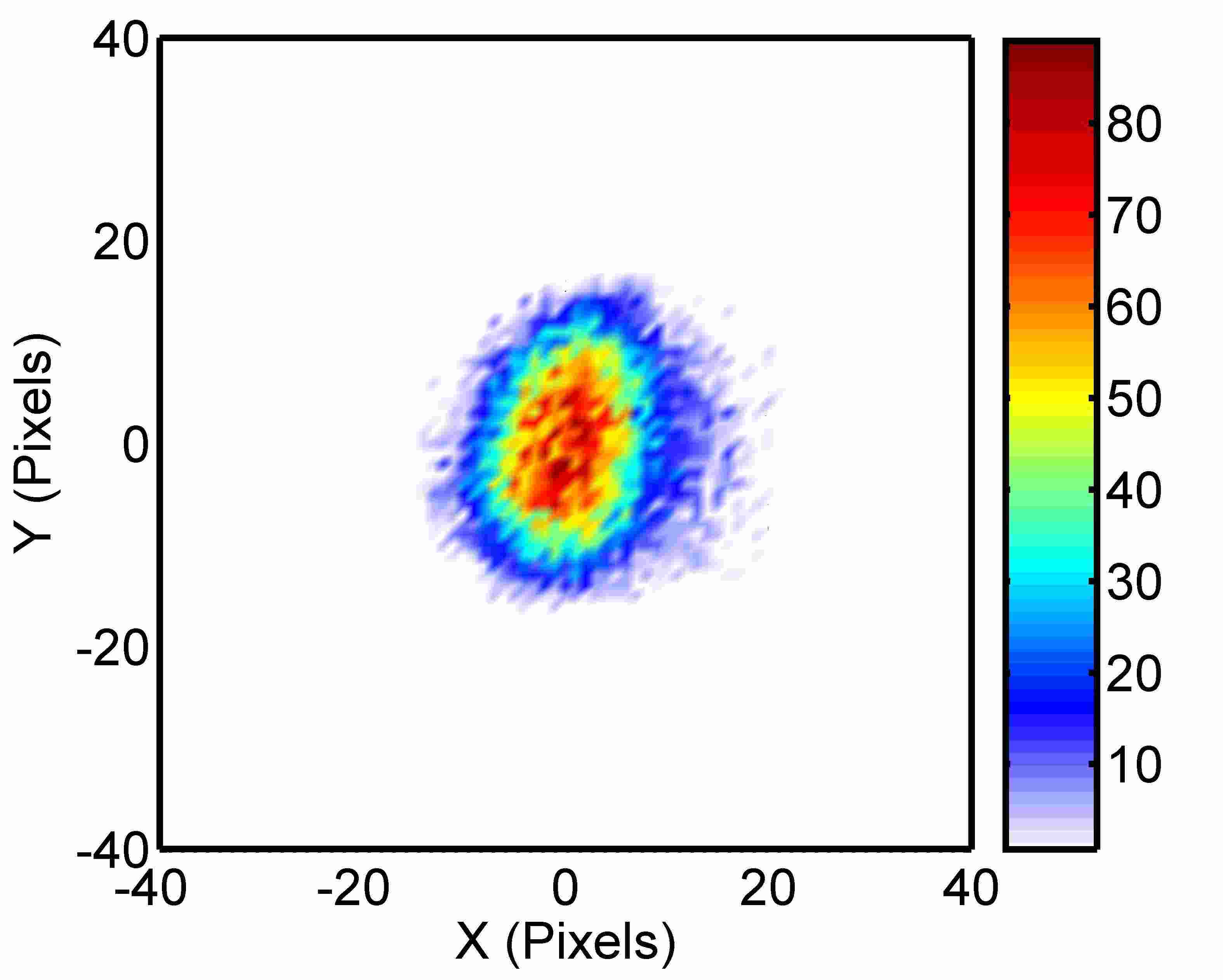}}
  \subfloat[\ce{O-} at 9 eV]{\includegraphics[width=0.25\columnwidth]{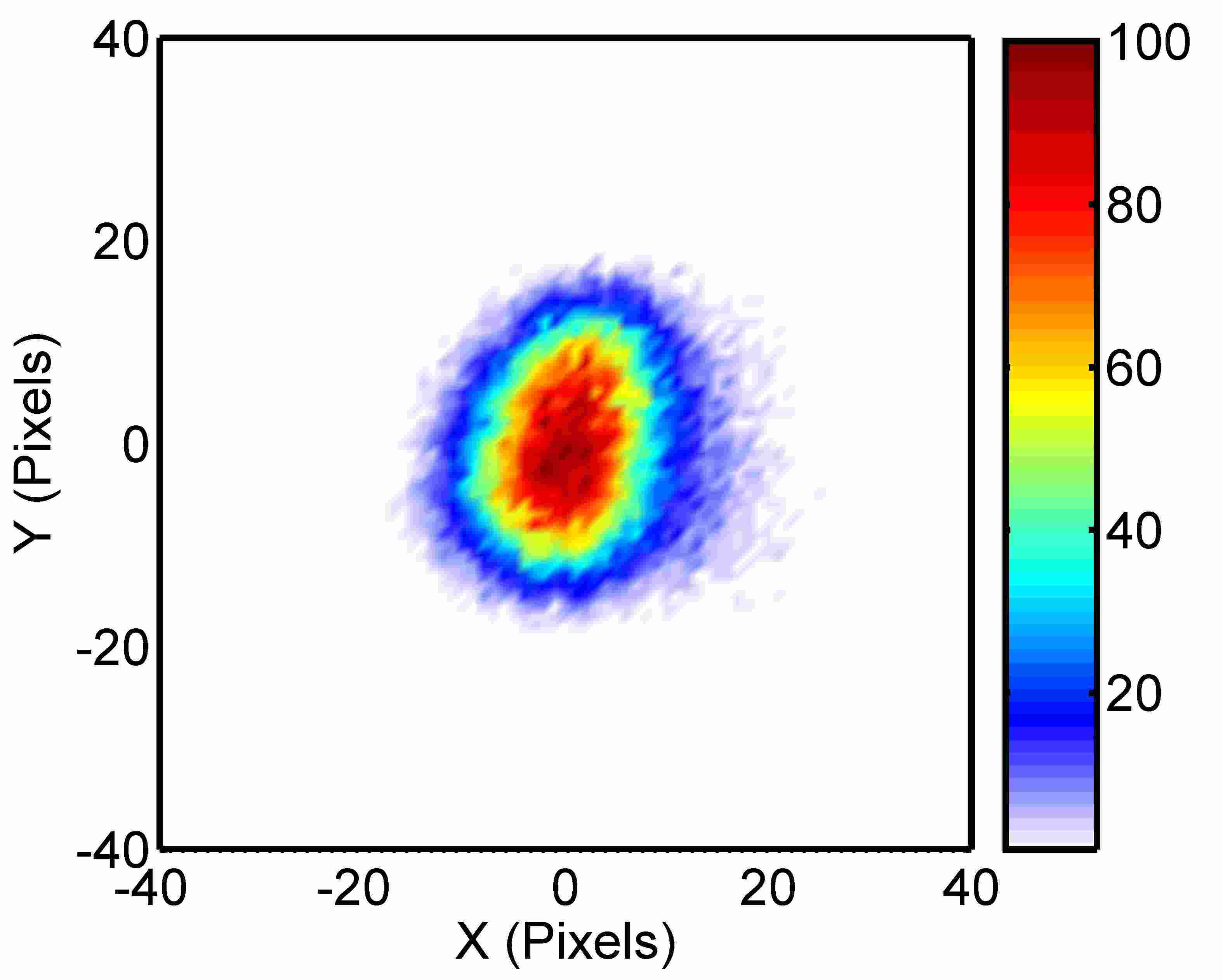}}\\
  \subfloat[\ce{O-} at 10 eV]{\includegraphics[width=0.25\columnwidth]{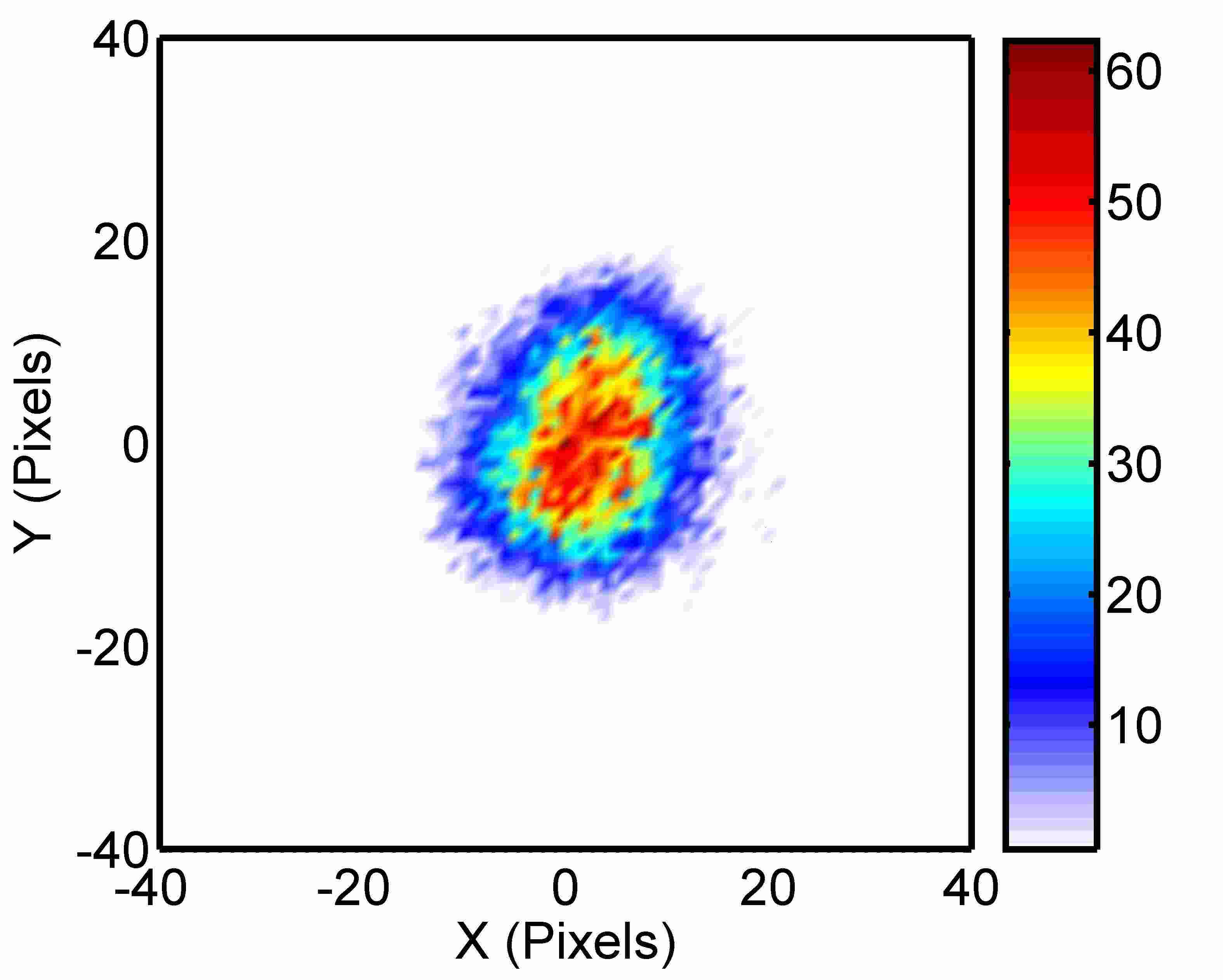}}
  \subfloat[\ce{O-} at 11 eV]{\includegraphics[width=0.25\columnwidth]{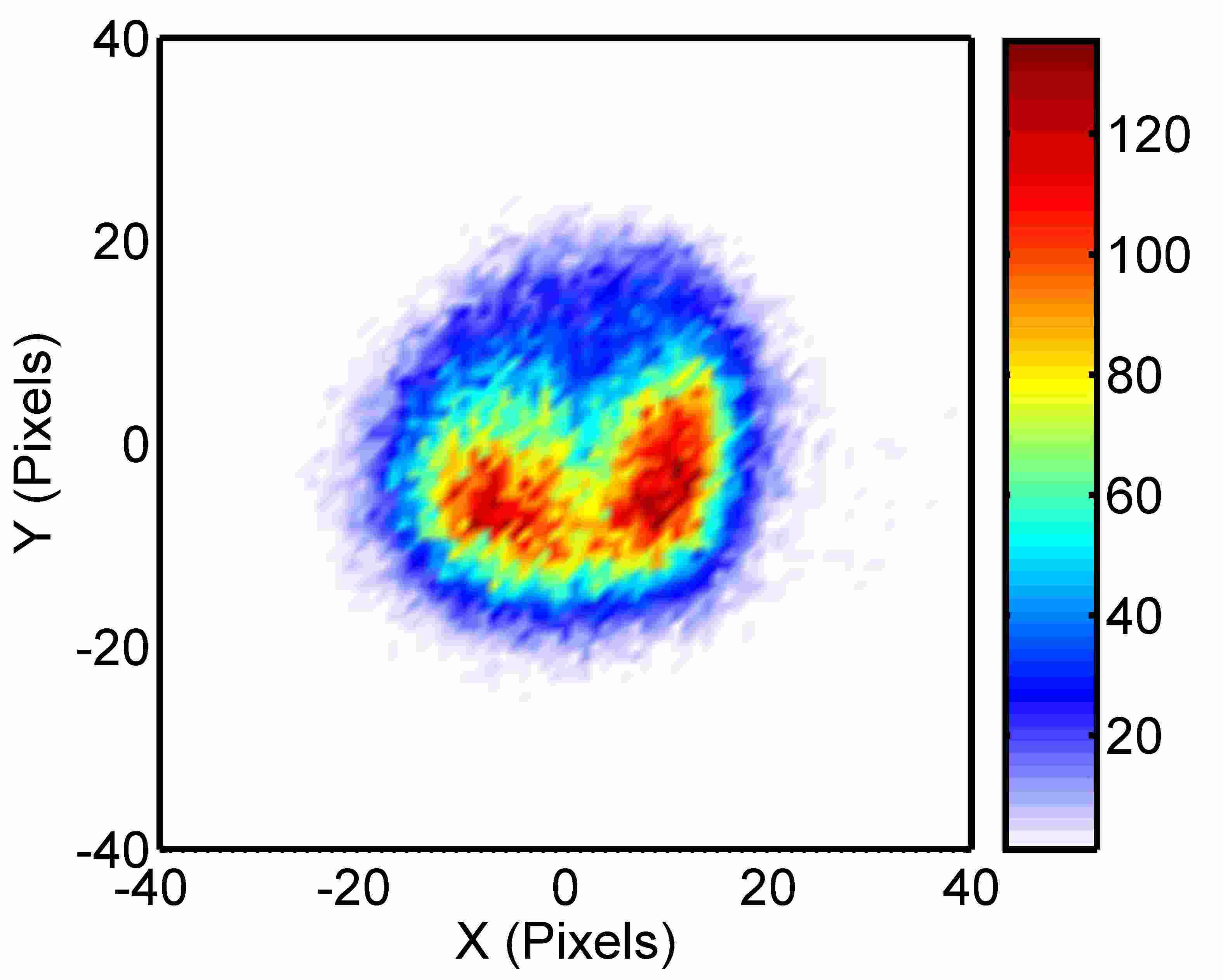}}
  \subfloat[\ce{O-} at 12 eV]{\includegraphics[width=0.25\columnwidth]{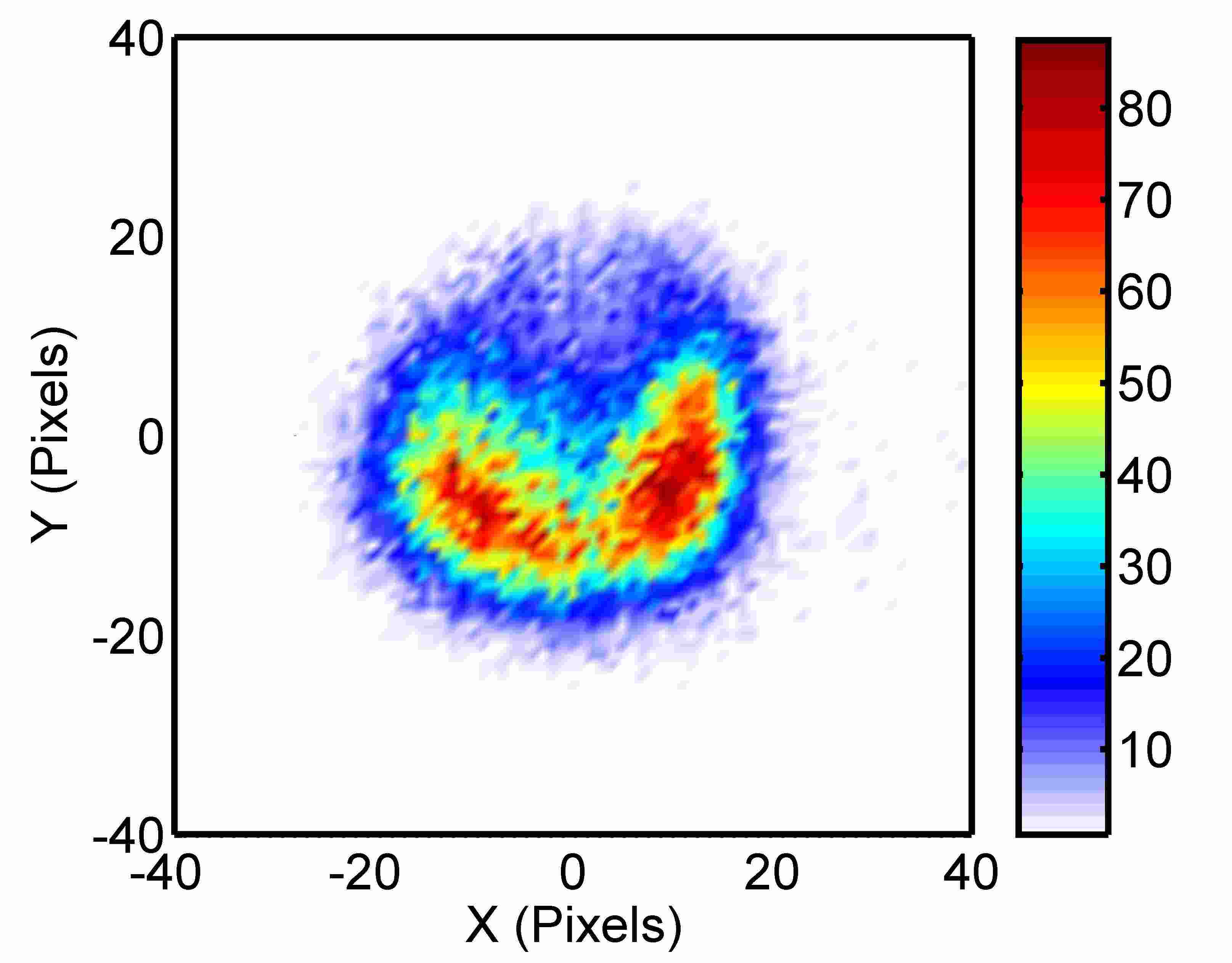}}
  \subfloat[\ce{O-} at 13 eV]{\includegraphics[width=0.25\columnwidth]{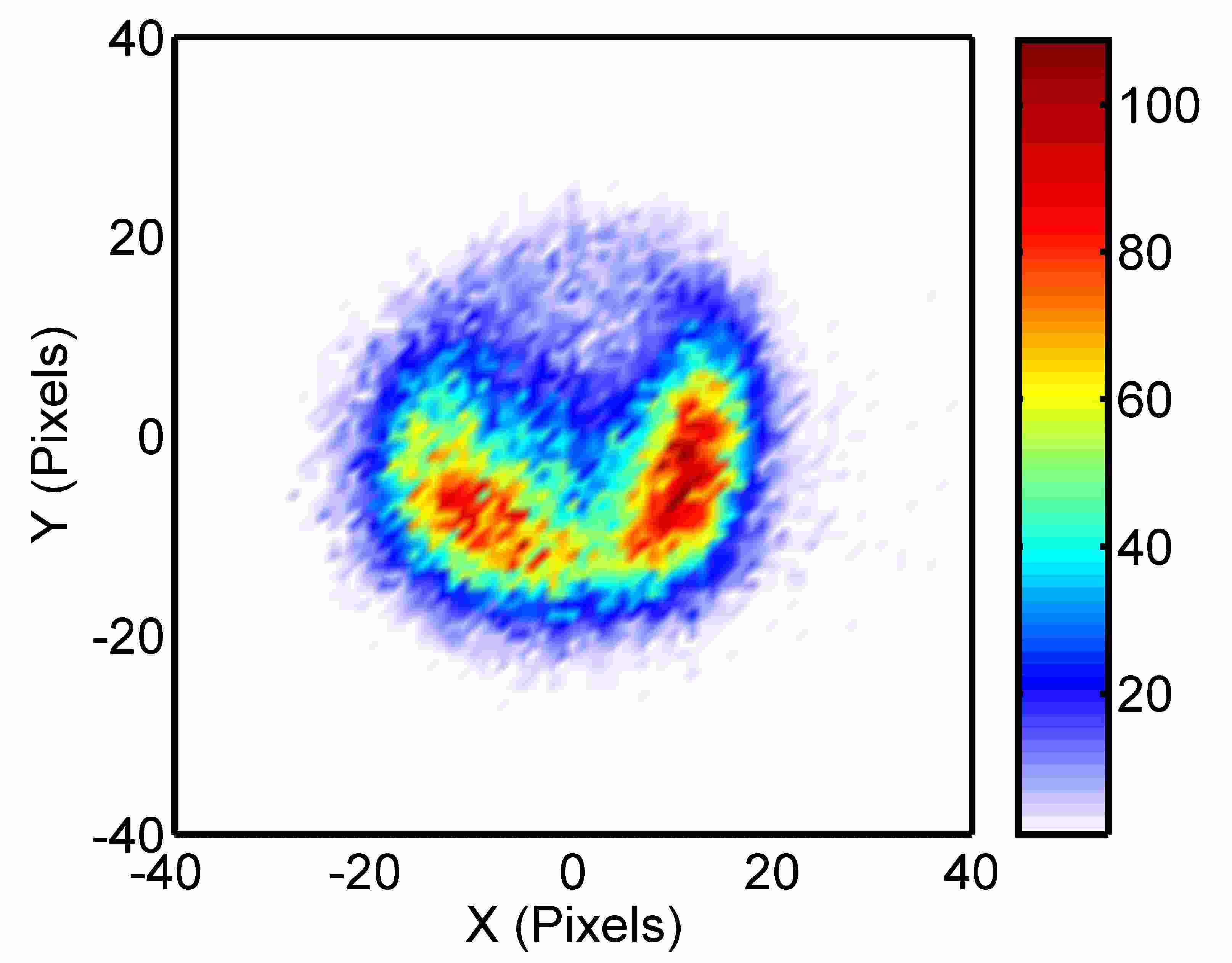}}
 \caption{Velocity images of \ce{O-} ions from DEA to \ce{H2O} at various electron energies}
 \label{fig3.4}
 \end{figure}

\begin{figure}[!h]
\centering
  \subfloat[\ce{D-} at 5.5 eV]{\includegraphics[width=0.25\columnwidth]{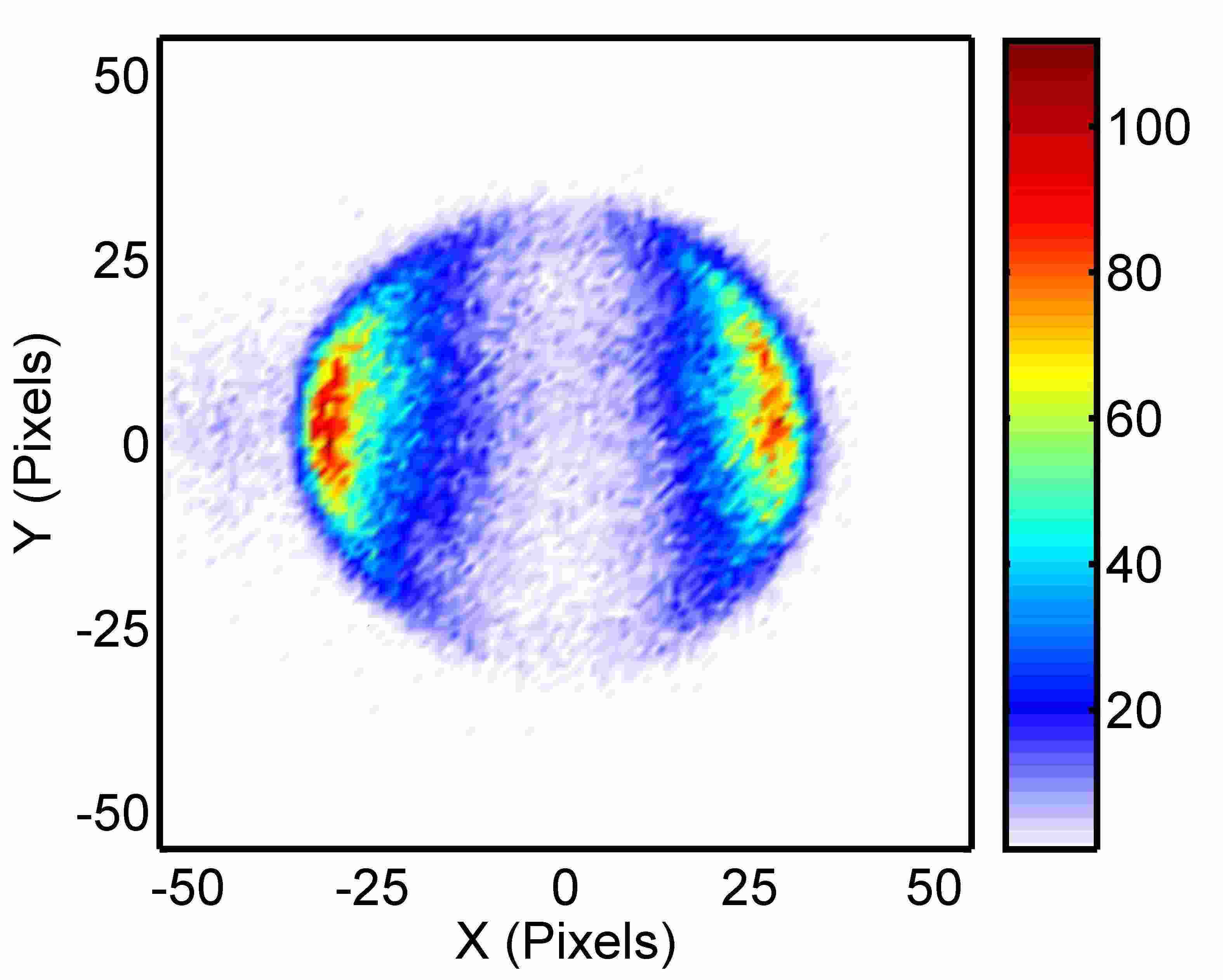}}
  \subfloat[\ce{D-} at 6.5 eV]{\includegraphics[width=0.25\columnwidth]{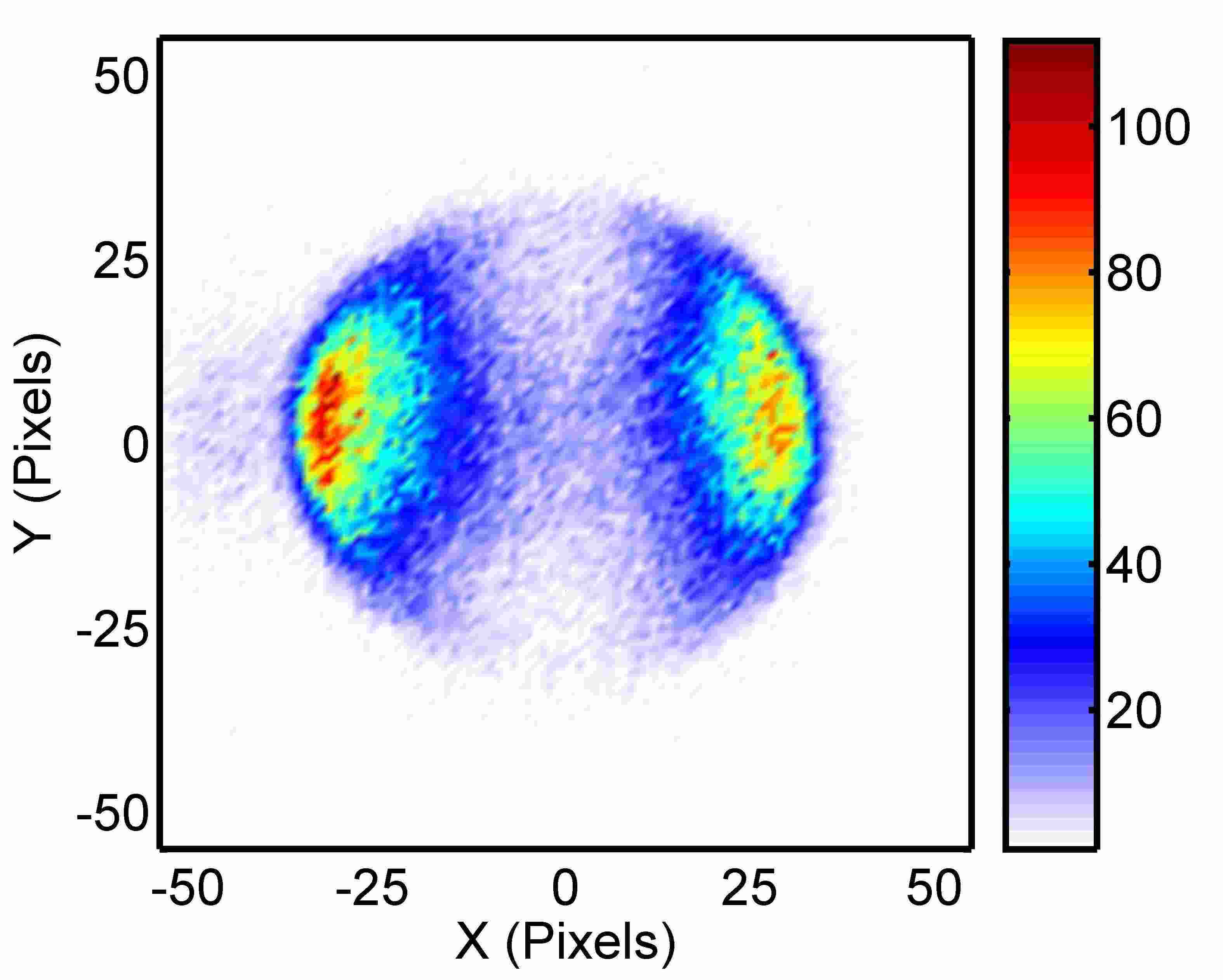}}
  \subfloat[\ce{D-} at 7.5 eV]{\includegraphics[width=0.25\columnwidth]{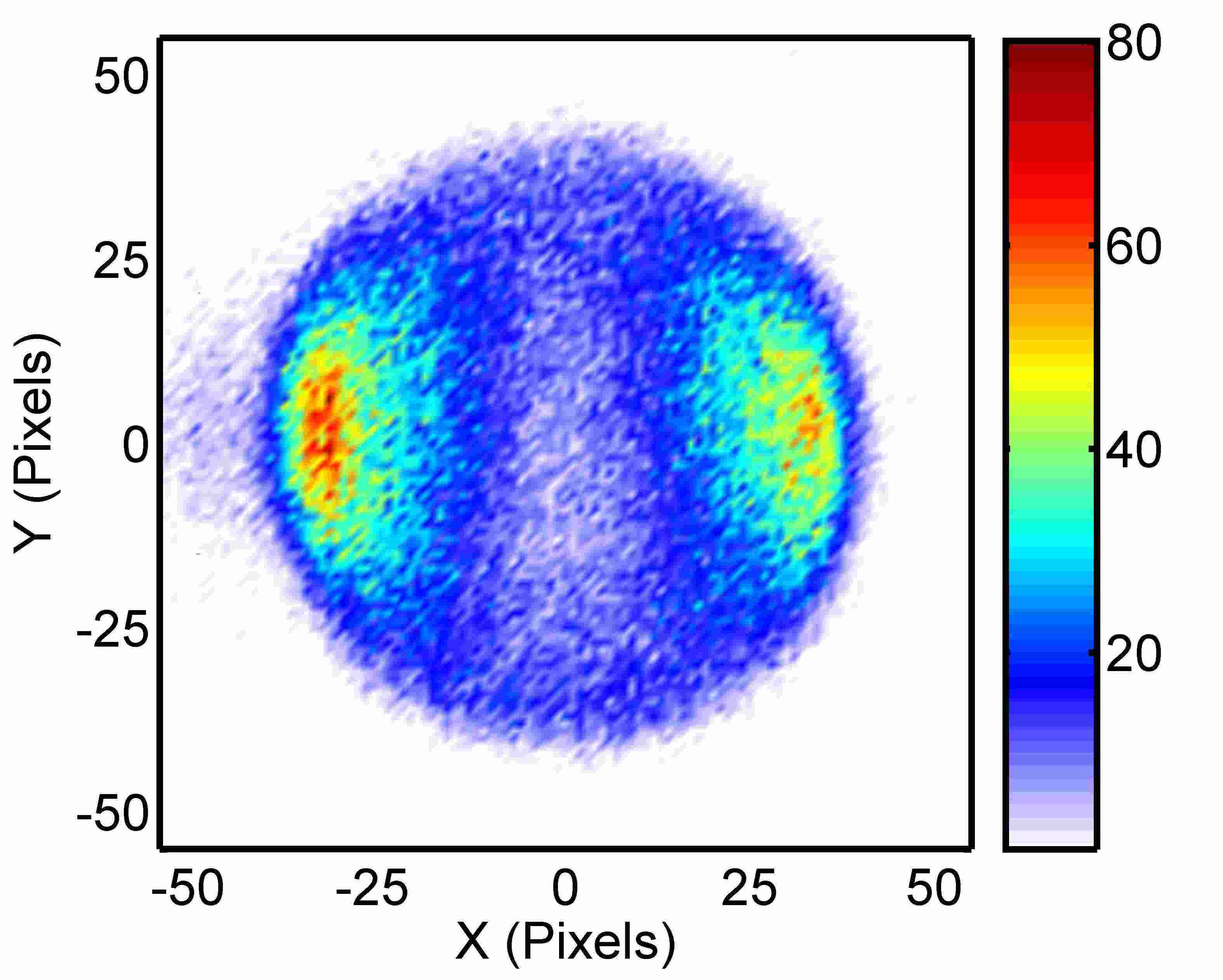}}
  \subfloat[\ce{D-} at 8.5 eV]{\includegraphics[width=0.25\columnwidth]{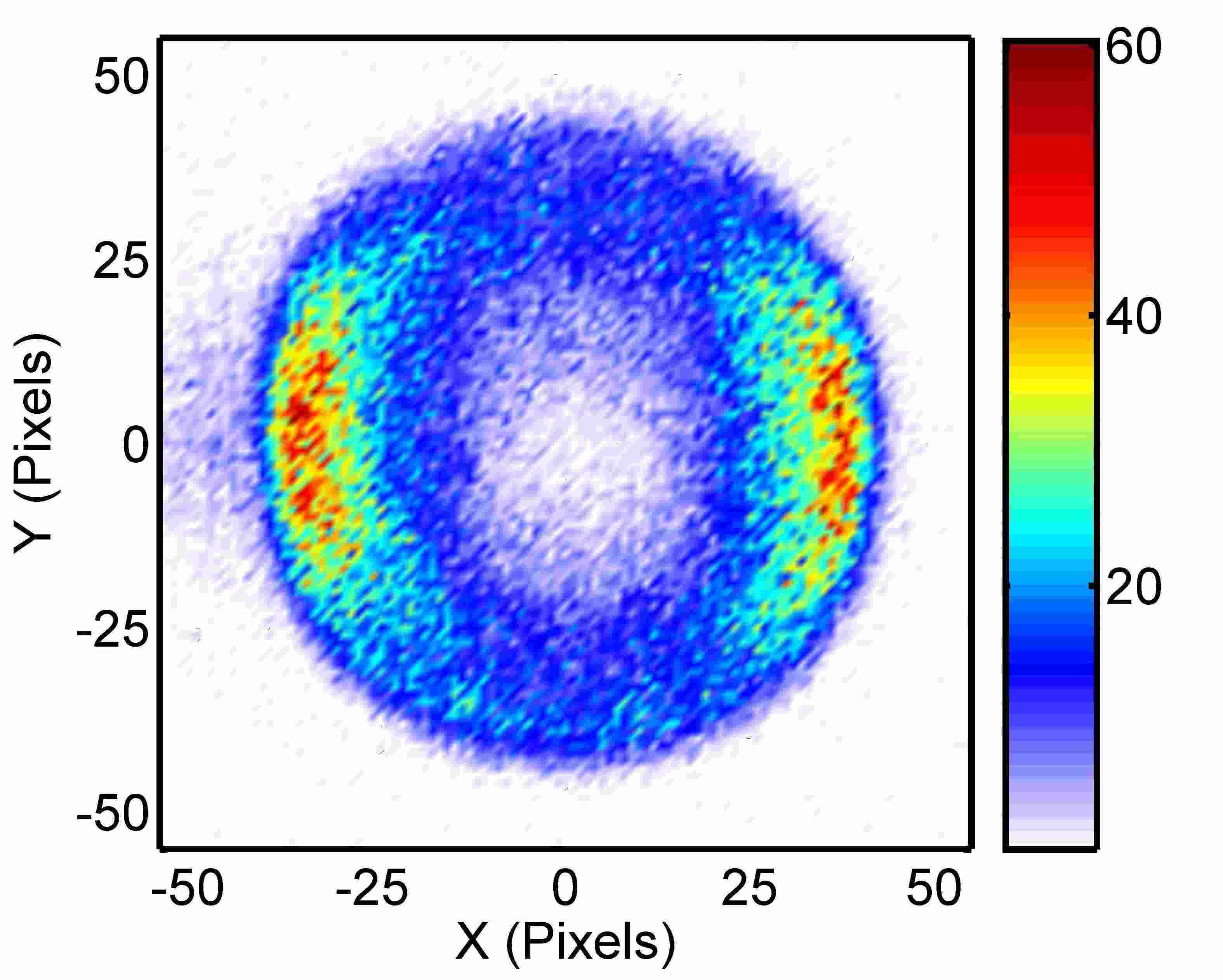}}\\
  \subfloat[\ce{D-} at 9.5 eV]{\includegraphics[width=0.25\columnwidth]{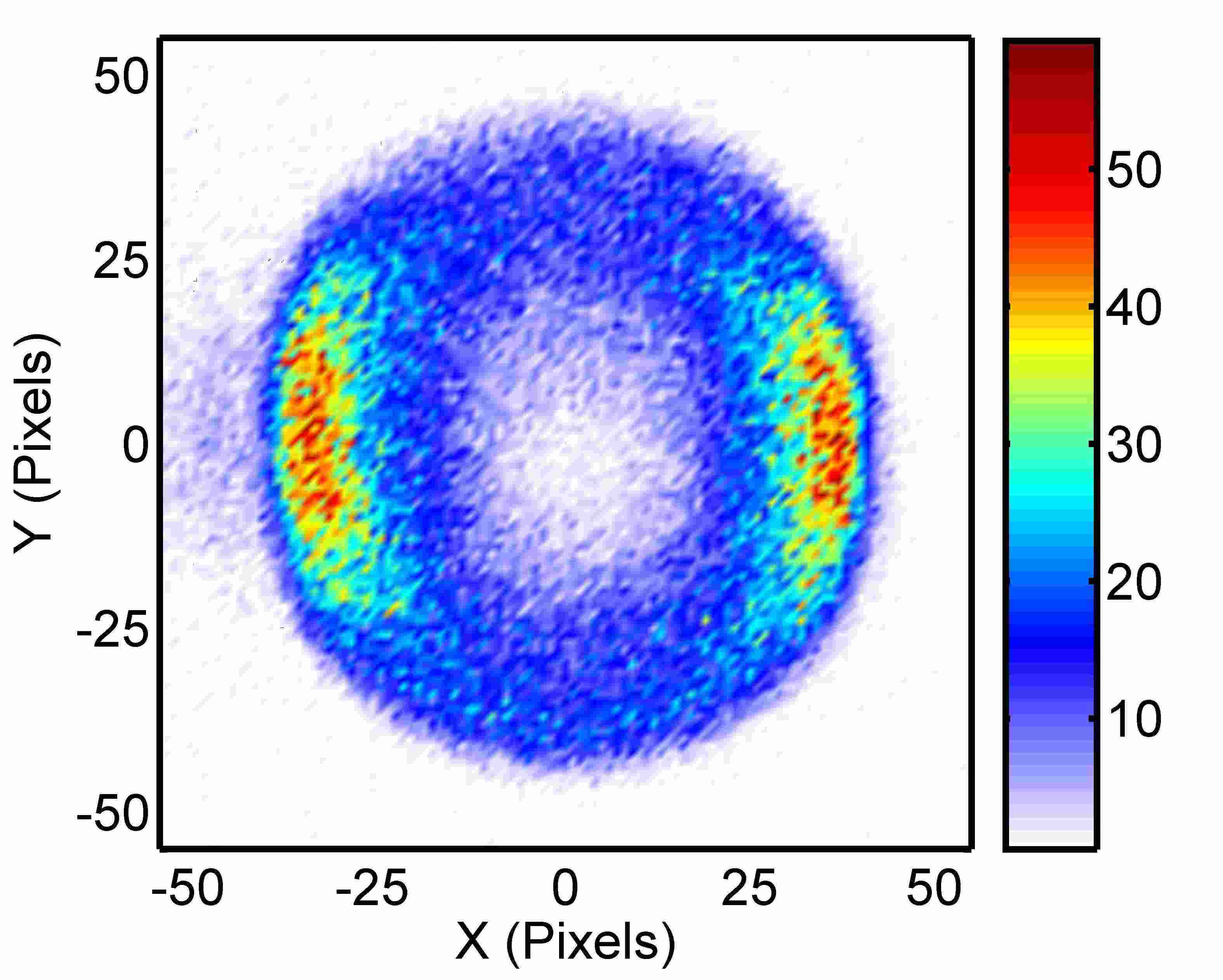}}
  \subfloat[\ce{D-} at 10.8 eV]{\includegraphics[width=0.25\columnwidth]{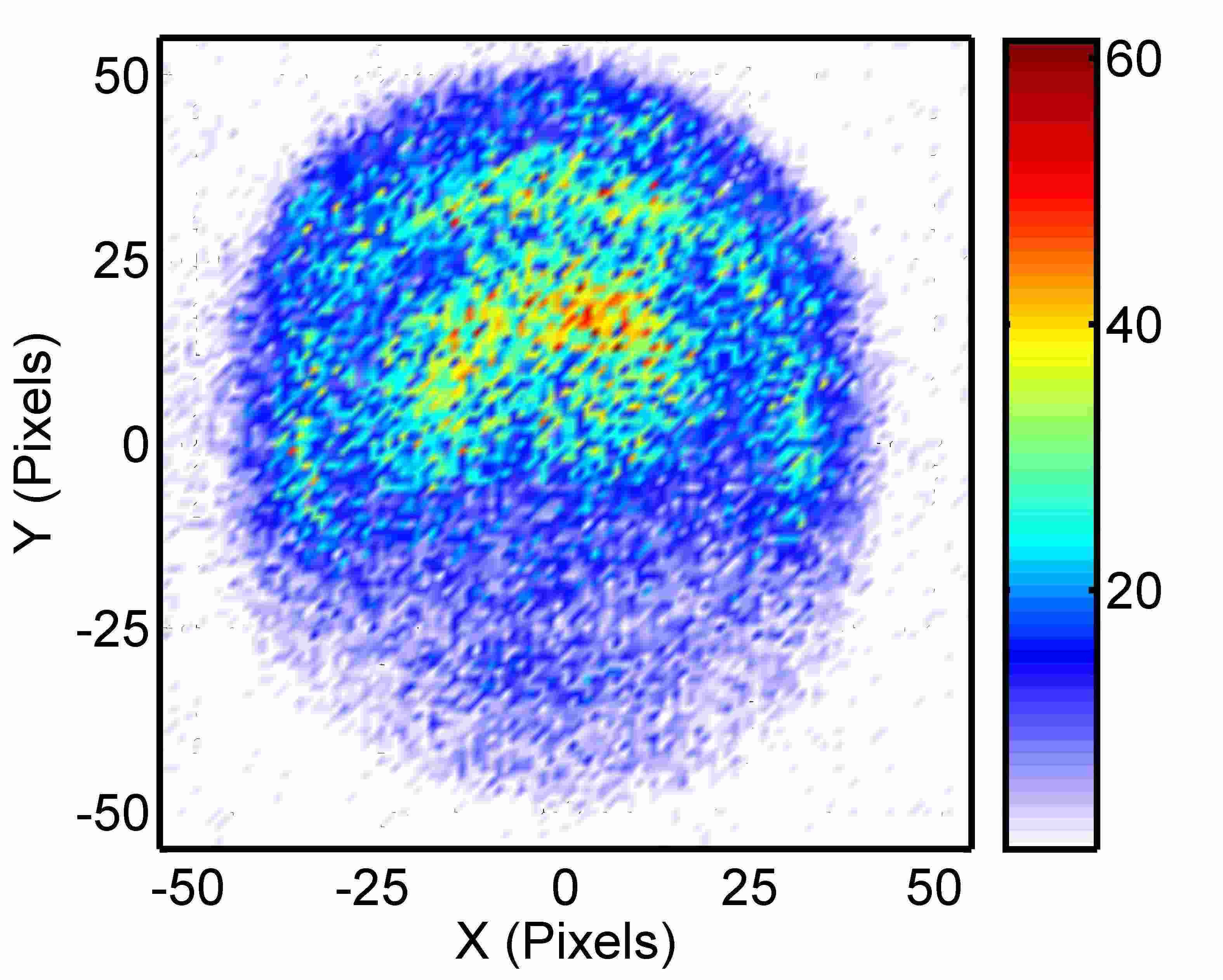}}
  \subfloat[\ce{D-} at 11.8 eV]{\includegraphics[width=0.25\columnwidth]{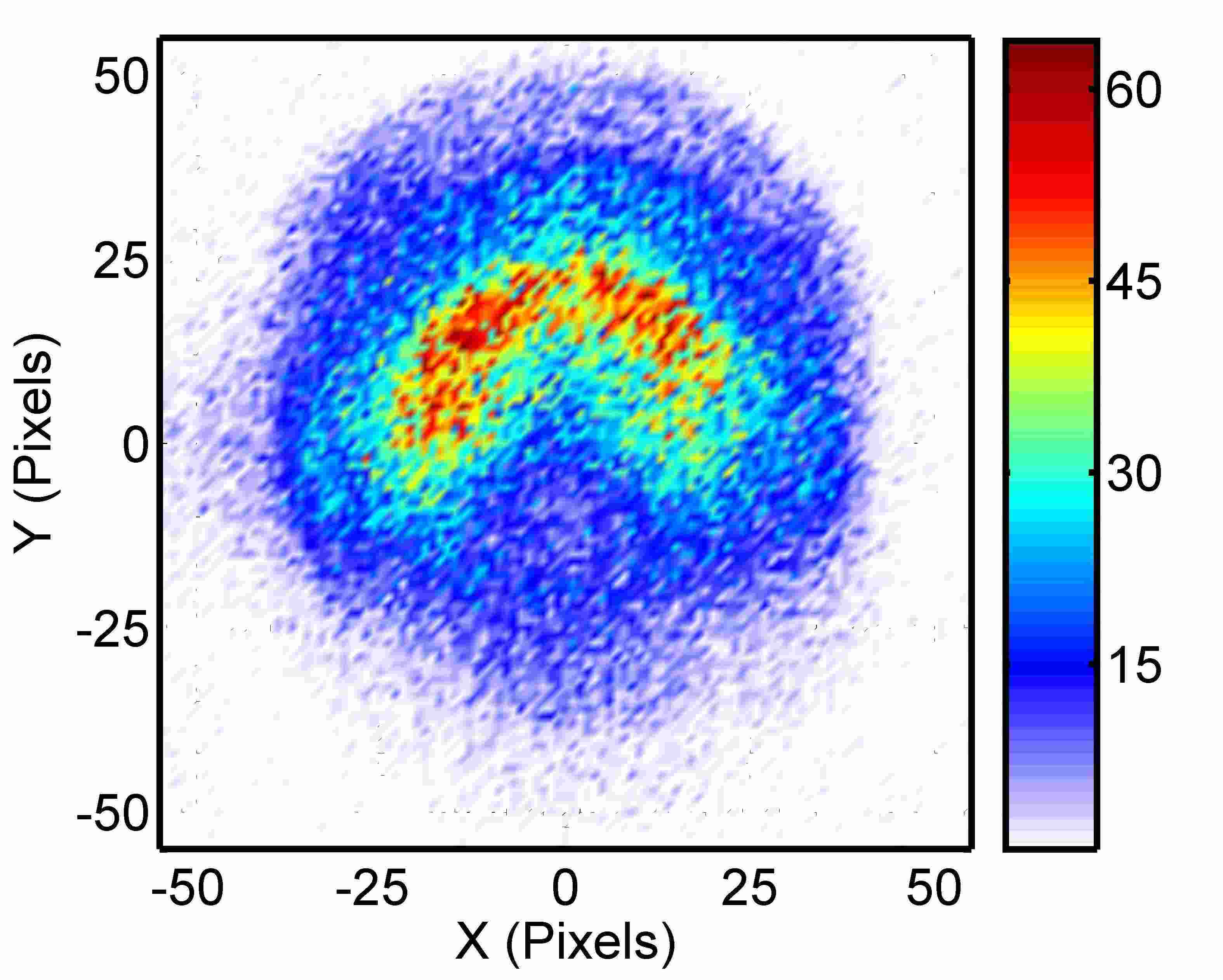}}
  \subfloat[\ce{D-} at 12.8 eV]{\includegraphics[width=0.25\columnwidth]{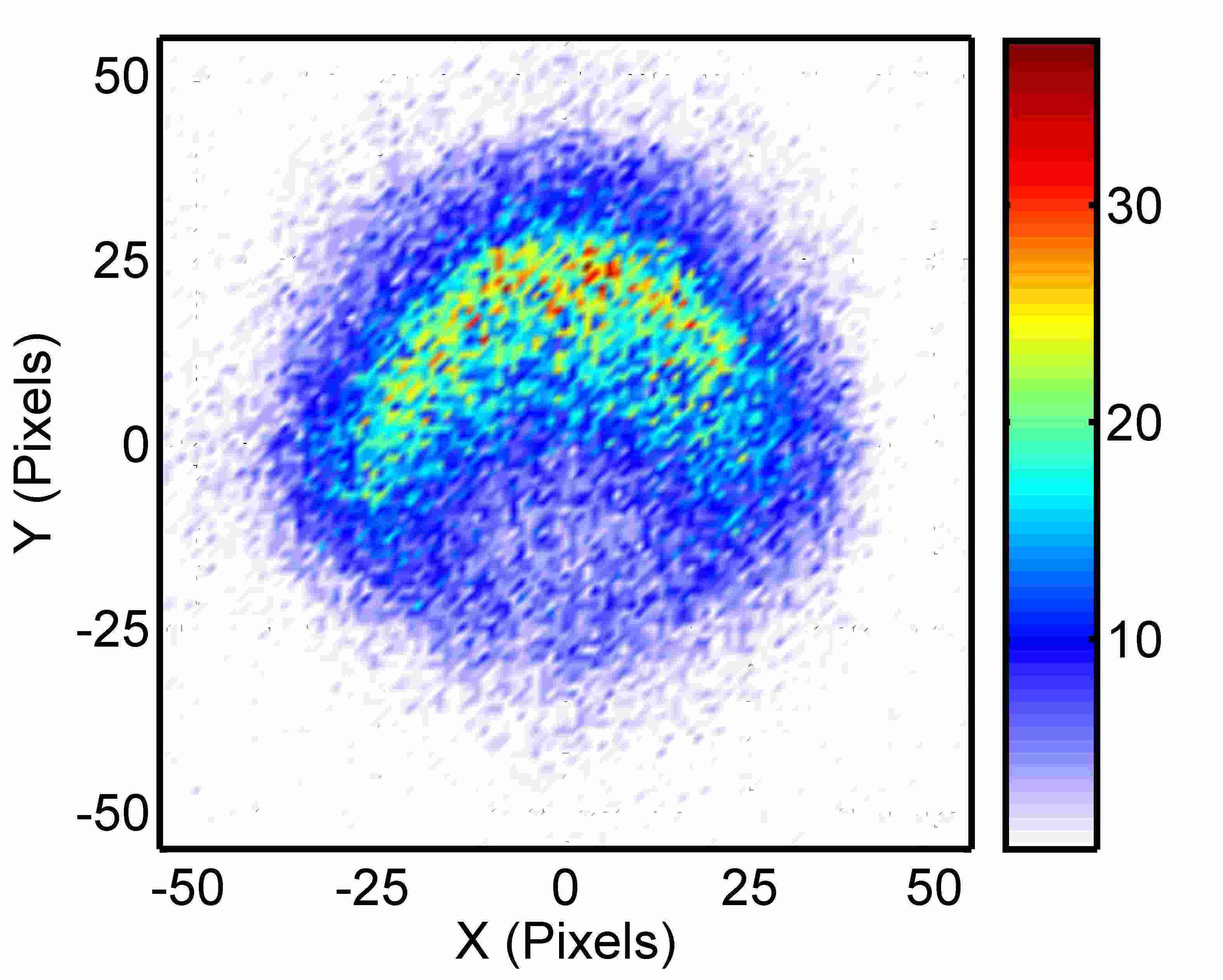}}
 \caption{Velocity images of \ce{D-} ions from DEA to \ce{D2O} at various electron energies}
 \label{fig3.5}
 \end{figure}

\begin{figure}[!htbp]
\centering
  \subfloat[\ce{O-} at 6 eV]{\includegraphics[width=0.25\columnwidth]{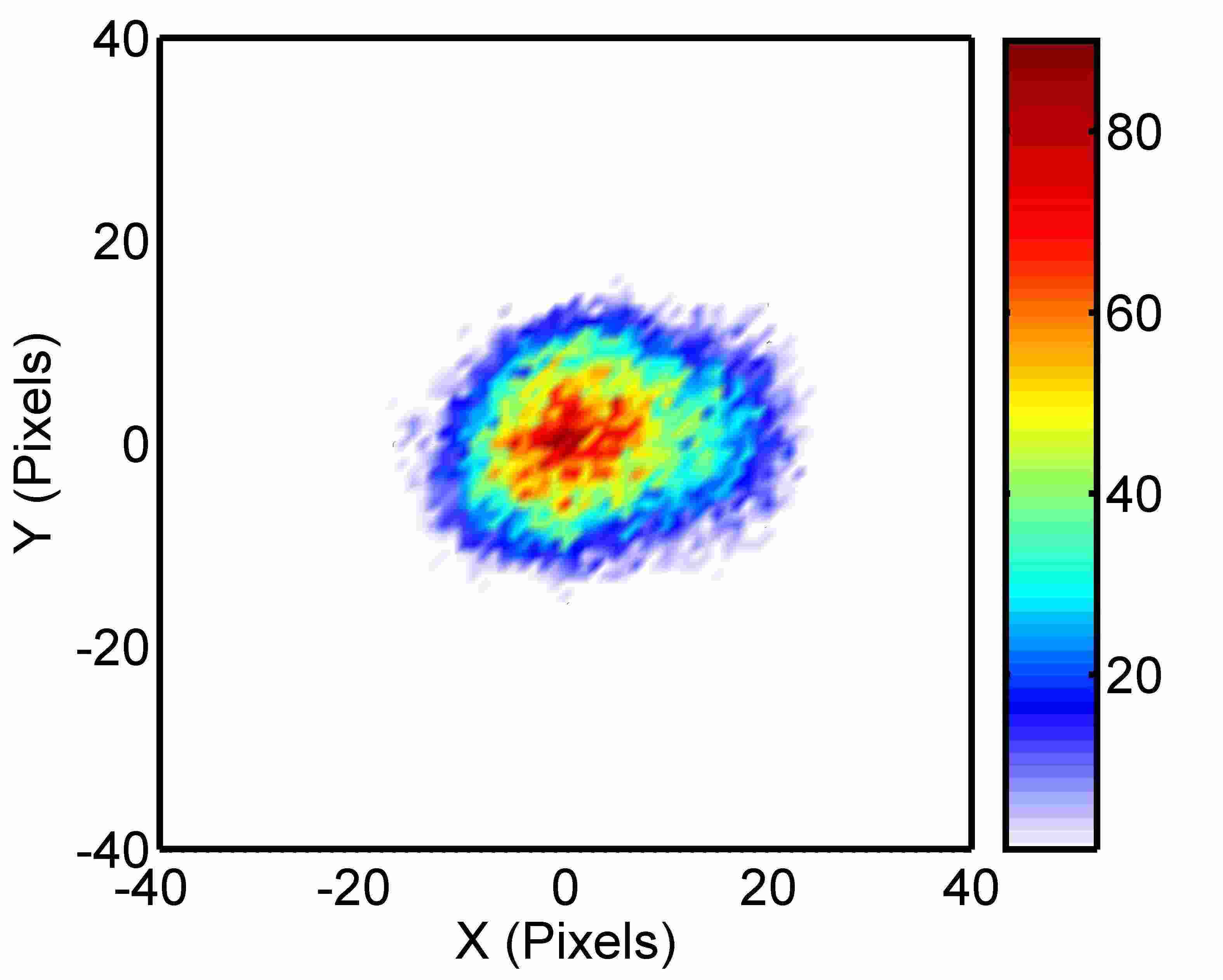}}
  \subfloat[\ce{O-} at 7 eV]{\includegraphics[width=0.25\columnwidth]{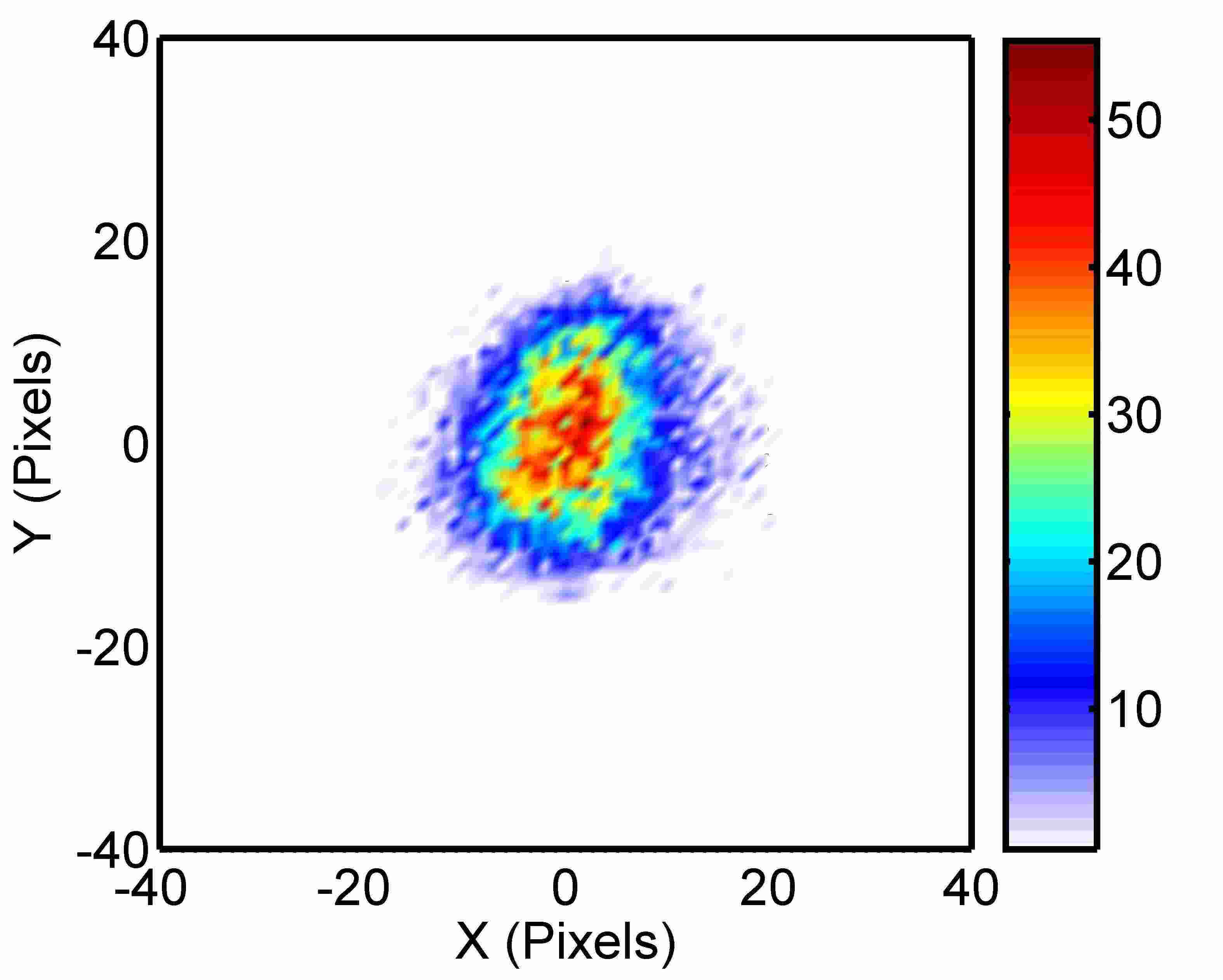}}
  \subfloat[\ce{O-} at 8 eV]{\includegraphics[width=0.25\columnwidth]{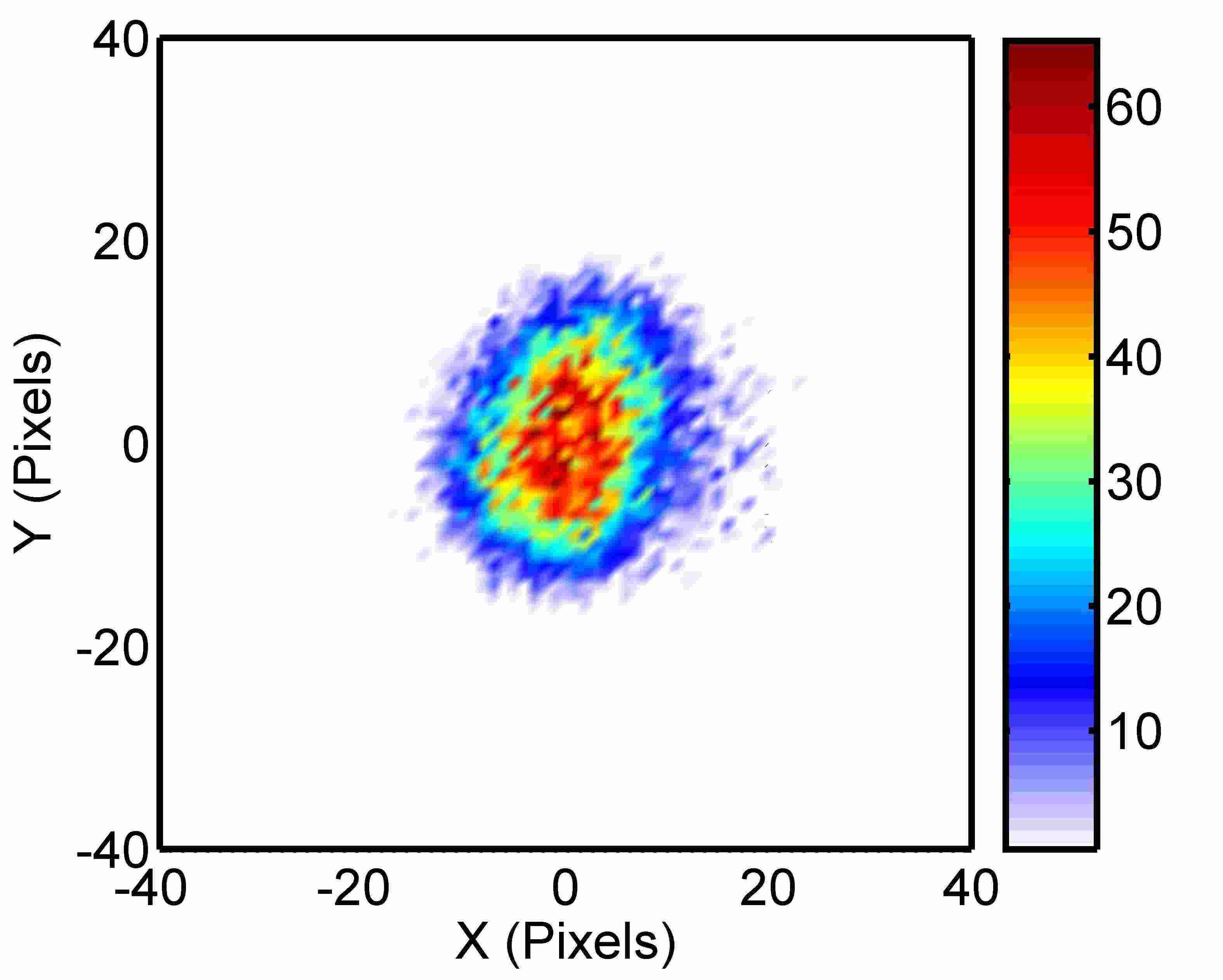}}
  \subfloat[\ce{O-} at 9 eV]{\includegraphics[width=0.25\columnwidth]{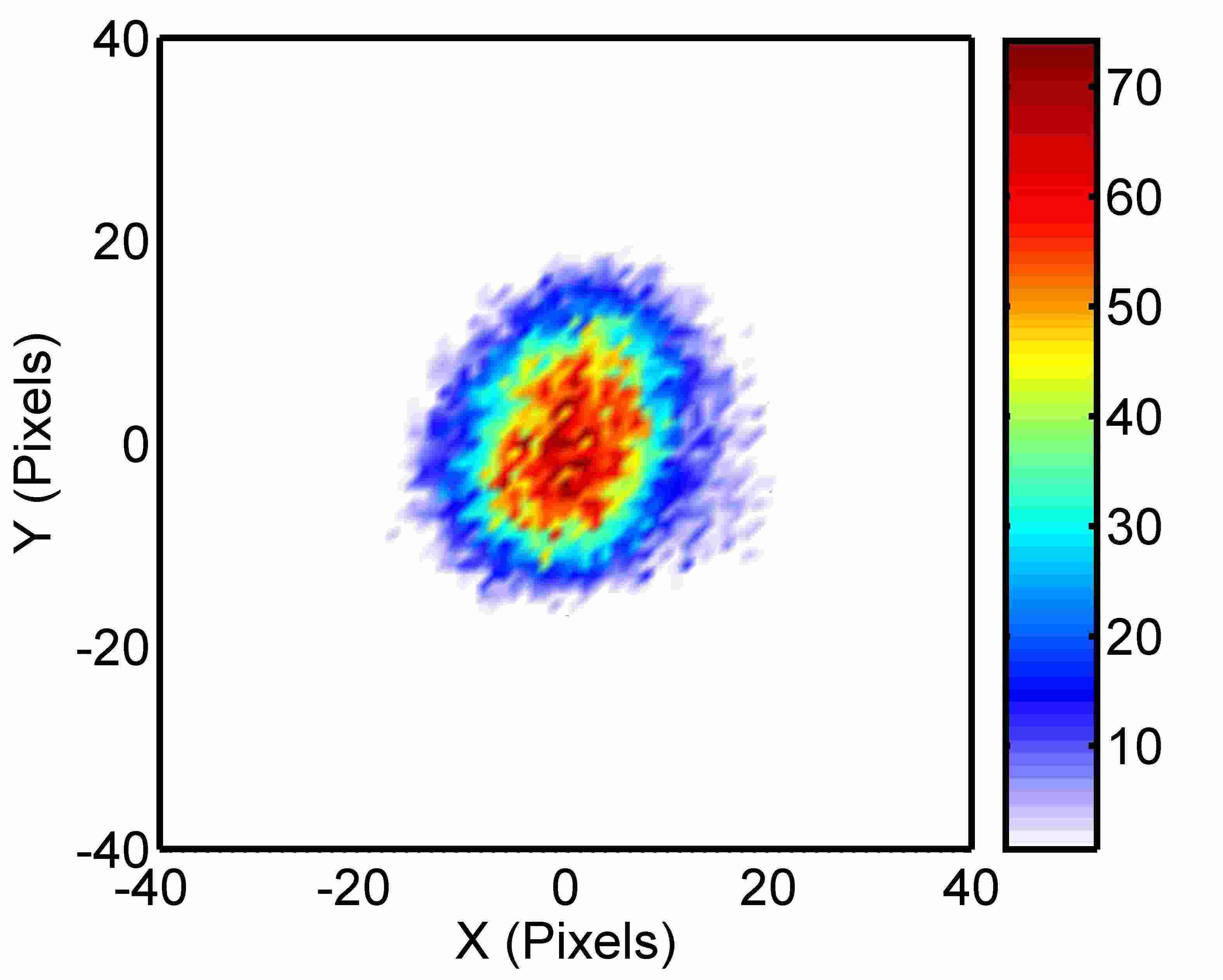}}\\
  \subfloat[\ce{O-} at 10 eV]{\includegraphics[width=0.25\columnwidth]{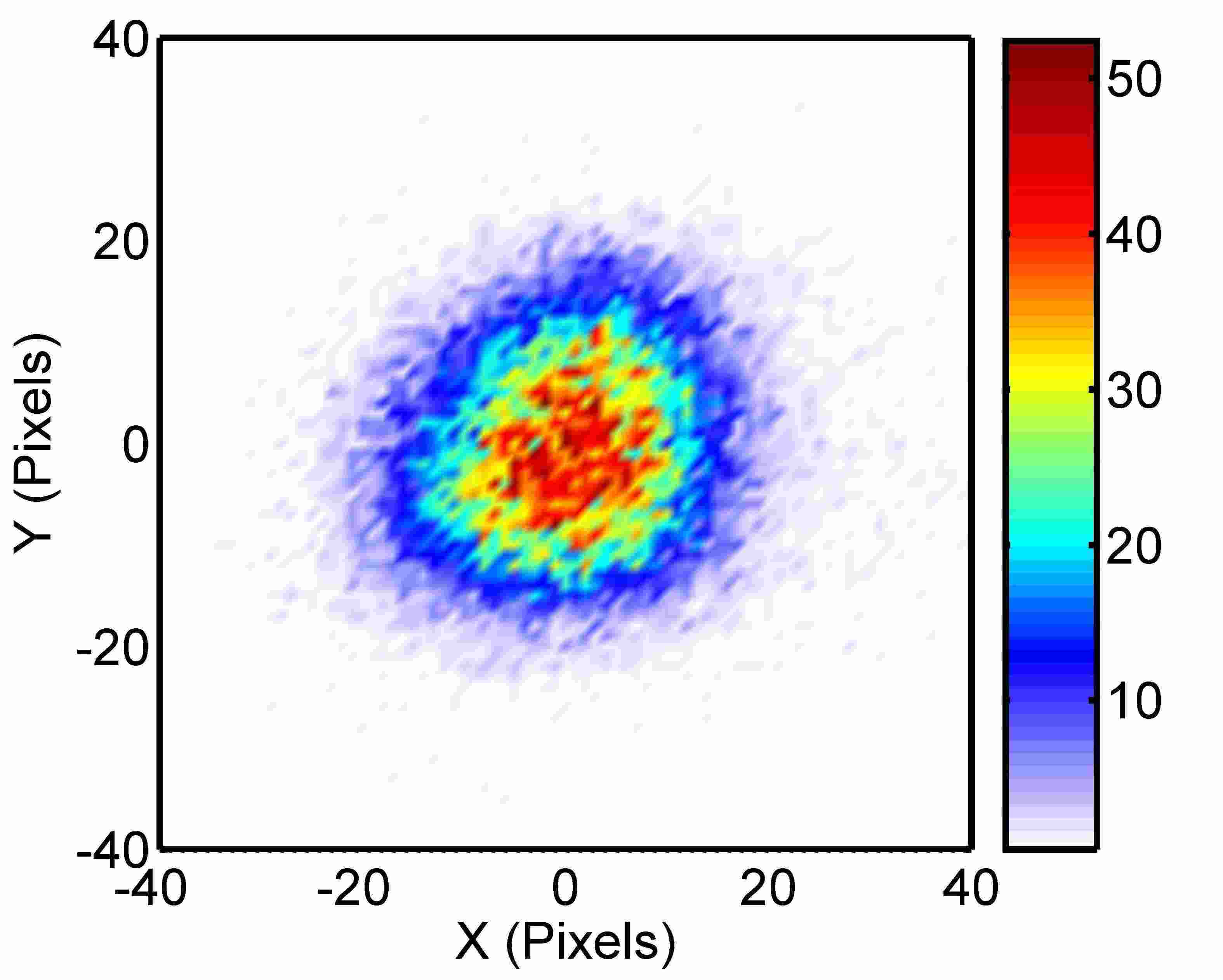}}
  \subfloat[\ce{O-} at 11 eV]{\includegraphics[width=0.25\columnwidth]{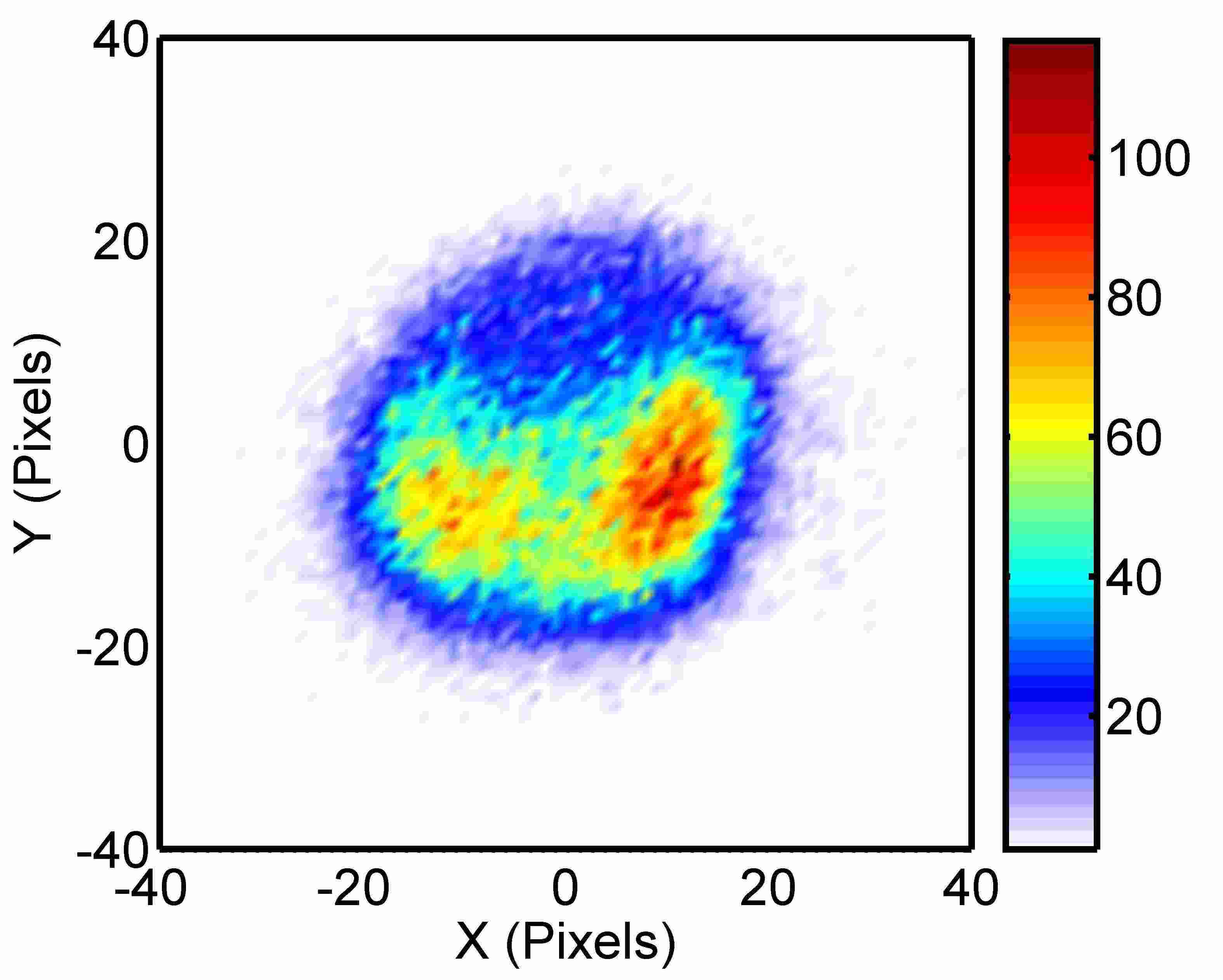}}
  \subfloat[\ce{O-} at 12 eV]{\includegraphics[width=0.25\columnwidth]{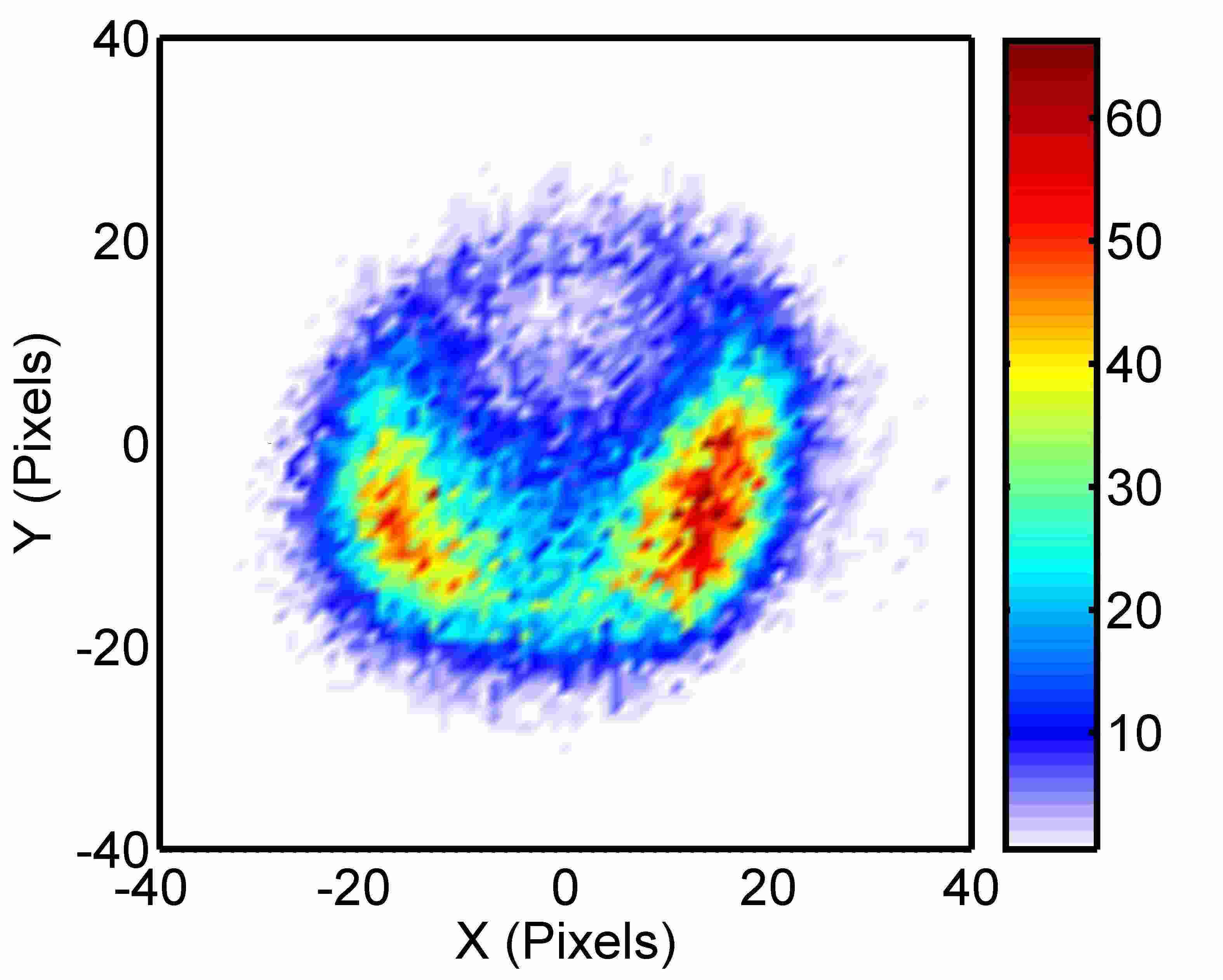}}
  \subfloat[\ce{O-} at 13 eV]{\includegraphics[width=0.25\columnwidth]{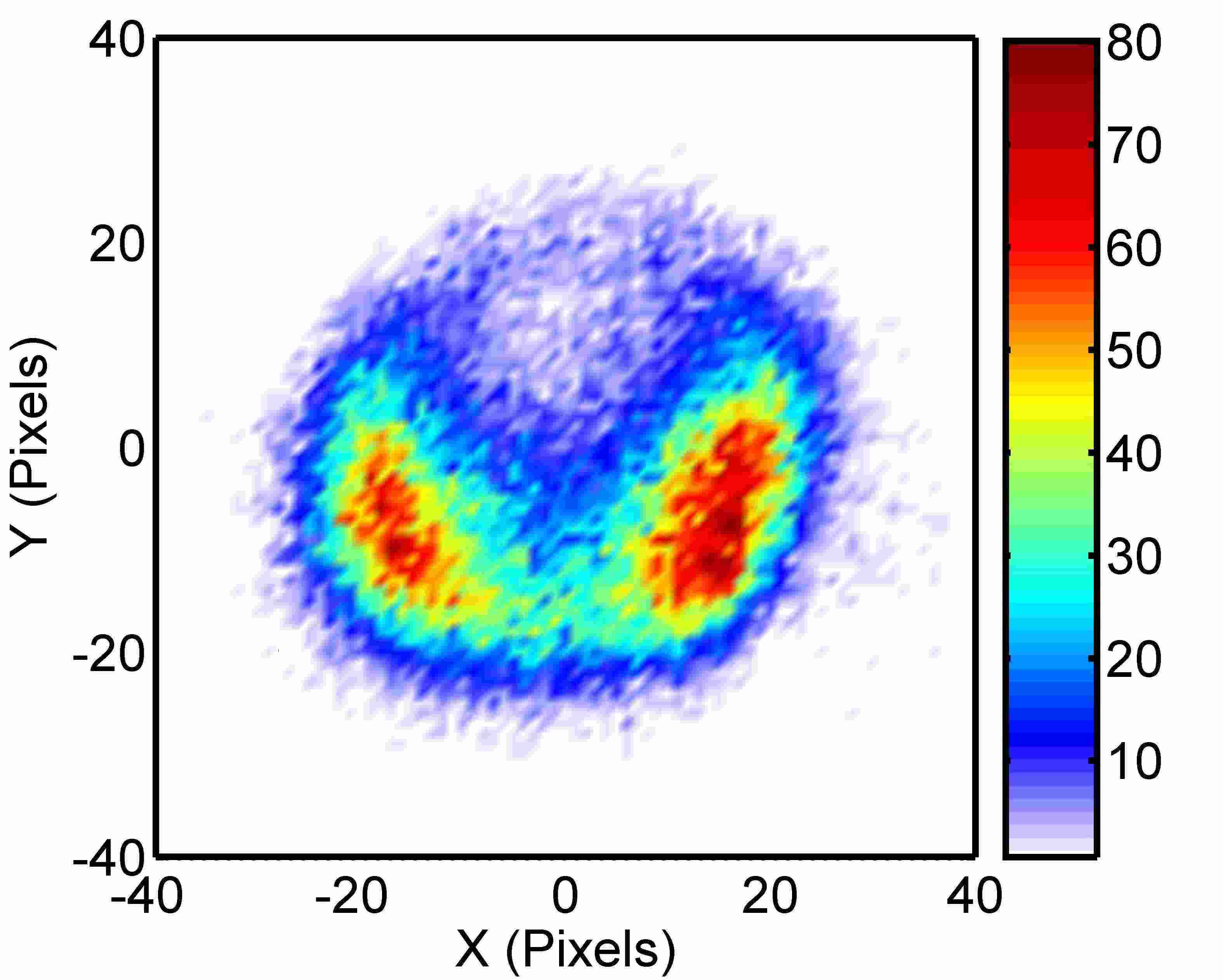}}
 \caption{Velocity images of \ce{O-} ions from DEA to \ce{D2O} at various electron energies}
 \label{fig3.6}
 \end{figure}

As discussed above, resonant electron attachment to water molecule appears to lead to the formation of transient \ce{H2O^{-*}} negative ion of \ce{B1}, \ce{A1} and \ce{B2} symmetry respectively in the energy range 5-13 eV. These resonances are seen as peaks in the cross sections of the fragment anions produced from the dissociation of the water anion. As shown in Figure \ref{fig3.1}, the peaks for \ce{H-} and \ce{D-} ions occur at 6.5 eV, 8.5 eV and 11.8 eV respectively, while those for \ce{O-} occur at 7 eV, 9 eV and 12 eV respectively. The velocity images of \ce{H-} and \ce{O-} ions from \ce{H2O} and \ce{D-} and \ce{O-} ions from \ce{D2O} in the 5-13 eV energy range covering the three resonances are shown in Figures \ref{fig3.3}, \ref{fig3.4}, \ref{fig3.5} and \ref{fig3.6} respectively. The electron beam direction is from top to bottom through the centre of every image. The radial distance from the centre of the image represents the magnitude of the velocity of the ions when they are formed. Thus, the images give differential cross sections in both energy and angle. 

To obtain the kinetic energy distribution from the velocity image, ion counts are integrated over the $2\pi$ angular range as a function of radius and normalized thereafter. The radial coordinate is calibrated in terms of kinetic energy (eV) using the data of \ce{O-} from DEA to \ce{O2} taken across its 6.5 eV resonance peak at the same experimental conditions as the fragment anion being imaged. \ce{O-} produced from DEA to \ce{O2} serves as a good calibration source as all the excess energy is carried away by the fragments as kinetic energy with no internal excitation. The expected kinetic energy of \ce{O-} for electron energies 5-8 eV ranges from 0.65 to 2.15 eV. The measured radius of the \ce{O-} images at these electron energies is converted into kinetic energy and then used to calibrate the radial coordinate to obtain the kinetic energy distribution of \ce{H-} and \ce{O-}. As the size of the velocity map image on the detector is independent of the mass of the ion that is imaged and depends only on the initial kinetic energy of the ions, the scale calibrated for the \ce{O-} ion cloud can be used for the hydride ions for the same voltage conditions. We also use the data of \ce{O-} from DEA to NO for calibration especially when kinetic energies are less than 1 eV. The angular distribution is obtained by integrating ion counts over a range of the radius and in the angular range $\theta$-5 to $\theta$+5 degrees covering scattering angles from $0^{\circ}$ to $360^{\circ}$ in steps of $10^{\circ}$.

\subsection{First resonance process peaking at 6.5 eV}

\begin{figure}[!h]
\centering
  \subfloat[]{\includegraphics[height=5cm,width=6cm]{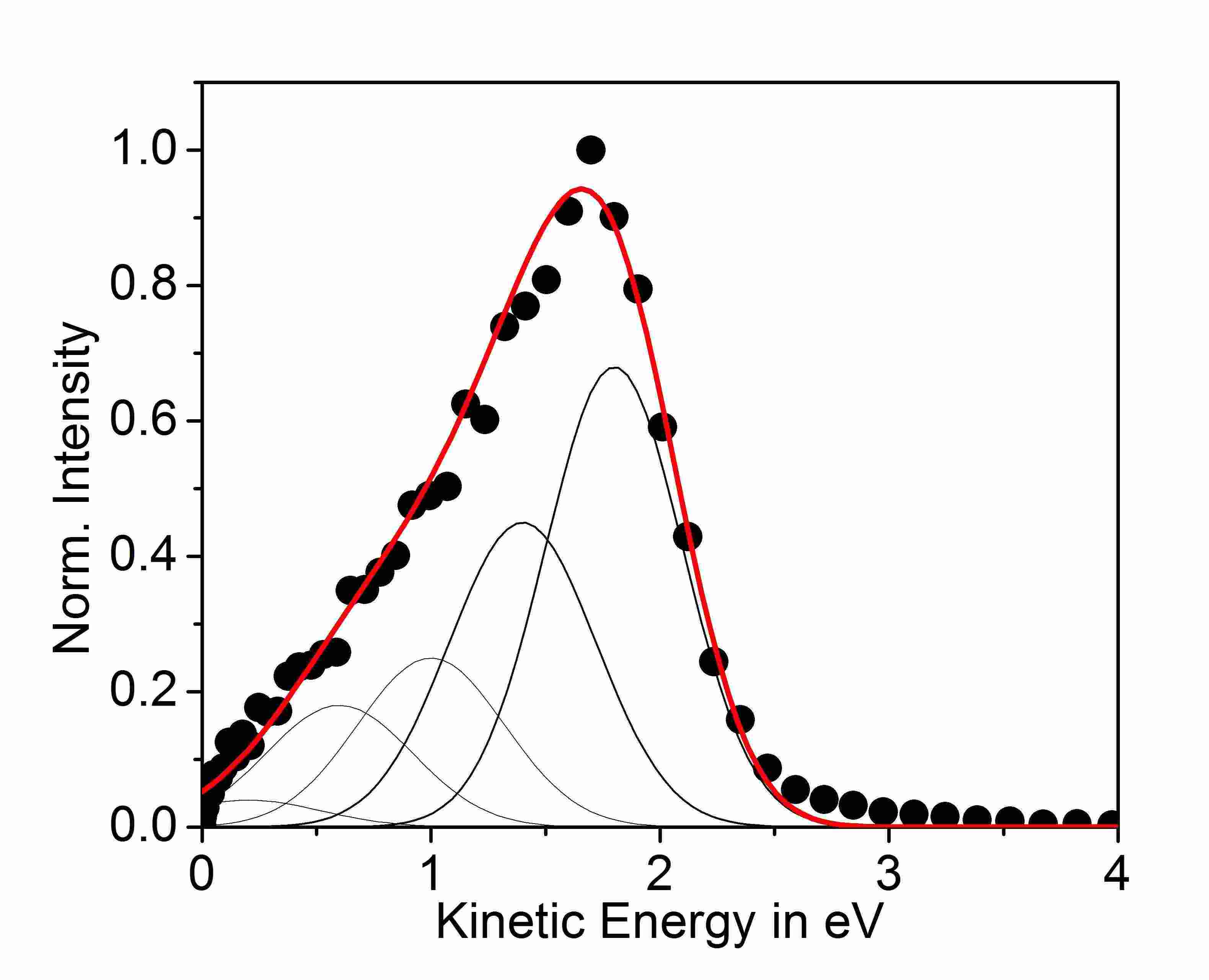}}
  \subfloat[]{\includegraphics[height=5cm,width=6cm]{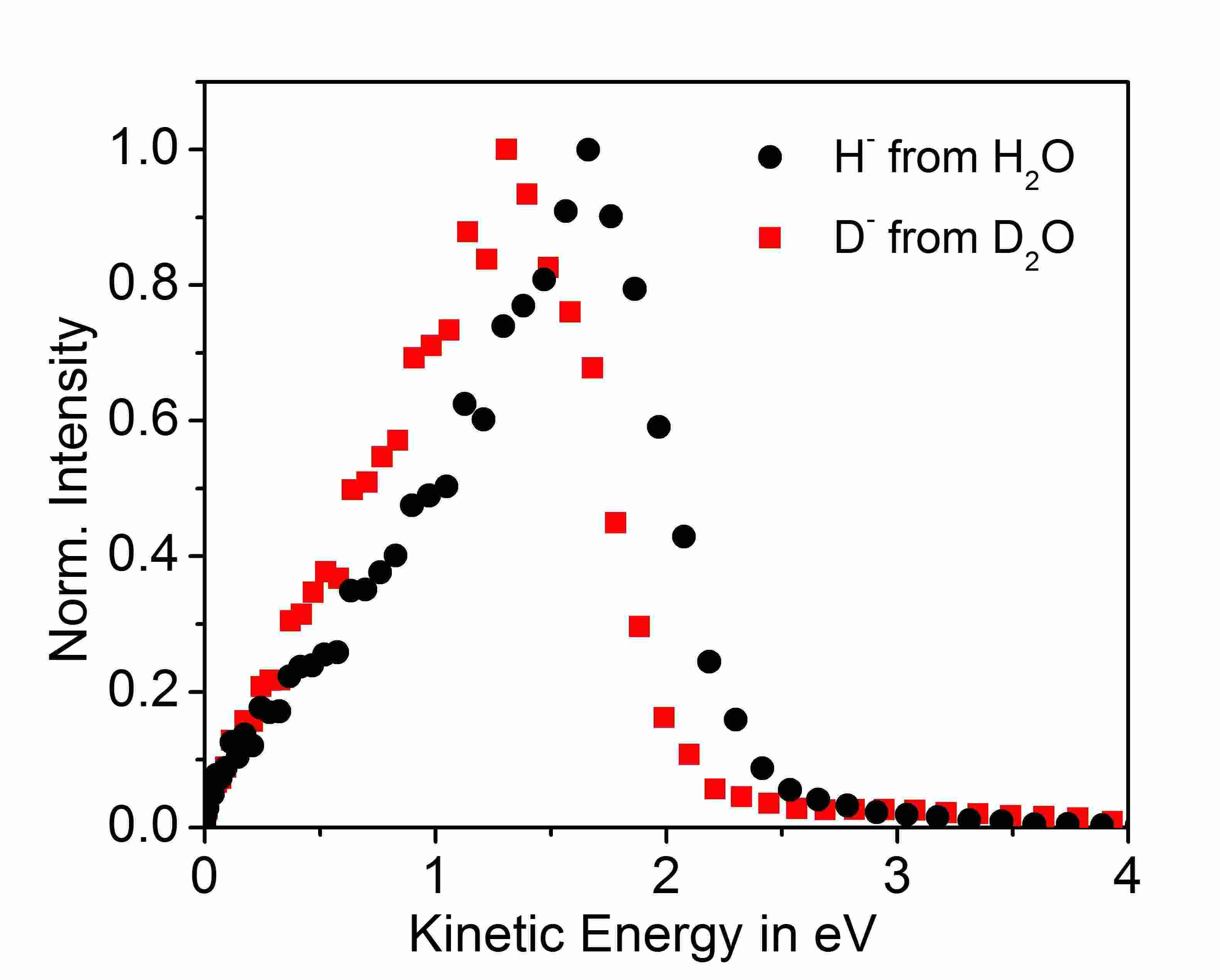}} \\
  \subfloat[]{\includegraphics[height=5cm,width=6cm]{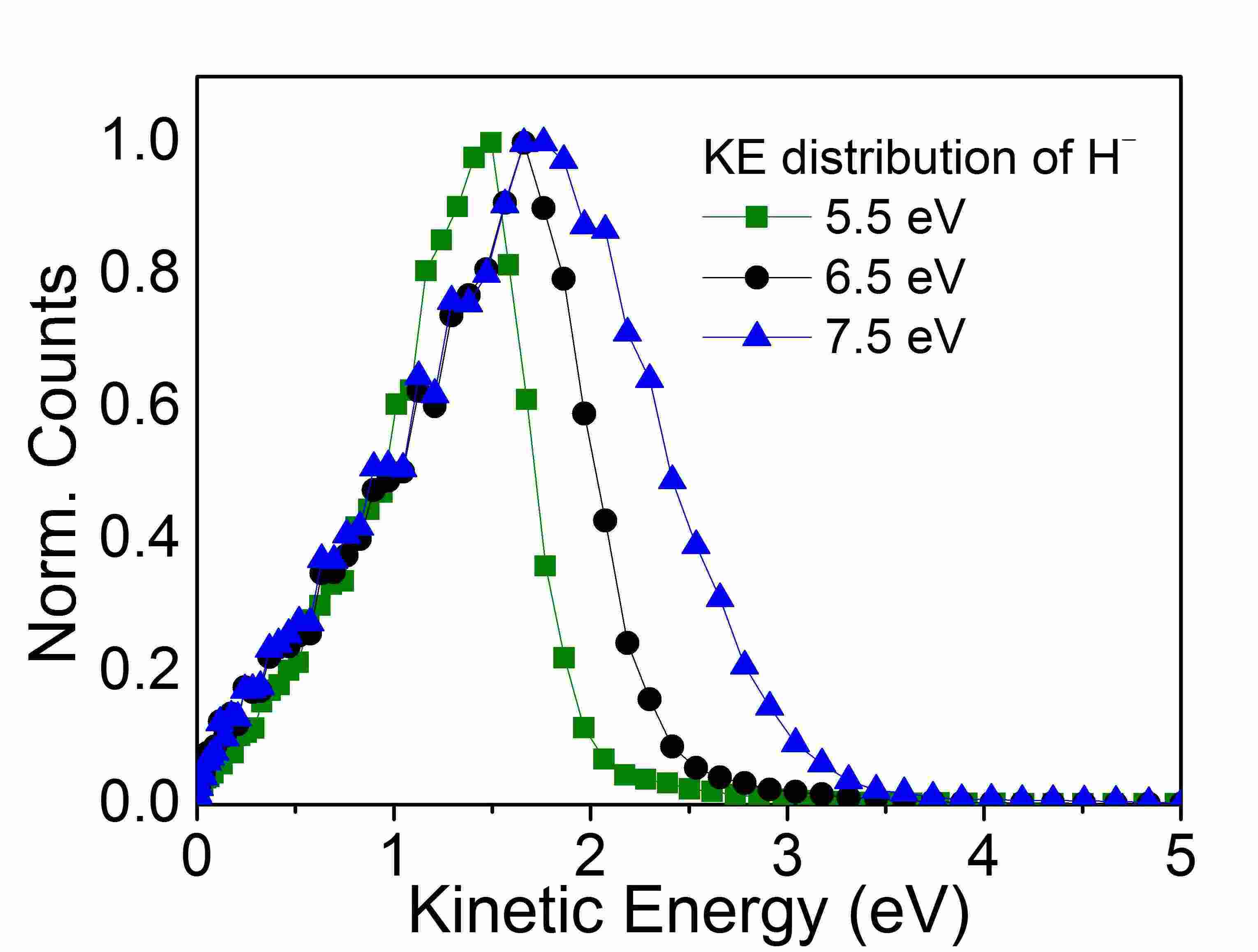}}
  \subfloat[]{\includegraphics[height=5cm,width=6cm]{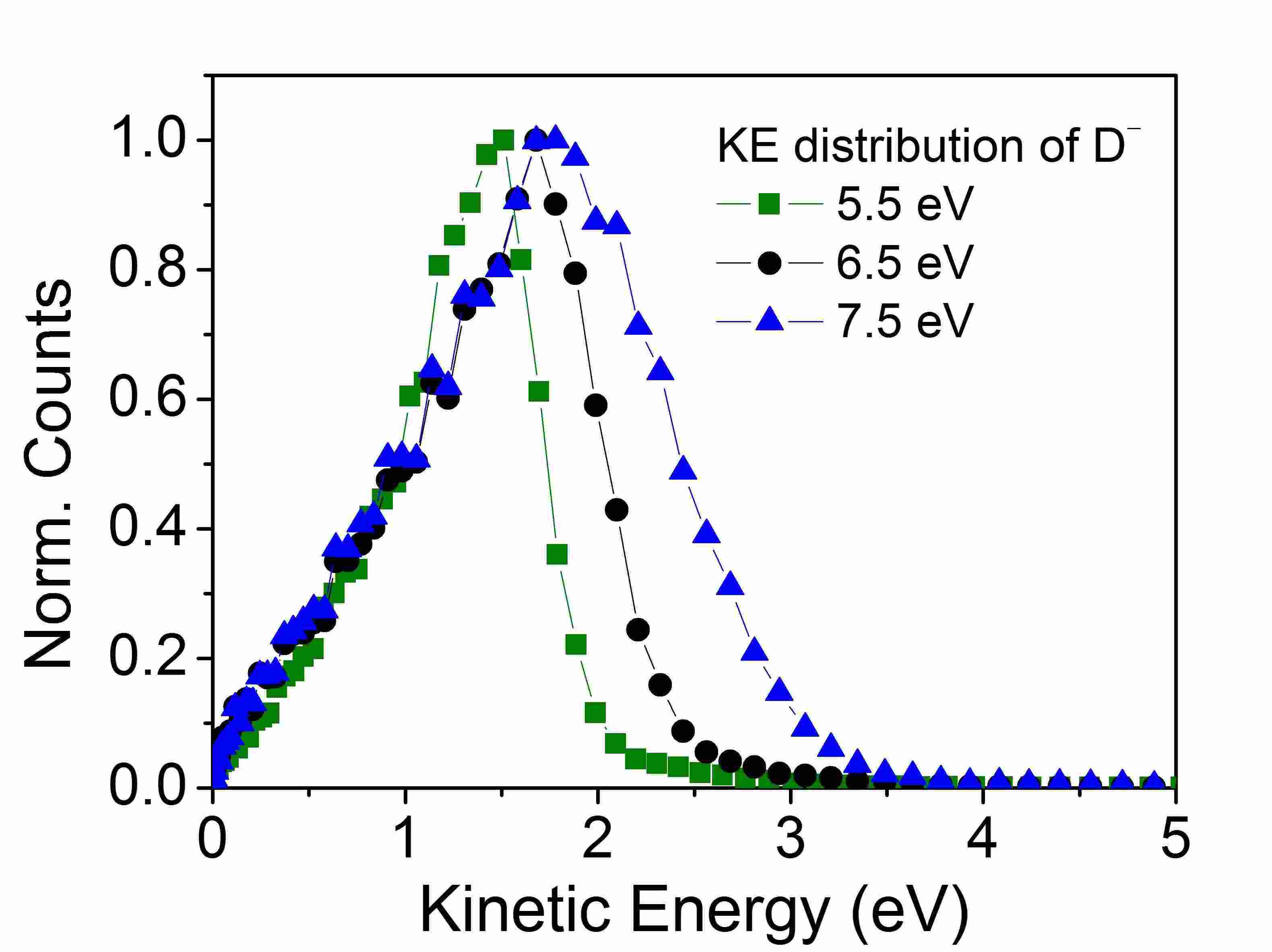}}
\caption{(a) Kinetic energy distribution of \ce{H-} ions at 6.5 eV. The inner curves are the individual Gaussian fits obtained corresponding to vibrational excitations from v = 0 (rightmost) to v = 4 (leftmost) of the OH radical. (b) Kinetic energy distribution of \ce{H-} ions and \ce{D-} ions at 6.5 eV for comparison. (c) and (d) show variation KE distribution of \ce{H-} and \ce{D-} ions across the first resonance process at electron energies 5.5 eV, 6.5 eV and 7.5 eV}
 \label{fig3.7}
 \end{figure}

The \ce{H-} (and \ce{D-}) ions produced from the first resonance process (as seen in Figure \ref{fig3.3}-(a), (b) and (c) and Figure \ref{fig3.5}-(a), (b) and (c)) are ejected perpendicular to the electron beam direction with a continuous distribution of intensity along the radial direction which appear to increase with radius. The kinetic energy distribution of the \ce{H-} ions at 6.5 eV is plotted in Figure \ref{fig3.7}(a). Figure \ref{fig3.7}(b) shows \ce{H-} and \ce{D-} spectra in the same plot for comparison. Figure \ref{fig3.7}(c) and (d) show the variation of KE distribution of \ce{H-} and \ce{D-} ions respectively at electron energies 5.5 eV, 6.5 eV and 7.5 eV sweeping across the first resonance.

As seen in Figure \ref{fig3.7}(a), the kinetic energy of \ce{H-} ions range from 0 to about 2.5 eV with a peak at 1.8 eV. For electron energy of 6.5 eV, the maximum kinetic energy observed is 2.5 eV. Thus, about 4 eV energy is used in the dissociation process and the dissociation channel consistent with observed kinetic energies is \ce{H- + OH(^{2}$\Pi$)} with threshold energy 4.35 eV. This is in agreement with the observations of Belic et al. \cite{c3belic}. The broad KE distribution is an indication of vibrational and rotational excitation of the OH fragment. In this context, one would expect discrete peaks in the kinetic energy distribution corresponding to these vibrational and rotational excitations. Belic et al. \cite{c3belic} have reported observing clear peaks corresponding to the vibrational excitation of the OH fragment. However, we are unable to see distinctly the rings corresponding to these in the present VMI data due to the poorer energy resolution of the electron beam, which is about 0.5 eV. Though the energy resolution is not sufficient to clearly see the vibrational levels, we attempted to retrieve the intensity distribution of vibrational excitation by fitting for contributions corresponding to excitation of various energetically allowed vibrational levels of the OH fragment. The contributions of allowed vibrational excitations are taken as Gaussians with the same width, but different heights. The width and the heights are used as free parameters and the best fit for them is obtained. The fit shows occupation of vibrational levels $\nu$=0 to $\nu$=4. The relative intensities of the vibrational levels obtained this way are given in Table \ref{tab3.3} along with the results of Belic et al. \cite{c3belic} and the calculations of Haxton et al. \cite{c3h2}. The present results show relatively larger intensities for the higher vibrational levels as compared to the previous measurement and the theoretical calculations. One reason for the difference with the previous measurement could be in the inadequate calibration of the transmission function of the energy analyzer used by Belic et al. \cite{c3belic} who optimized their spectrometer for relatively energetic ions around 1.7 eV. They mention that the transmission function of their kinetic energy analyzer was constant within 20\% in the range of 1 to 3 eV. No information is given for the transmission function for their analyzer below 1 eV which is expected to go down as with any conventional electrostatic analyzer. One of the advantages of a VMI spectrometer is its constant transmission throughout the energy range starting from zero, except for energies large enough to leave the image out of bound of the detector. The kinetic energy distribution of \ce{D-} ions is also similar to that of \ce{H-} ions except that it is scaled down due to the higher mass of \ce{D-} (by a factor 0.95) and slightly higher threshold of 4.52 eV as seen in Figure \ref{fig3.7}(b). The KE distribution of \ce{H-} and \ce{D-} ions across the resonance as shown in Figures \ref{fig3.7}(c) and (d) show the maximum kinetic energy increase with increase in electron energy from 5.5 eV to 7.5 eV. Also the distribution is broadened indicating excitation of higher vibrational states of OH (OD).  
 
\begin{table}[h]
\caption{Comparison of relative intensities of various vibrational levels of OH}
\begin{center}
\begin{tabular}{cccc}
\hline
\\
Vibrational Level & Theory$^{a}$ & Earlier Expt.$^{b}$ & Present Expt.\\
\\
\hline
\\
0 & 1 & 1 & 1 \\
\\
1 & 1.08 & 0.83 & 0.82 \\
\\
2 & 0.60 & 0.30 & 0.57 \\
\\
3 & 0.04 & 0.07 & 0.32 \\
\\
4 & - & - & 0.25 \\
\\
\hline
\end{tabular}
\end{center}
\hspace{3cm}$^{a}$Haxton et al. \cite{c3h2};$^{b}$Belic et al. \cite{c3belic}
\label{tab3.3}
\end{table} 

\begin{figure}[!ht]
\centering
  \subfloat[]{\includegraphics[height=3.9cm,width=5.1cm]{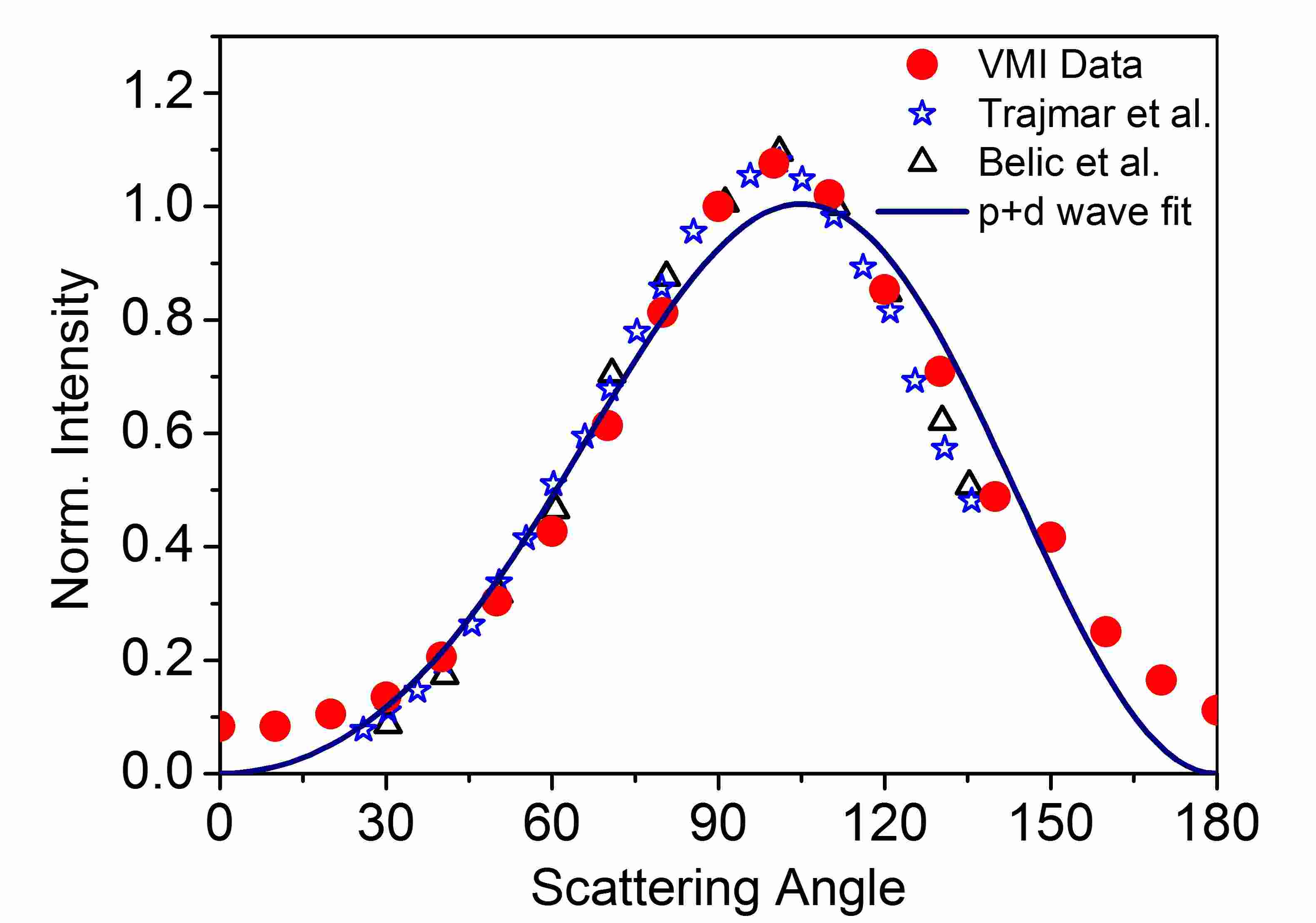}}
  \subfloat[]{\includegraphics[height=4cm,width=5cm]{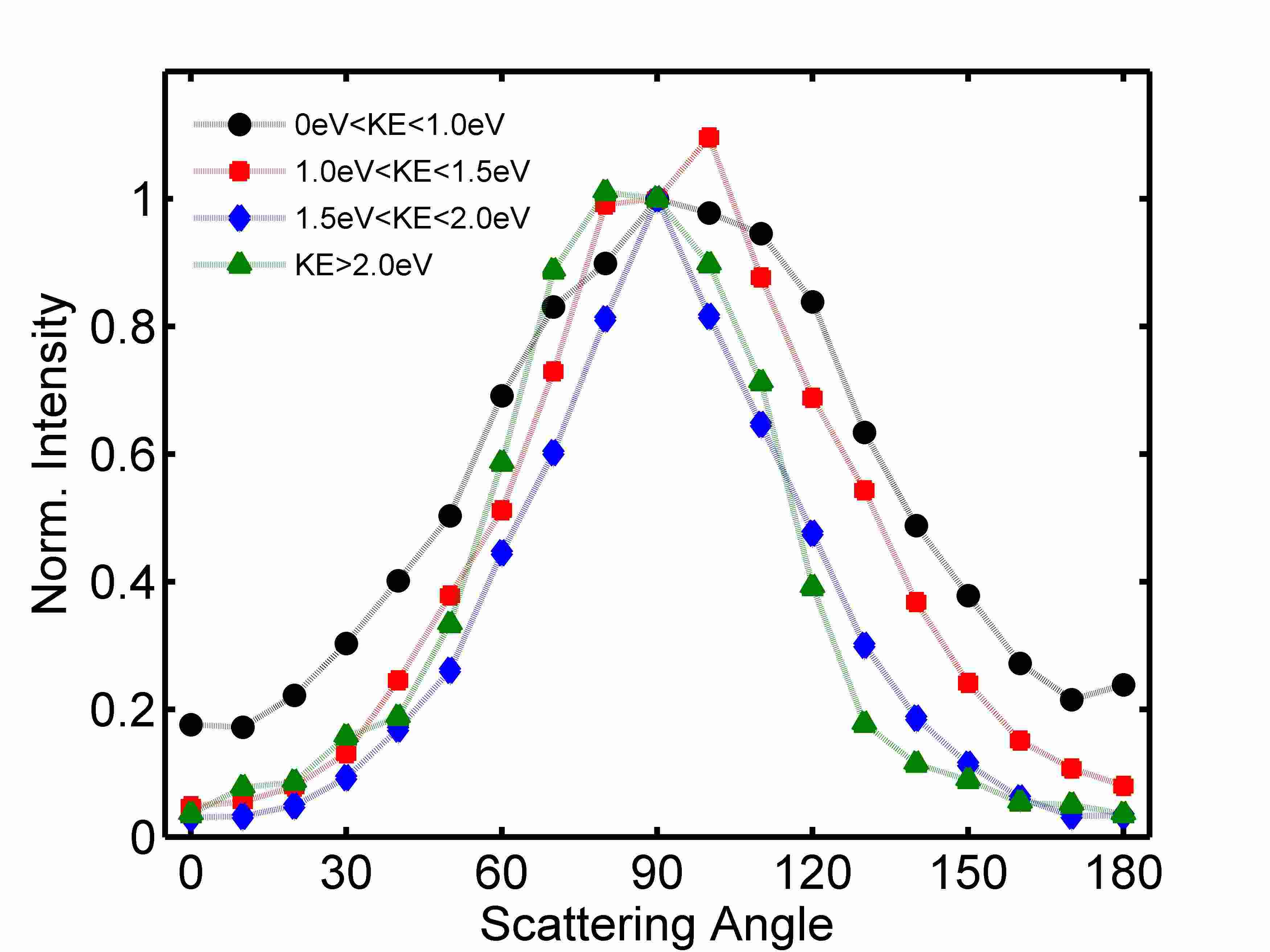}}
  \subfloat[]{\includegraphics[height=4cm,width=5cm]{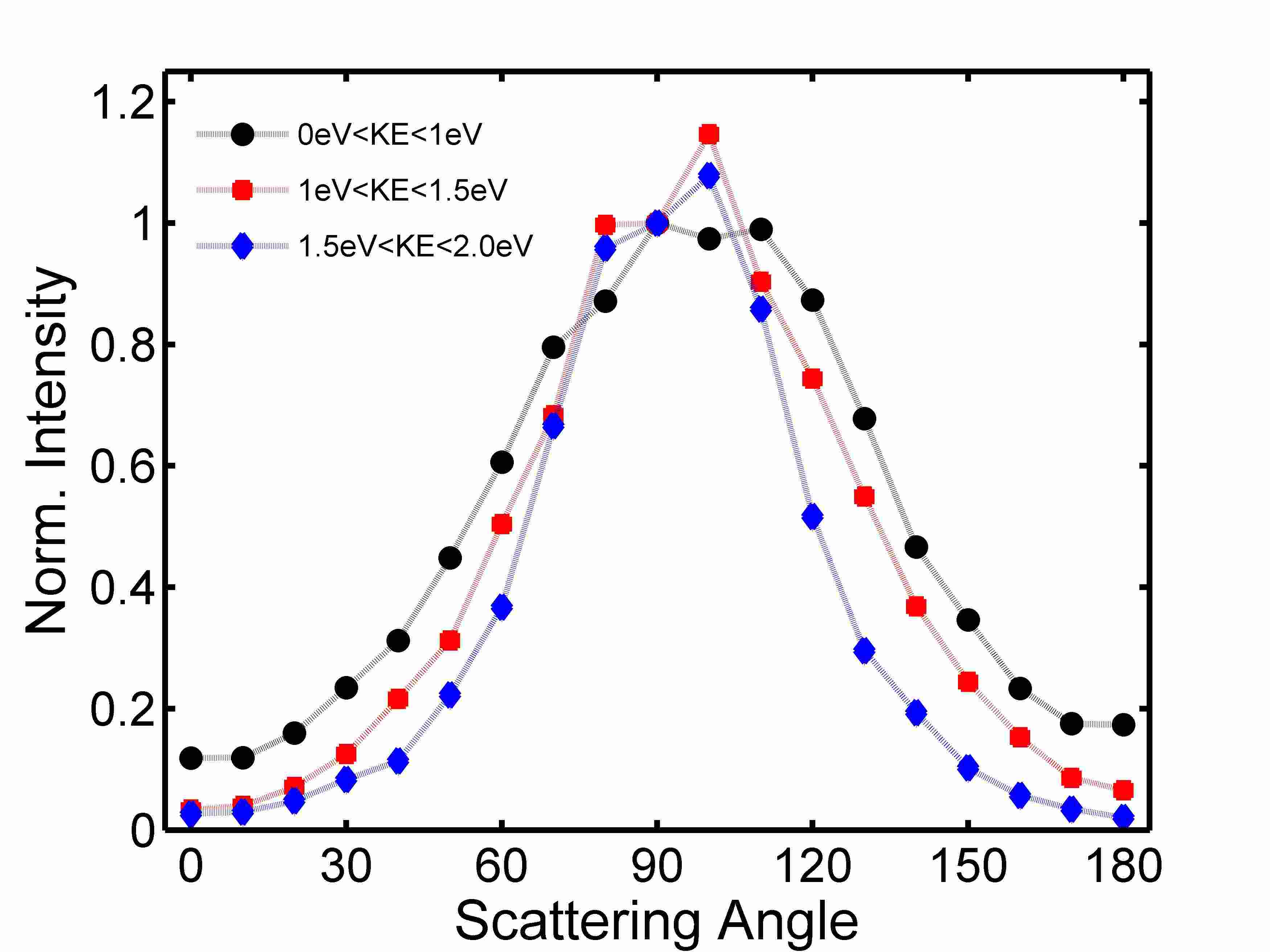}}
  \caption{(a) \ce{H-} angular distribution (integrated over all kinetic energies) at 6.5 eV (circles - present data, stars - Trajmar and Hall \cite{c3trajmar}, triangles - Belic et al.\cite{c3belic}, solid line - fit for \ce{B1} symmetry using p and d partial waves). Angular distribution plots of (b) \ce{H-} and (c) \ce{D-} ions as a function of kinetic energy at 6.5 eV.}
\label{fig3.8}
\end{figure}

The angular distribution of \ce{H-} ions at 6.5 eV over the entire kinetic energy distribution is shown in Figure \ref{fig3.8}(a). The intensity distribution for \ce{H-} appears to peak at an angle of about $100^{\circ}$ with respect to the electron beam direction. This is consistent with the data of Trajmar et al. \cite{c3trajmar} and Belic et al. \cite{c3belic} as shown in Figure \ref{fig3.8}(a). From the observed angular distribution, we conclude the symmetry of the resonance to be \ce{B1} in agreement with earlier reports. We fit our data with the functional form given in equation 3.4 for \ce{B1} symmetry taking the lowest allowed partial waves - $p$ and $d$. Resultant curve is shown in Figure \ref{fig3.8} (a) as a solid line. We find the ratio of the $d$ wave to $p$ wave amplitude to be 0.31 with a phase difference of 3.139 (close to $\pi$) radians. A similar procedure has been used by Haxton et al. \cite{c3h4} earlier to show that the observed asymmetry in the angular distribution at this resonance is due to the contribution from the $d$ wave with a different phase. The angular distribution of \ce{D-} ions from \ce{D2O} at this resonance is exactly similar to \ce{H-} ions.

Figures \ref{fig3.8}(b) and (c) show angular distributions of \ce{H-} and \ce{D-}  ions plotted over smaller intervals of the kinetic energy distribution. All the plots peak close to $90^{\circ}$ and fall off in intensity at lower and higher angles. We see no drastic change in the angular distributions except for a few observations. The plot for ions in the 0 to 1 eV interval (shown by black dots) shows finite intensity at $0^{\circ}$ and $180^{\circ}$ compared to other plots. This is attributed to the fact that the ions with these kinetic energies are incident over a very small radial range (about 20 pixels) compared to ions with 2 eV that hit at about 40 pixels distance. Considering energy resolution of about 0.5 eV of the electron beam and an extended interaction volume, poorer image resolution over this small radial range can lead to comparable intensities at forward-backward angles. Further, the angular distribution seems to be more symmetric about $90^{\circ}$ at higher kinetic energies (or OH fragment in ground vibrational state) than at lower energies (OH fragment in higher vibrational states).
 
The images for \ce{O-} across the first resonance at electron energies 6eV, 7eV and 8 eV from \ce{H2O} are shown in Figure \ref{fig3.4}(a), (b) and (c) and from \ce{D2O} in Figure \ref{fig3.6}(a), (b) and (c) respectively. They show a small blob suggesting very little kinetic energy which we are unable to resolve. The dissociation channel leading to the formation of \ce{O-}  is \ce{O- + H2} (threshold: 3.56 eV). The maximum kinetic energy expected for \ce{O-} is only 1/9th of the available excess energy. If the \ce{H2} molecule is formed in the vibrational ground state, \ce{O-} will have energy close to 0.4 eV. In such a situation, we should be able to see a resolved ring for \ce{O-} as seen at the resonance at 12 eV (as seen in Figures \ref{fig3.4}(g) and (h)). From the size of the image we estimate the kinetic energy of \ce{O-} to be about 0.05 eV. This low energy implies that there is very little excess energy available as kinetic energy, with most of it going into vibrational excitation of the \ce{H2}. This is easily understandable since the formation of H-H bond and the dissociation of the two O-H bonds are occurring simultaneously and most likely this will leave the \ce{H2} molecule with large internuclear separation and hence large vibrational energy. In the case of \ce{O-} from \ce{D2O}, the expected maximum kinetic energy is 0.7 eV. We measure the kinetic energy to be 0.1 eV indicating that the \ce{D2} formed at this resonance will be in high vibrational states. Our results indicate that though low in cross section \cite{c3rawat}, the DEA to water around the 7 eV resonance produces \ce{D2} and \ce{H2} in very high vibrational levels. 

\subsection{Second resonance process peaking at 8.5 eV} 

\begin{figure}[!ht]
\centering
  \subfloat[]{\includegraphics[height=6cm,width=8cm]{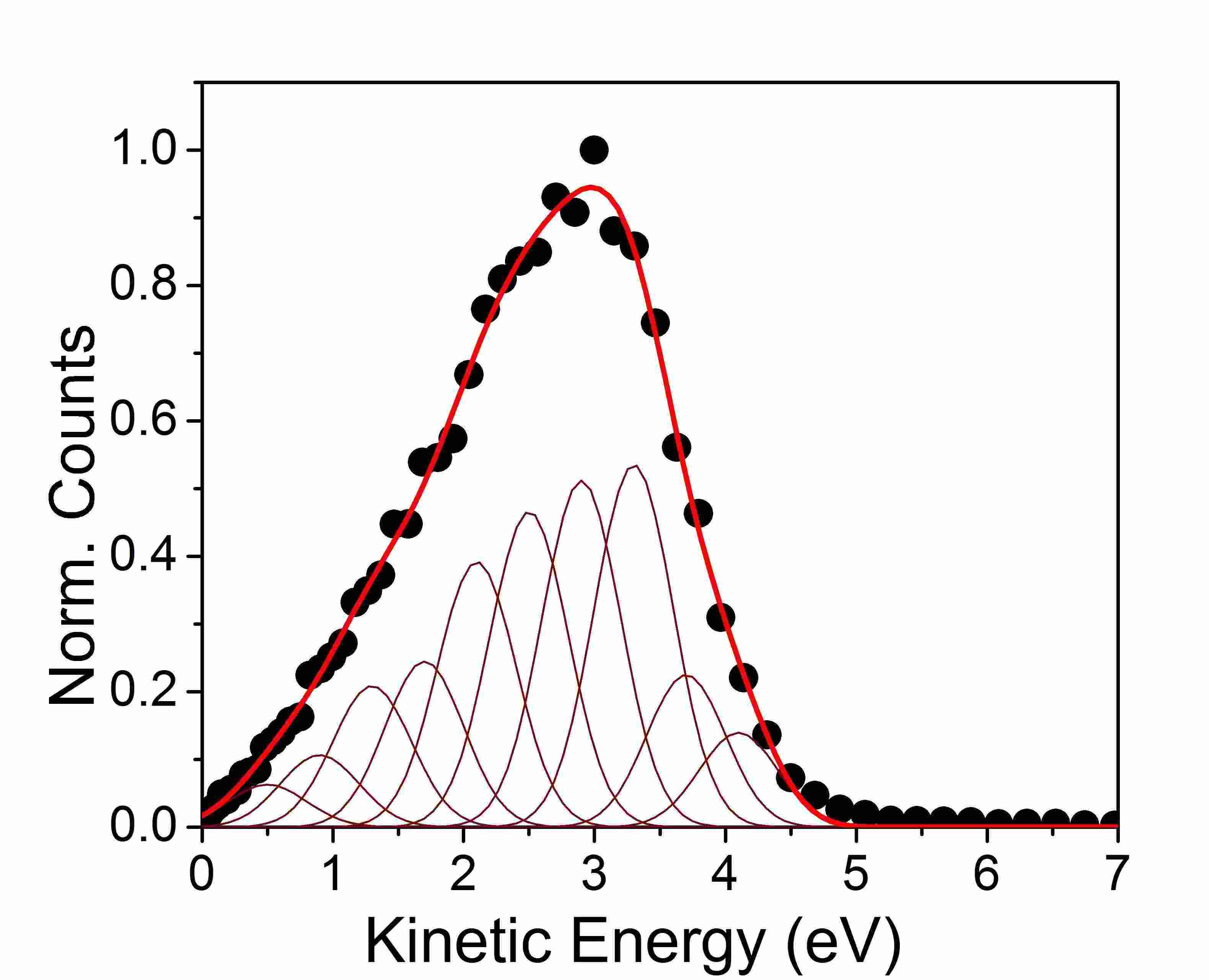}}
  \subfloat[]{\includegraphics[height=6cm,width=8cm]{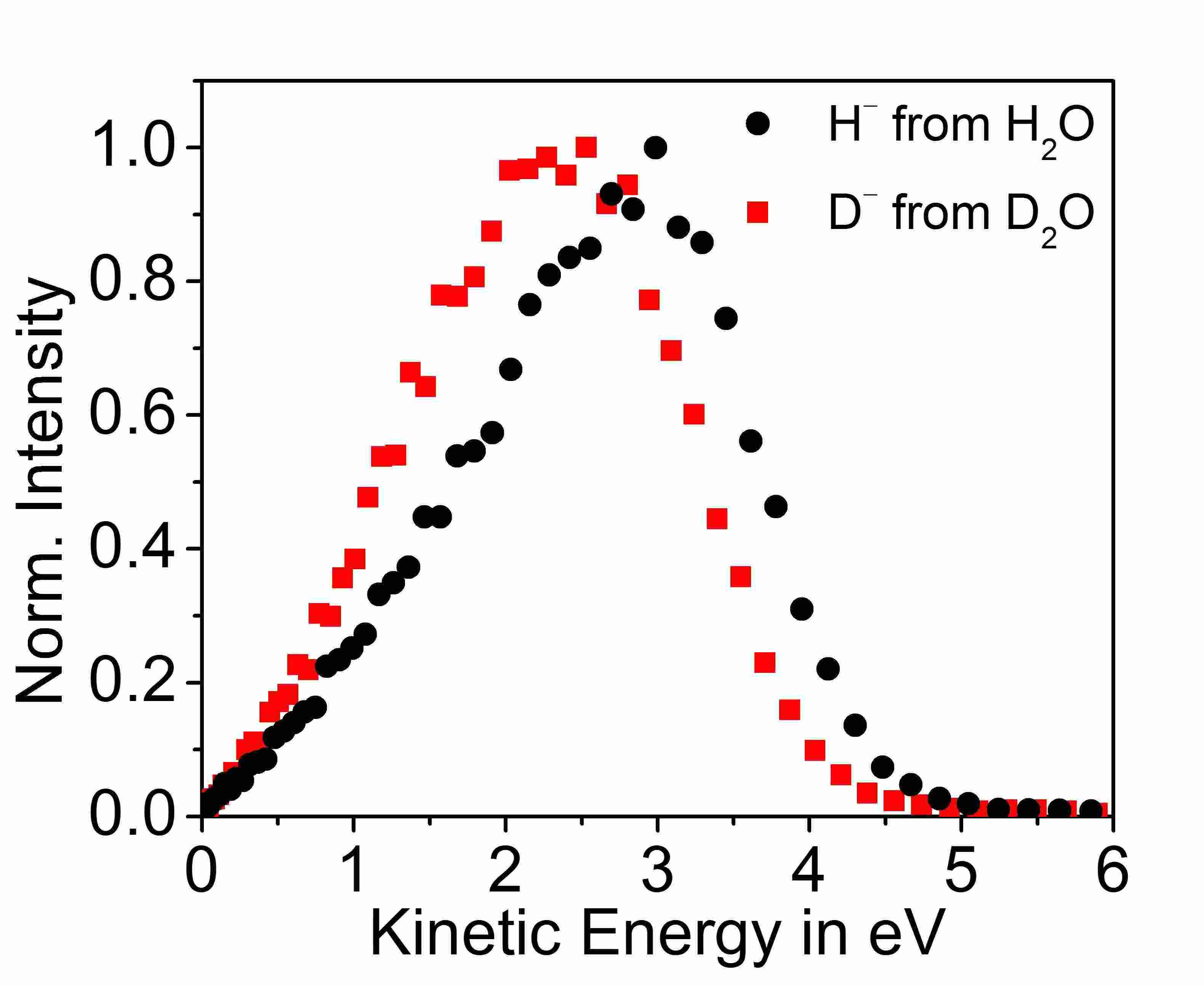}} \\
  \subfloat[]{\includegraphics[height=6cm,width=8cm]{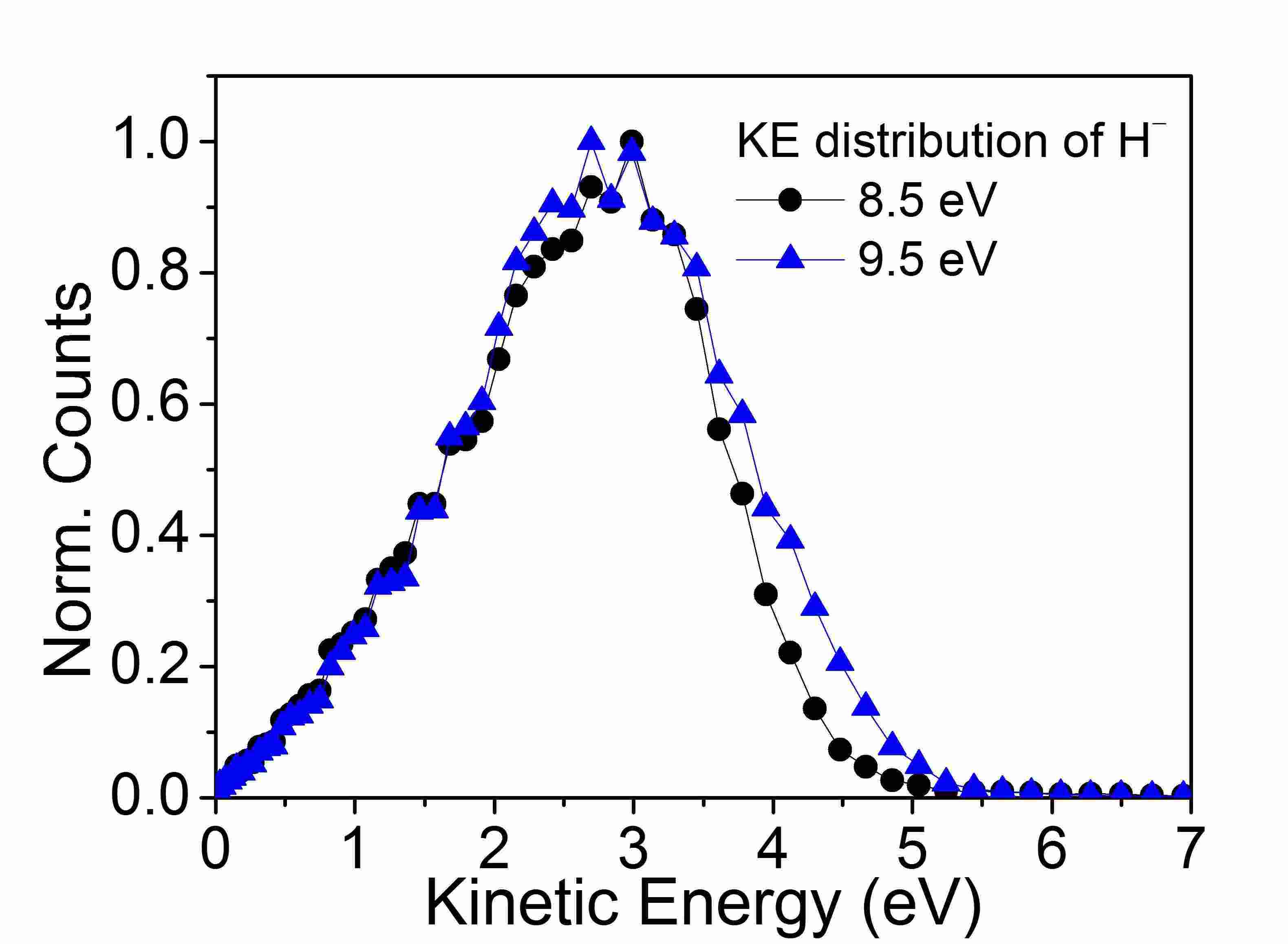}}
  \subfloat[]{\includegraphics[height=6cm,width=8cm]{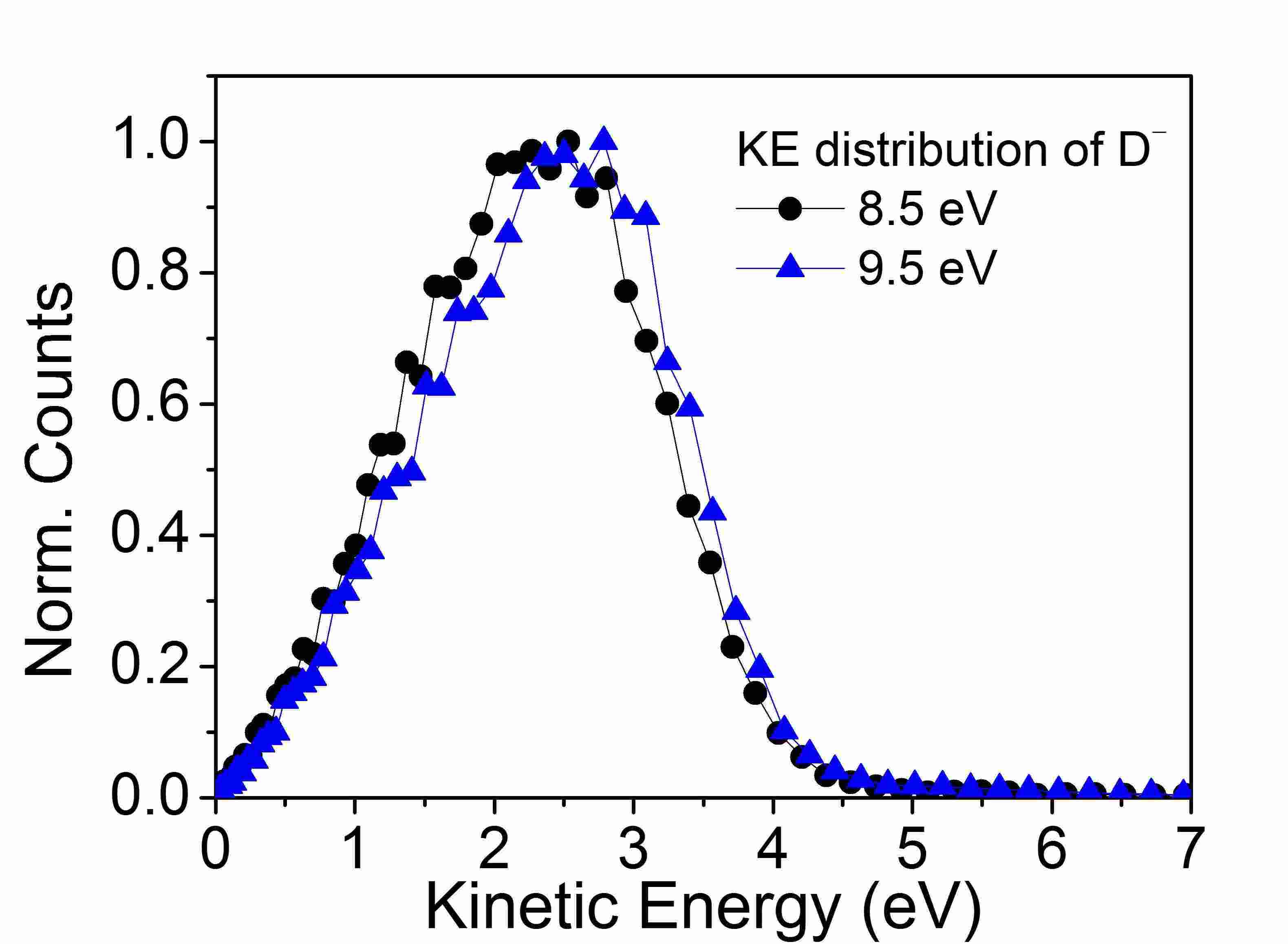}}
\caption{(a) Kinetic energy distribution of \ce{H-} ions at 8.5 eV. The inner curves are the individual Gaussian fits obtained corresponding to vibrational excitations from $\nu$ = 0 (rightmost) to $\nu$ = 9 (leftmost) of the OH radical. (b) Kinetic energy distribution of \ce{D-}  ions from \ce{D2O} compared with that of \ce{H-} from \ce{H2O} at 8.5 eV. Kinetic energy distribution of (c) \ce{H-} from \ce{H2O} and (d) \ce{D-} from \ce{D2O} at 8.5 eV and 9.5 eV electron energies.}
 \label{fig3.9}
 \end{figure}

The second resonance process peaking at 8.5 eV is understood to be a Feshbach resonance where an electron attaches to the \ce{(3a1)^{-1} (4a1)^{1}} excited state of the neutral water molecule and forms a \ce{H2O^{-*}} anion of \ce{^{2}A1} symmetry. From the velocity images taken at 8.5 and 9.5 eV (see Figures \ref{fig3.3}(d), \ref{fig3.3}(e), \ref{fig3.5}(d) and \ref{fig3.5}(e)), \ce{H-} and \ce{D-} ions are seen to be ejected mostly perpendicular to the electron beam (like in first resonance) with finite intensity in other angles as well. The kinetic energy distribution of \ce{H-} ions at 8.5 eV is given in Figure \ref{fig3.9}(a). Here, the kinetic energy ranges from 0 to above 4 eV. With incident electron energy of 8.5 eV, the only channel that allows maximum kinetic energy of \ce{H-} ions to about 4 eV is the \ce{H- + OH(X ^{2}$\Pi$)} channel with threshold energy of 4.35 eV. This is the same dissociation channel producing \ce{H-} ions as at the first resonance. Like in the case of first resonance, we tried to retrieve the vibrational state intensities by fitting contributions at the allowed vibration levels. We find contributions of ten vibrational levels from $\nu$ = 0 to $\nu$ = 9 are needed to fit the observed energy distribution. The relative intensities of these are given in Table \ref{tab3.4} along with those obtained by Belic et al \cite{c3belic}. Though the maximum probability appears to be around $\nu$ = 2 in both, the relative intensity distribution among the various levels appears to be different. The relative intensity distribution seems to go down faster in the data of Belic et al \cite{c3belic} with the increase in vibrational levels. The difference in the distributions could be attributed to the focusing of the energy analyser at 3 eV by Belic et al \cite{c3belic}. This may have led to discrimination of the lower energy ions which correspond to higher vibrational levels. The \ce{H-} kinetic energy spectra at 8.5 eV resonance as calculated by Haxton et al \cite{c3h6}, shows the distribution peaking close to 3.5 eV which corresponds to $\nu$ = 1 vibrational state in the OH radical. The kinetic energy distribution we measured for \ce{D-} ions is similar to \ce{H-} ions at this resonance except for scaling down as seen in Figure \ref{fig3.9}(b) due to difference in masses and slightly higher threshold of the \ce{D- + OD (^{2}$\Pi$)} channel i.e. 4.52 eV. Looking at the variation of KE distribution, the plots at 8.5 eV and 9.5 eV for \ce{H-} and \ce{D-} ions in Figure \ref{fig3.9}(c) and \ref{fig3.9}(d) show a slight increase in the maximum kinetic energy but there is no appreciable change in the width of the distribution. 

\begin{table}[h]
\caption{Relative intensities of OH vibrational levels at 8.5 eV}
\begin{center}
\begin{tabular}{cccc}
\hline
\\
Vibrational Level & Theory$^{a}$ & Earlier Expt.$^{b}$ & Present Expt.\\
\\
\hline
\\
0 & 0.98 & 0.40 & 0.26 \\
\\
1 & 1.20 & 0.93 & 0.42 \\
\\
2 & 1.00 & 1.00 & 1.00 \\
\\
3 & 0.72 & 0.67 & 0.96 \\
\\
4 & 0.51 & 0.41 & 0.88 \\
\\
5 & 0.36 & 0.28 & 0.74\\
\\
6 & 0.26 & 0.24 & 0.46\\
\\
7 & 0.18 & 0.21 & 0.39\\
\\
8 & 0.12 & 0.14 & 0.20\\
\\
9 & 0.08 & 0.09 & 0.11\\
\\
\hline
\end{tabular}
\end{center}
\hspace{3cm}$^{a}$Haxton et al. \cite{c3h6}; $^{b}$Belic et al. \cite{c3belic}
\label{tab3.4}
\end{table} 

\begin{figure}[!ht]
\centering
  \subfloat[]{\includegraphics[width=0.3\columnwidth]{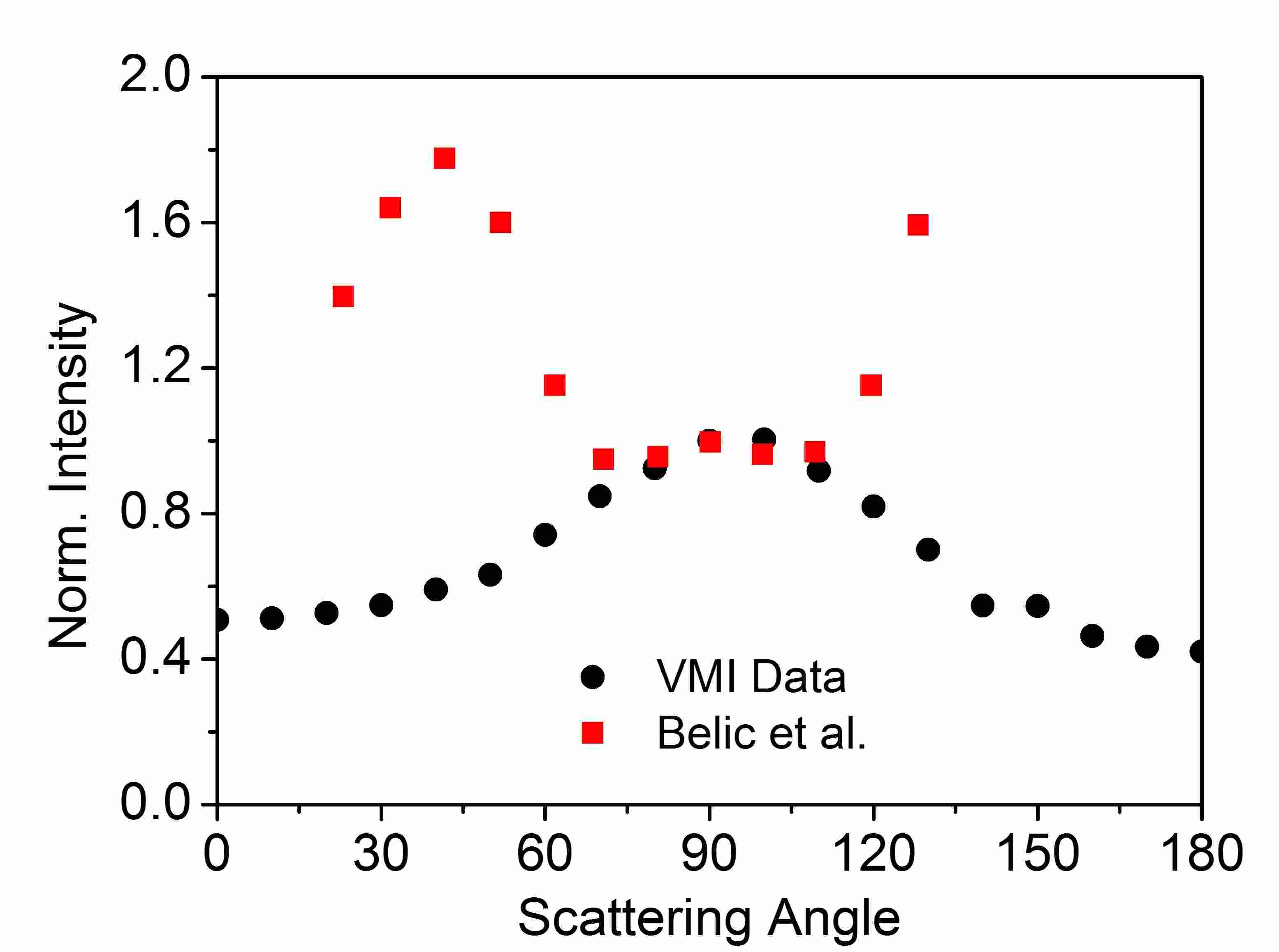}}
  \subfloat[]{\includegraphics[width=0.3\columnwidth]{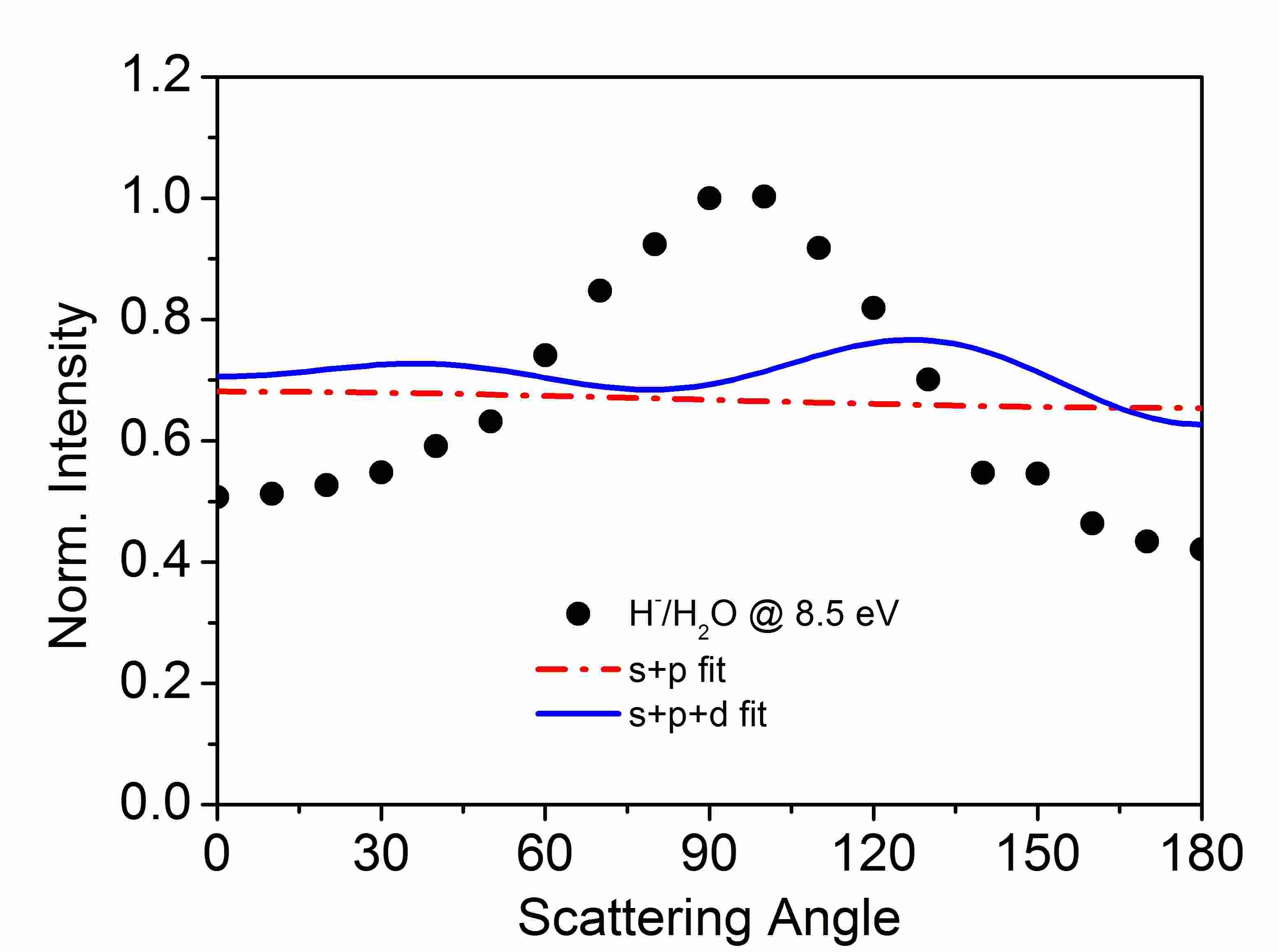}}
  \subfloat[]{\includegraphics[width=0.3\columnwidth]{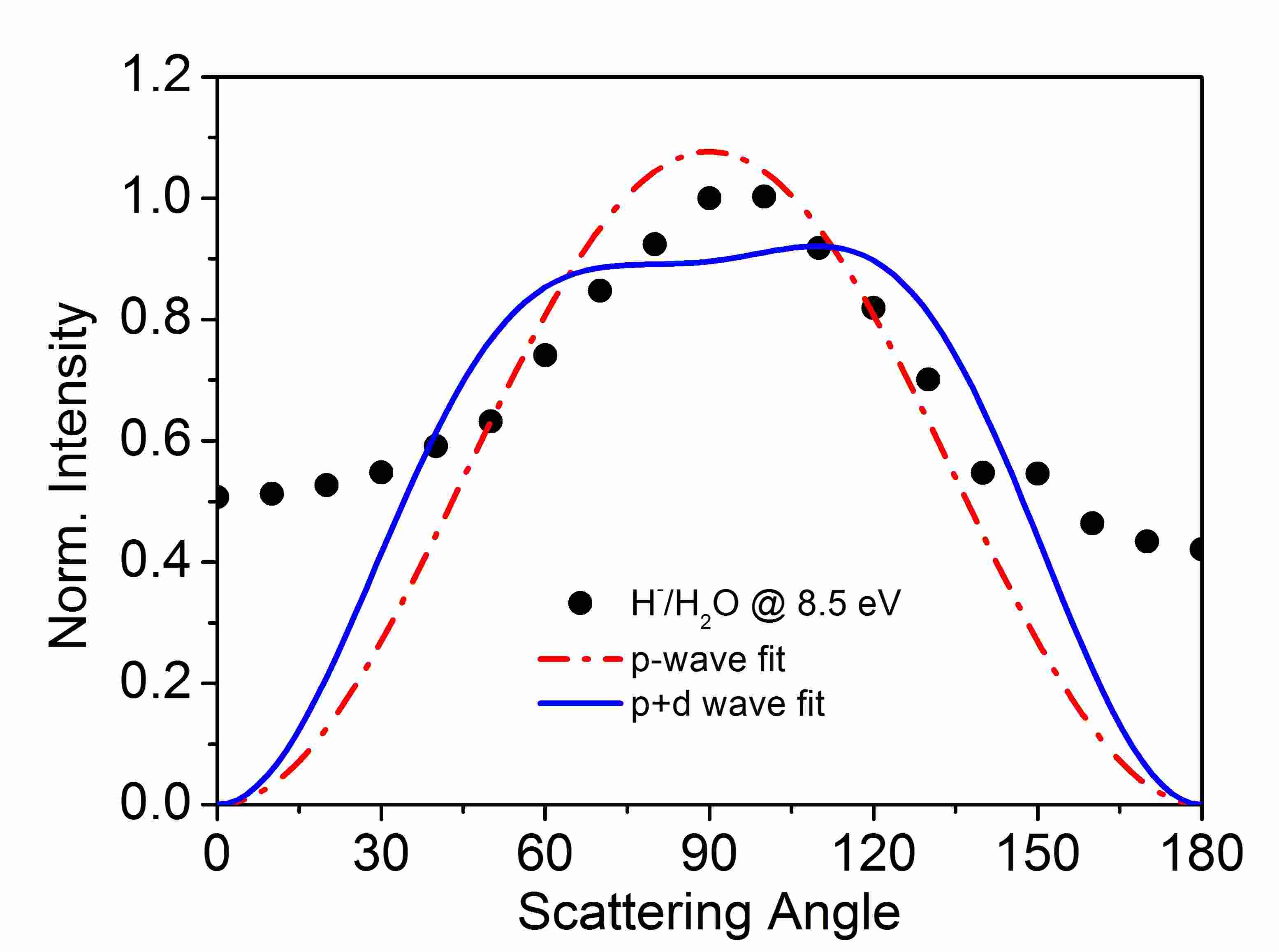}} \\
  \subfloat[]{\includegraphics[width=0.3\columnwidth]{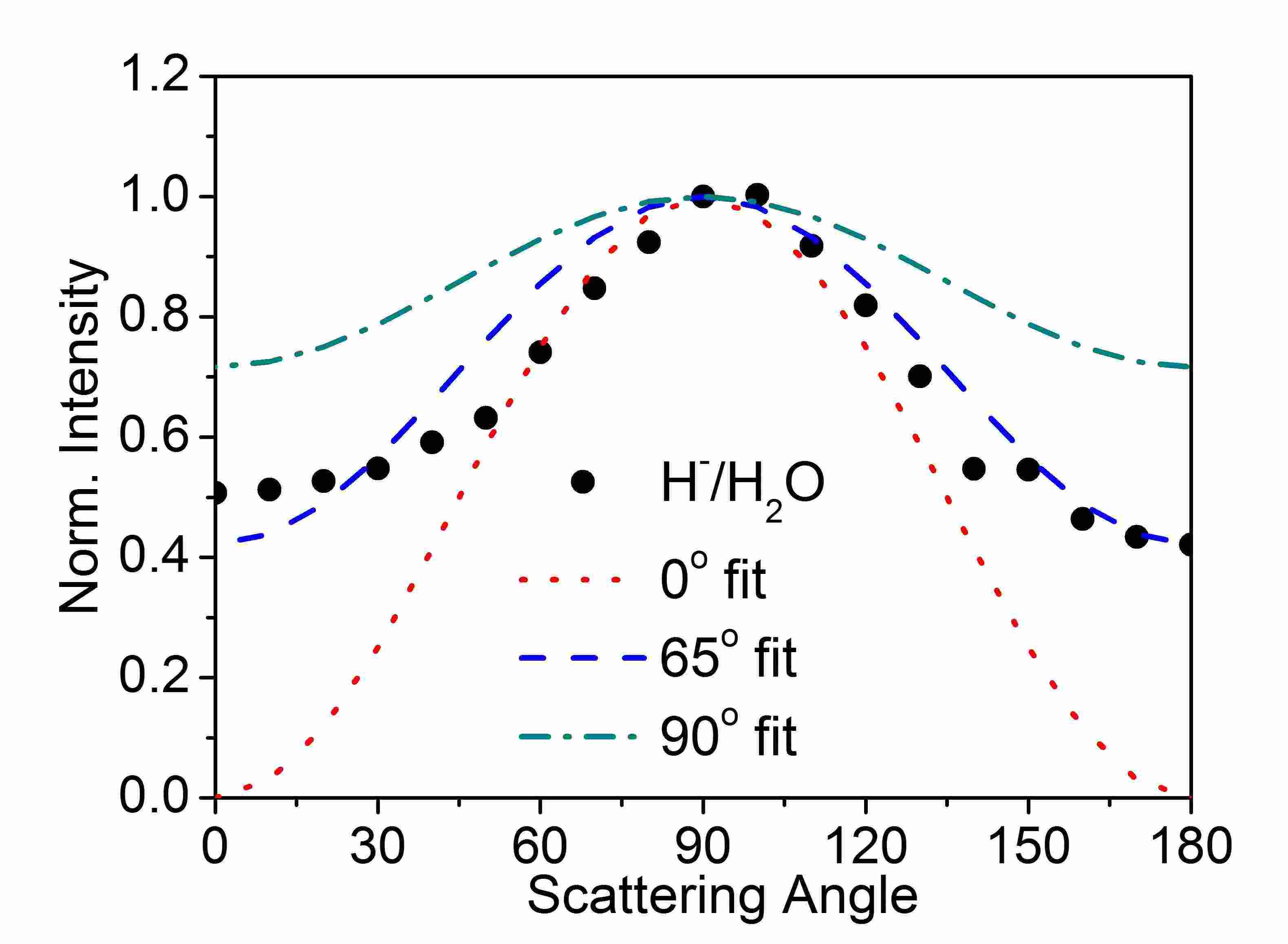}}
  \subfloat[]{\includegraphics[width=0.3\columnwidth]{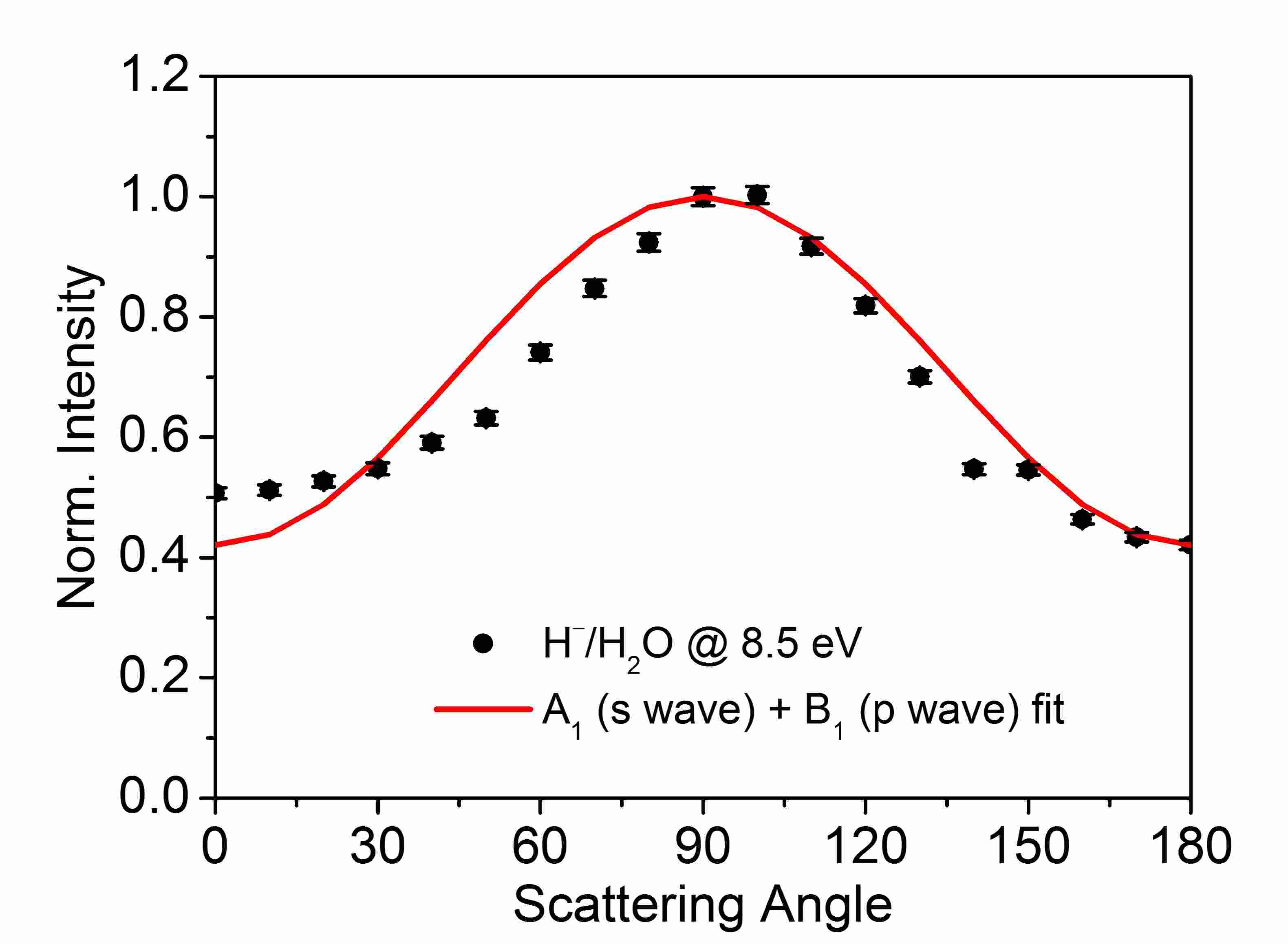}}
\caption{(a) Comparison of \ce{H-} data at 8.5 eV from our setup and Belic et al. \cite{c3belic} (b) \ce{A1} and (c) \ce{B1} symmetry fits to our angular distribution data (d) Comparison with simulated angular distribution curves for the \ce{B1} symmetry, taking into account rotational effects up to different angles - $0^{\circ}$ (dotted curve), $65^{\circ}$ (dashed line) and $90^{\circ}$ (dash dot curve). (e) The red solid line is the fit with an $s$-wave contribution from a state of \ce{A1} symmetry and a $p$-wave contribution from a state of \ce{B1} symmetry.}
 \label{fig3.10}
 \end{figure}

The angular distribution of \ce{H-} ions at 8.5 eV obtained by integrating over the entire kinetic energy distribution is shown in Figure \ref{fig3.10}(a) along with the previous results from Belic et al \cite{c3belic}. The angular distribution that we have obtained has an isotropic contribution along with a bell-shaped distribution peaking at $90^{\circ}$. The data for \ce{D-} are very similar to that of \ce{H-}. As shown in Figure \ref{fig3.10}(a), the angular distribution measured by Belic et al \cite{c3belic} peaks at $45^{\circ}$ and $135^{\circ}$ respectively with finite intensity at $90^{\circ}$ and corresponds to \ce{H-} ions produced with OH in vibrational ground state ($\nu$=0). They claimed that the peak at $90^{\circ}$ due to \ce{H-} ions coming from the first resonance process and hence to minimise the contribution from the first resonance, they report the angular distribution of \ce{H-} ions produced with maximum kinetic energy. Based on these results, they concluded that the resonance to be \ce{^{2}A1} state. The angular distribution from our velocity imaging measurement also peaks at $90^{\circ}$, but contribution from the first resonance arising due to the finite energy resolution of the electron beam (about 0.5 eV) was ruled out. The \ce{H-} angular distribution at this energy is distinctly different from that at the first resonance and measurement at 9.5 eV (where the contribution from the first resonance is negligible) is also found to be similar to that at 8.5 eV. 

Initially, we obtained the angular distribution from the velocity image by integrating over the entire radial range of the image (as shown in Figure \ref{fig3.10}(a)) and proceeded to explain our result in the following way.  To determine the symmetry of the resonance, we fit our data for states of different symmetry, using the calculations for angular distribution as described in Section 3.3. The results for the \ce{A1} symmetry are given in Figure \ref{fig3.10}(b). The three partial waves with lowest $l$-values which could contribute to the \ce{A1} state are the $s$, $p$ and $d$ waves. The probability that higher partial waves will contribute is increasingly small. It is clear from the figure that the fit obtained using these three partial waves are quite inadequate to explain our results. Fits for the \ce{B1} symmetry using the $p$-wave and a combination of $p$ and $d$ waves are given in Figure \ref{fig3.10}(c). Of these two, the lone $p$-wave fits well the peak seen at $90^{\circ}$. However, this does not agree with the finite contribution that we see in the forward and backward directions. It might appear that contribution from an $s$-wave should be able to explain this discrepancy. However, based on the symmetry, an $s$-wave contribution to a \ce{B1} state is ruled out. Other possible symmetries like \ce{A2} and \ce{B2} also do not explain the observed angular distribution. To summarize, the \ce{B1} symmetry appeared to be the closest choice for explaining the results except for the finite contribution in the forward and backward directions.

Next, we tried to account for this finite intensity in the forward-backward angles in the following way. The angular distribution fits discussed above is based on the axial recoil approximation in which it is assumed that the dissociation takes place in a time scale before the molecule could undergo rotation or structural changes. The formation of the \ce{B1} resonance requires that the electron approach the molecule normal to its plane. Thus, even if there are changes in the molecular geometry (due to bending mode vibration in this case) before the dissociation happens, the angular distribution should correspond to the $p$-wave cross sections shown in Figure \ref{fig3.10}(c). However, if the lifetime of the state is long enough that the molecule undergoes rotation before it dissociates, then the angular distribution would get smeared out. We attempted a simulation of such a process by taking into account the effect of rotation before the molecules dissociate. We found that rotation about the principal axis by an angle of $65^{\circ}$ would give a fairly good fit as seen in Figure \ref{fig3.10}(d). It may be noted that the bending mode vibration coupled with rotation could smear out the distribution further. The rotation angle of $65^{\circ}$ in this case corresponds to a lifetime of $1.2 \times 10^{-13}$ sec for this resonance. Jungen et al \cite{c3jungen} calculated the lifetime of this state to be $6 \times 10^{-14}$ sec, based on the \ce{A1} symmetry. That the resonance dissociates on a slightly longer timescale points to the fact that it may initially be formed in a bound potential energy surface and the dissociation occurs through coupling with other surfaces. We wish to point out that there is a slight asymmetry between $0^{\circ}$ and $180^{\circ}$ in the angular distribution data, which we could not account for. In Figure \ref{fig3.10}(e), we also give a fit using contribution from both \ce{A1} and \ce{B1} states. Here, we consider only the lowest possible partial waves - $s$-wave for \ce{A1} and $p$-wave for \ce{B1}. The relative contributions from the \ce{A1} and \ce{B1} states were found to be 1:1.1. We find that this fit is in reasonable agreement with the data. Based on this, the presence of the \ce{A1} state cannot be ruled out at this energy. However, the Wigner-Witmer rule and the near-absence of \ce{OH-} at this resonance seem to rule out the contribution from an \ce{A1} state. 

One of the puzzles in the DEA to water has been the very low cross section for the formation of \ce{O-} and \ce{OH-} as compared to that for \ce{H-} despite their lower thresholds and the larger electron affinities for O and OH as compared to H. The low cross section for the formation of \ce{O-} may be explained since it involves considerable structural changes including the breaking of two O-H bonds and the formation of the H-H bond at the lower energy resonances. However, the formation of \ce{OH-} is a simple two-body breakup and there appeared no reason for the observed low cross section. Fluendy and Walker \cite{c3fluendy} argued that while the unexpectedly low cross sections for the formation of \ce{OD-} (or \ce{OH-}) at the 6.5 eV resonance could be ascribed to its \ce{B1} symmetry preventing the formation of the \ce{OH-} anion in the electronic ground state based on Wigner- Witmer rule and that at the 11.8 eV resonance due to the dissociation of the OH anion, there is no valid explanation for the low cross section at the 8.5 eV resonance, if it has an \ce{A1} symmetry. Hence, they felt that the observed low cross section for the formation of \ce{OD-} at the 8.5 eV resonance should make the assignment of the \ce{A1} symmetry to it doubtful. Their argument that the presence of a state with the \ce{A1} symmetry should lead to larger cross section for the formation of \ce{OH-} may be used to remove the ambiguity we observe in explaining the observed angular distribution in Figure \ref{fig3.10}(d). 

While the resonance at 8.5 eV is attributed to \ce{3a1 -> 3sa1} excitation giving rise to the \ce{A1} symmetry, there is some evidence supporting the \ce{B1} symmetry for this resonance. From the measurements of Chutjian et al \cite{c3chutjian}, we identify the resonance at 8.5 eV to be arising from the \ce{^{3}A2} (\ce{b1 -> 3pb2}) state. Their EEL data \cite{c3chutjian} on \ce{D2O} at an electron residual energy of 0.5 eV and a scattering angle of $90^{\circ}$ show a broad structure at 8.9/9.1 eV, which is identified as a dipole forbidden transition (\ce{b1 -> 3pb2}) causing the \ce{^{3,1}A2} state. With the extra electron captured in the \ce{3pb2} state, the overall symmetry would be \ce{B1} since \ce{A2 $\times$ B2 -> B1}. Following the work by Curtis and Walker \cite{c3curtis}, Fluendy and Walker \cite{c3fluendy} did classical trajectory calculations to relate to the formers' results. Interestingly, they could explain all the essential features of the experimental findings \cite{c3curtis} using only two anionic potential surfaces of symmetry \ce{^{2}B1} and \ce{^{2}B2}. They suggested that both the first and the second resonance in water be ascribed to the decay of a single \ce{^{2}B1} state but with different partial wave contributions at the higher energy. 

Thus, based on these arguments, we concluded that the observed angular distribution is due to the \ce{B1} state arising from a dipole forbidden transition in the neutral with the dissociation taking place at larger time scale so that rotational effects contribute \cite{c3nbrjpb}

\begin{figure}[!ht]
\centering
  \subfloat[\ce{H-} at 8.5 eV]{\includegraphics[width=0.4\columnwidth]{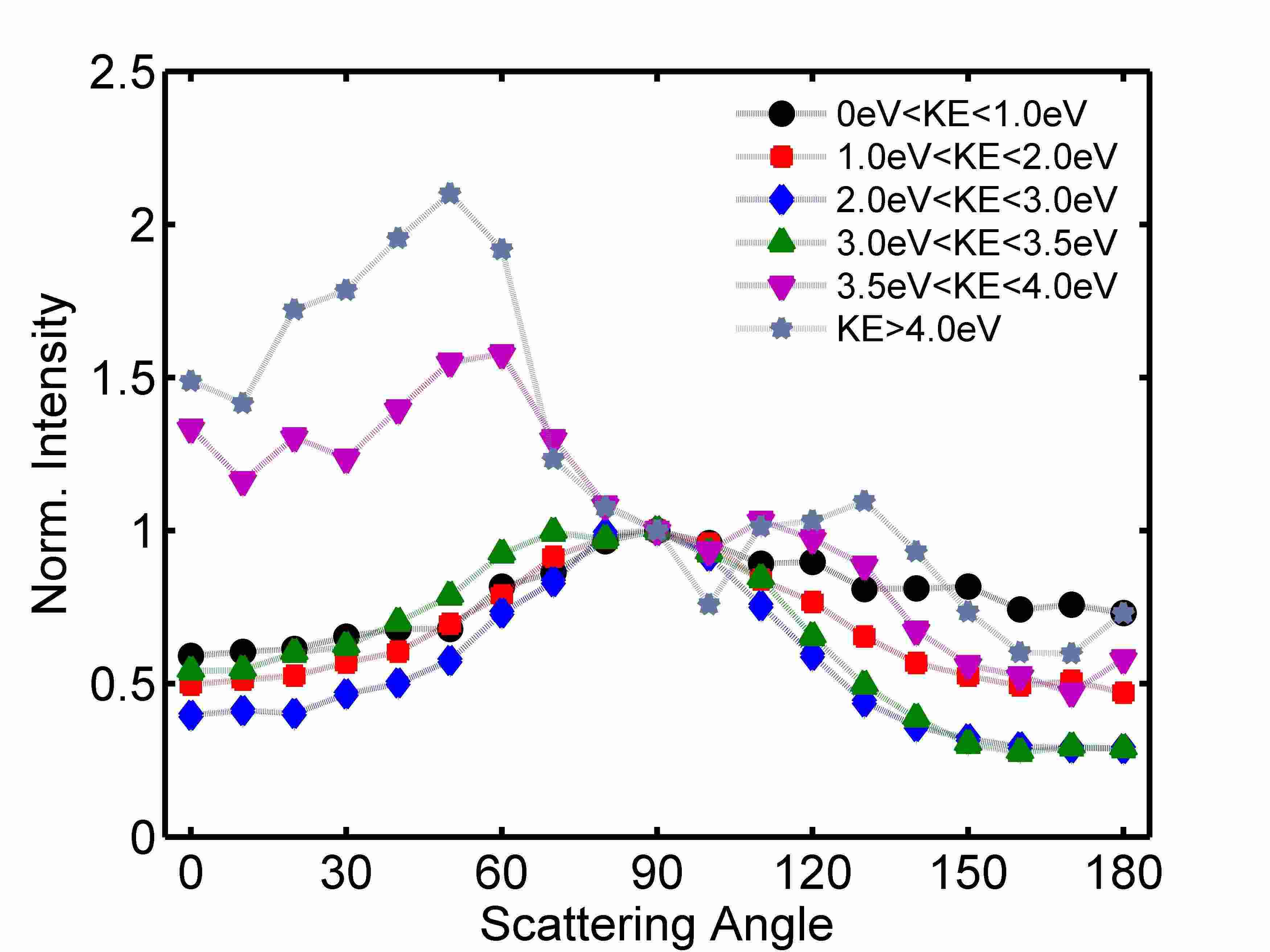}}
  \subfloat[\ce{H-} at 9.5 eV]{\includegraphics[width=0.4\columnwidth]{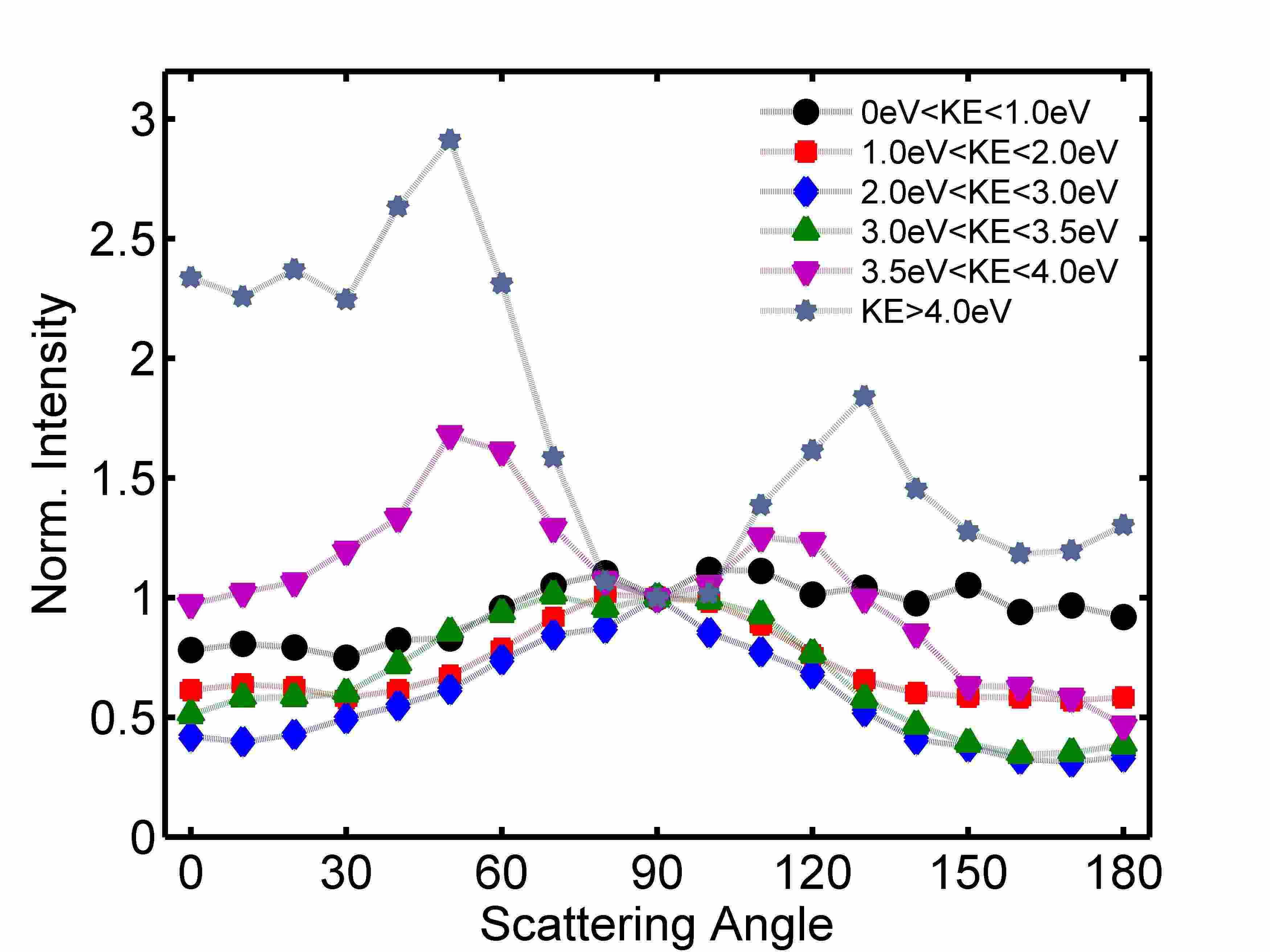}} \\
  \subfloat[\ce{D-} at 8.5 eV]{\includegraphics[width=0.4\columnwidth]{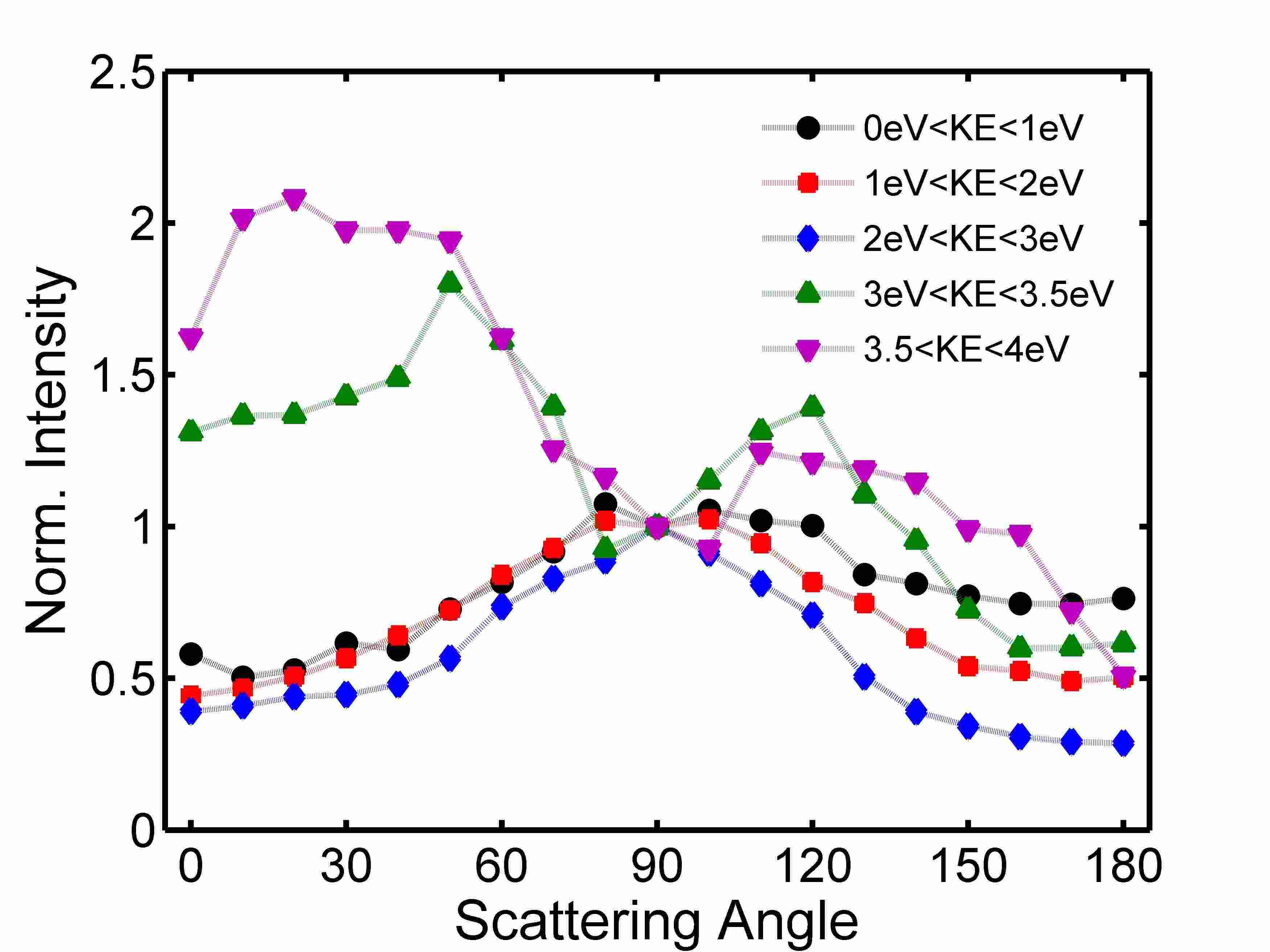}}
  \subfloat[\ce{D-} at 9.5 eV]{\includegraphics[width=0.4\columnwidth]{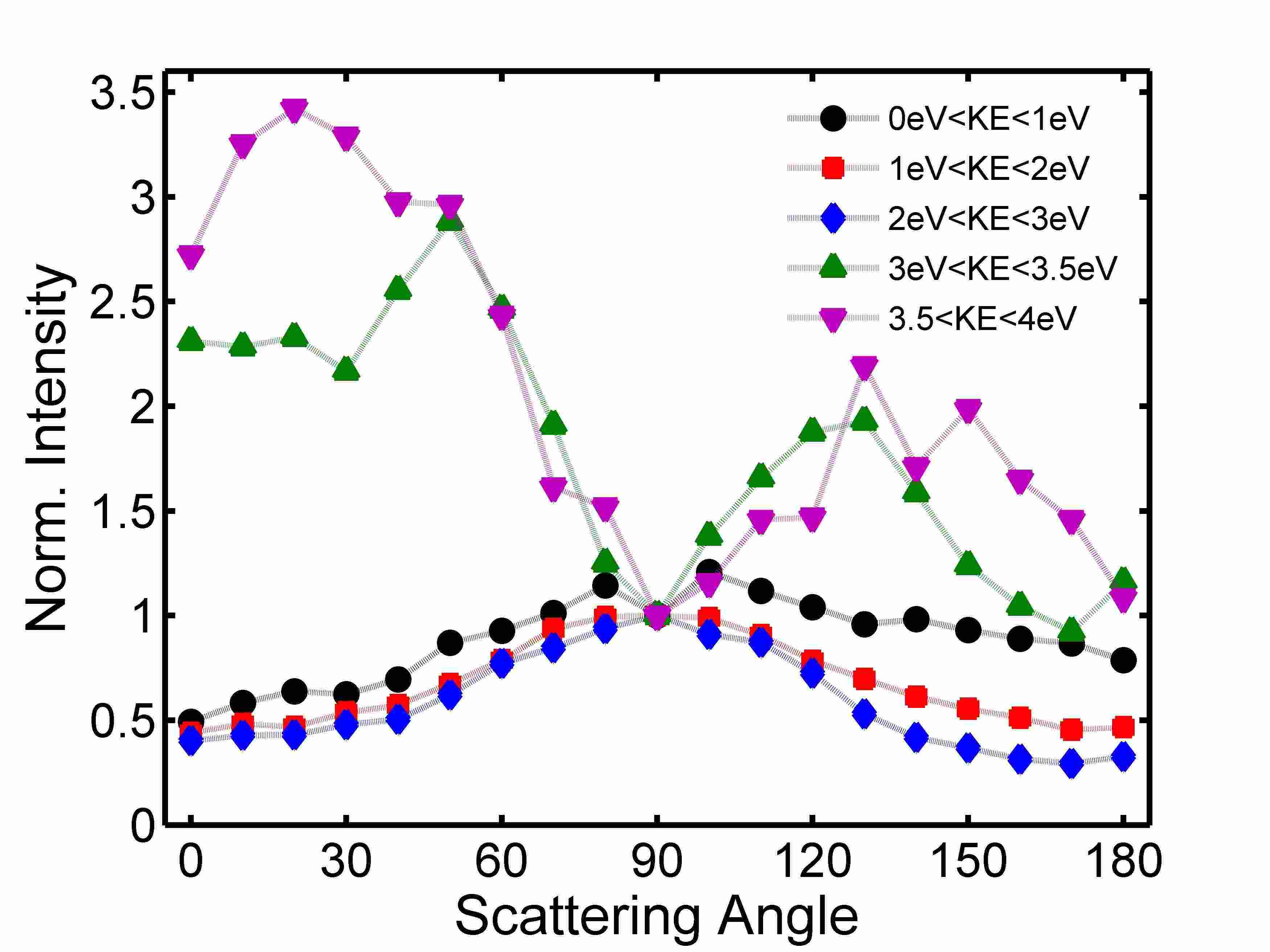}}
\caption{(Angular plots of \ce{H-} and \ce{D-} ions (normalized at $90^{\circ}$) from the second resonance process at 8.5 eV and 9.5 eV as a function of kinetic energy. These plots show variation in angular distribution as a function of kinetic energy suggesting structural changes of the water anion due to bending mode vibrations prior to dissociation.}
 \label{fig3.11}
 \end{figure}

However, we realised that the above conclusion is not true. On careful analysis of the \ce{H-} and \ce{D-} velocity images, we found that the angular distribution varies with kinetic energy (or the radius of the image). The angular plots as a function of kinetic energy are shown in Figure \ref{fig3.11} for 8.5 eV and 9.5 eV. We find that \ce{H-} (\ce{D-}) with higher kinetic energies (close to 4 eV; OH in ground vibrational state) peaks at $45^{\circ}$ and $135^{\circ}$ as observed by Belic et al \cite{c3belic} whereas ones with lower kinetic energies (OH in higher vibrational states) peak at $90^{\circ}$ with sufficient intensity at forward and backward angles. Recent report by Adaniya et al. \cite{c3adaniya} also mentioned that the \ce{H2O^{-*}} molecular negative ion at 8.5 eV once formed undergoes structural changes due to bending mode vibrations before the dissociation could take place, thereby distorting the angular distributions. Their classical trajectory calculations using the potential energy surfaces constructed for $^{2}A^{\prime}$ state showed that the H-O-H bond angle opens up quickly following electron attachment and reorients it towards and beyond linear geometry at $90^{\circ}$. Trajectories in which one of the H atoms takes more of the available kinetic energy are found to undergo less bending motion before dissociation. Thus, what we learn is, the dynamics of \ce{H-} ions taking away most of the available excess energy as kinetic energy can be described well by the axial recoil approximation and the \ce{H-} ions with lower kinetic energies ejected at $90^{\circ}$ as seen in the velocity images are due to water anion becoming linear due to bending mode vibrations and ejecting \ce{H-} ions in this geometry.

\begin{figure}[!ht]
\centering
  \subfloat[]{\includegraphics[width=0.4\columnwidth]{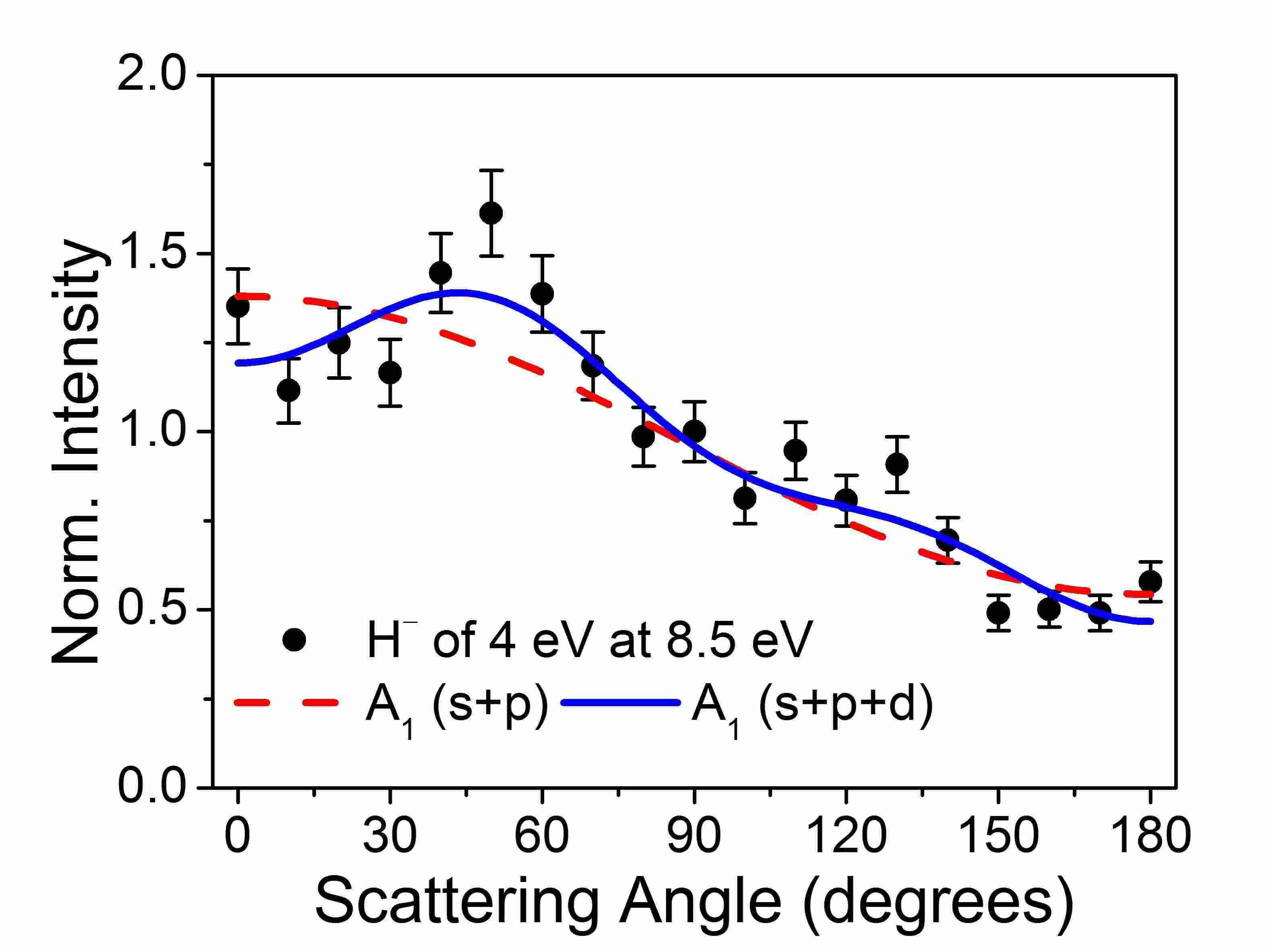}}
  \subfloat[]{\includegraphics[width=0.4\columnwidth]{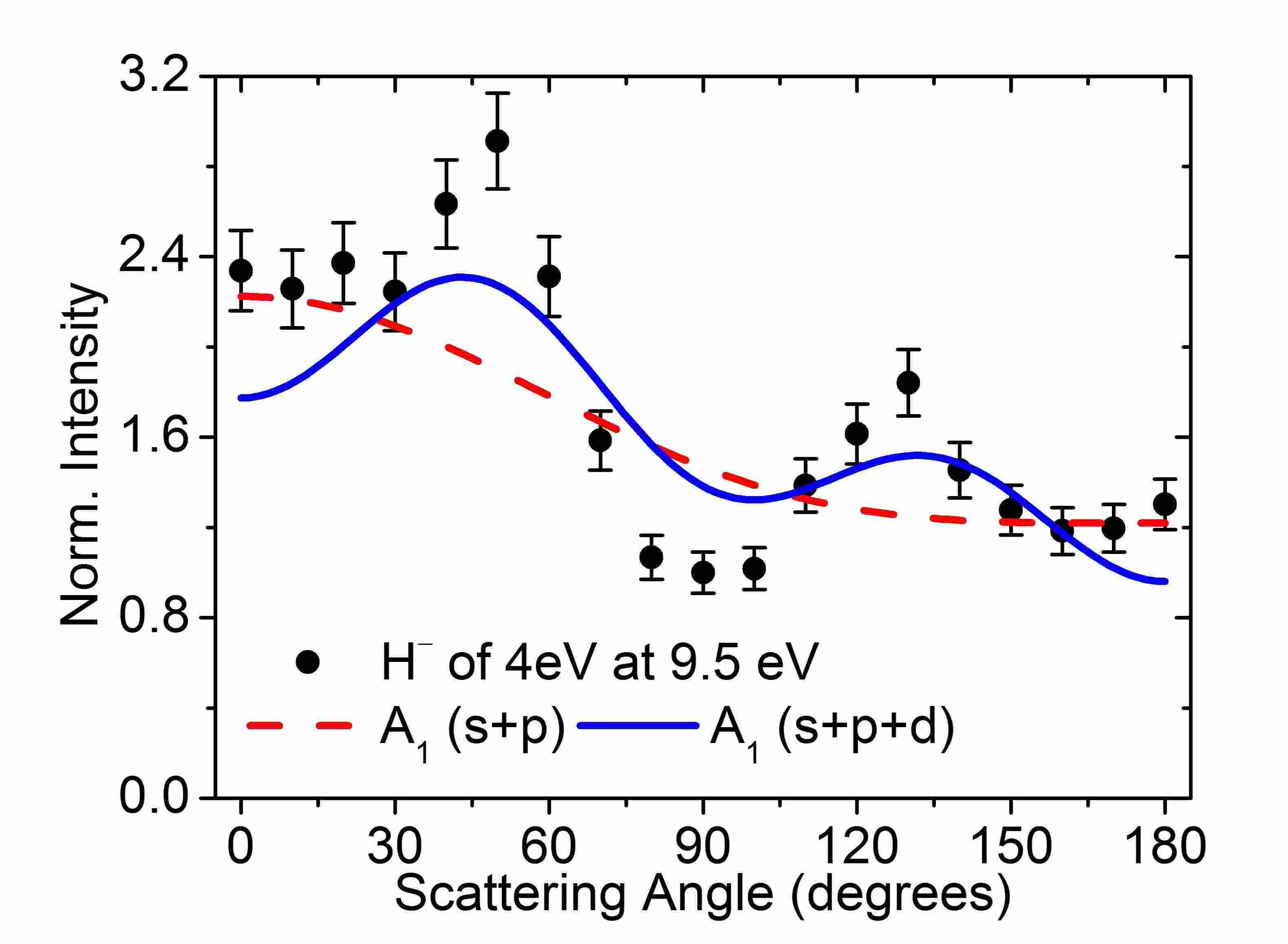}} \\
  \subfloat[]{\includegraphics[width=0.4\columnwidth]{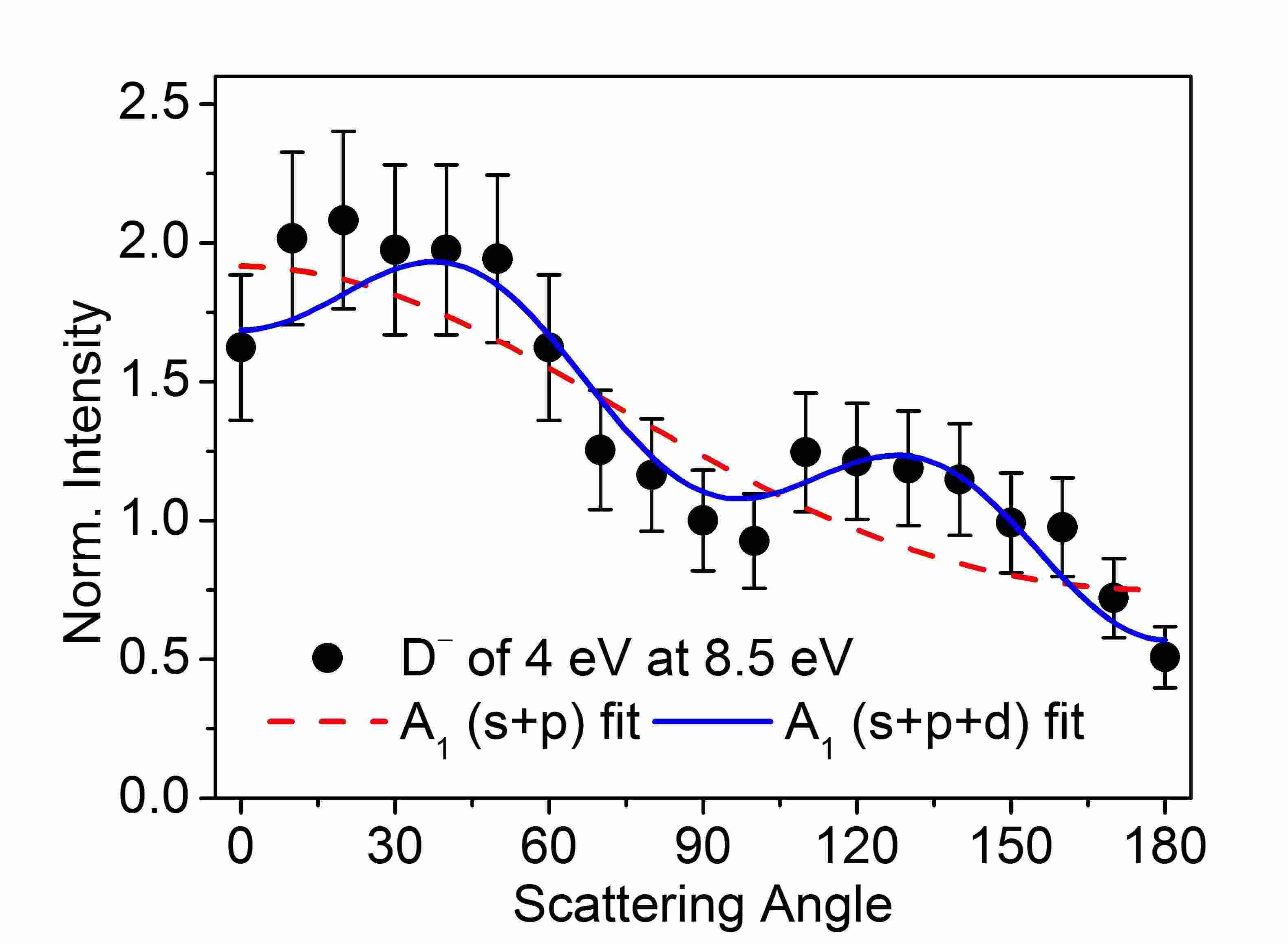}}
  \subfloat[]{\includegraphics[width=0.4\columnwidth]{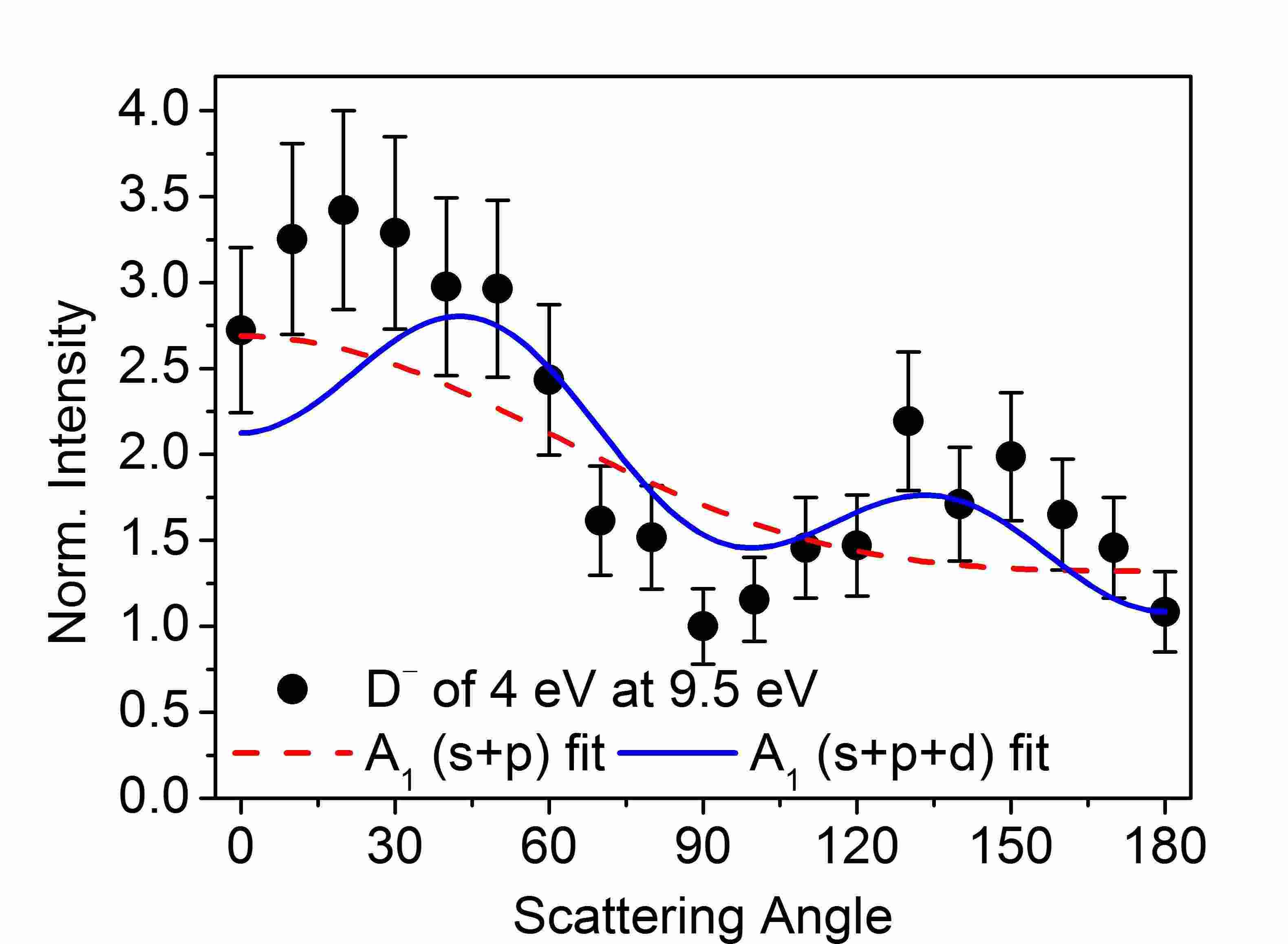}}
  \\
  \subfloat[]{\includegraphics[width=0.4\columnwidth]{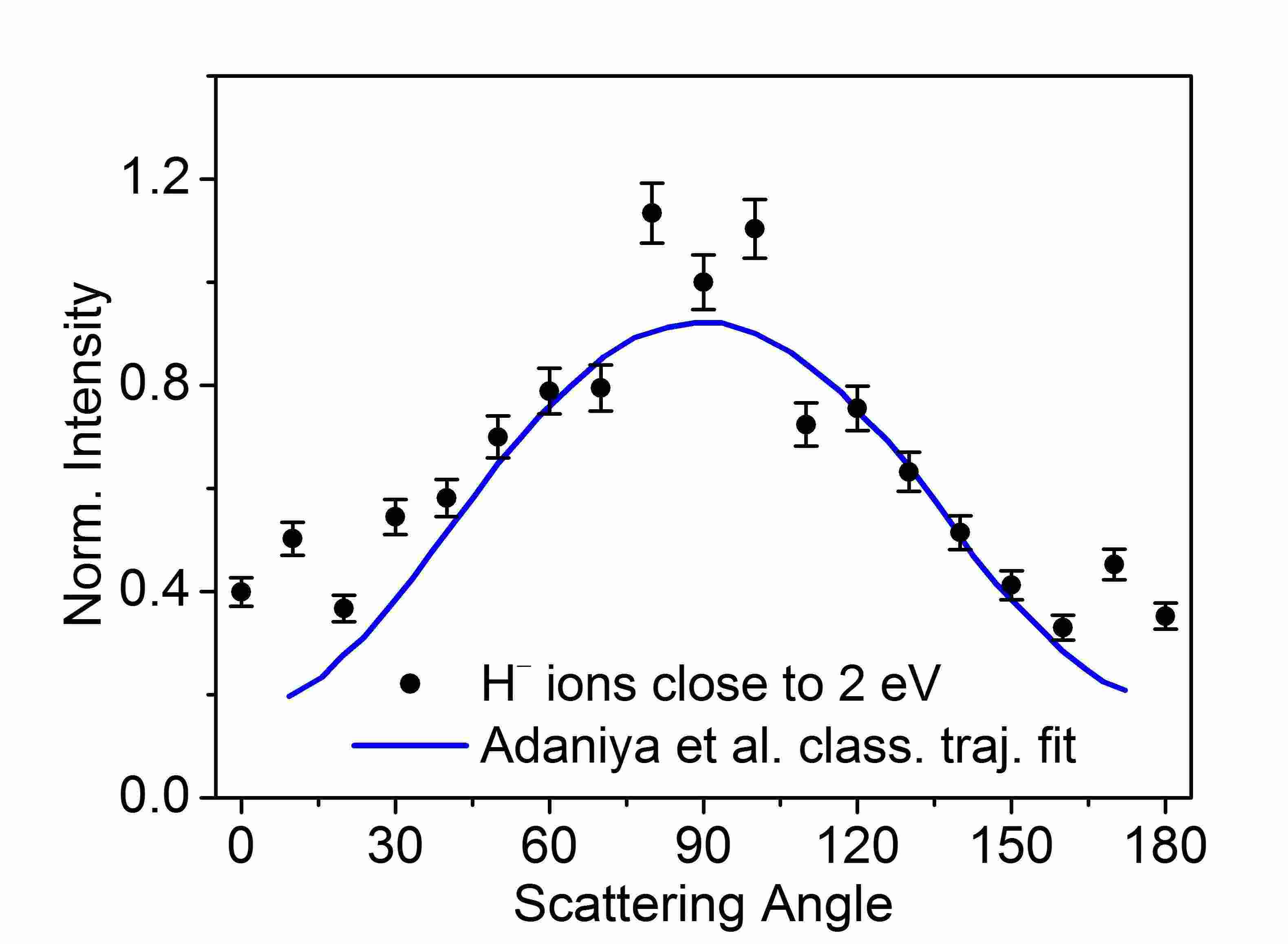}}
\caption{(a)-(d) \ce{A1} symmetry fits for \ce{H-} and \ce{D-} ions with 4 eV KER from the second resonance process at 8.5 eV and 9.5 eV. (e)\ce{H-} ions with KE 2 eV at 8.5 eV electron energy and comparison with classical trajectory curve given by Adaniya et al. \cite{c3adaniya}}
\label{fig3.12}
\end{figure}

On comparing our results with that of Adaniya et al. \cite{c3adaniya}, we find that their data for the kinetic energy release of 4 eV and the classical trajectory results are significantly different from the axial recoil model. In contrast, our results for 4 eV KER at 8.5 eV and 9.5 eV show peaks at $45^{\circ}$ and $135^{\circ}$ similar to what is expected for \ce{A1} resonance under axial recoil approximation, as shown by the fairly matching fits in Figure \ref{fig3.12}(a) - (d), taking the first three dominant partial waves ($s$, $p$ and $d$).  For smaller KER, the angular distribution tends to peak at $90^{\circ}$, as shown in Figure \ref{fig3.12}(e). Here again, our results are more consistent with possible deviation from axial recoil approximation as seen by the better agreement with the classical trajectory calculation than the experimental results in \cite{c3adaniya}.

The \ce{O-} formation from the second resonance process peaks at 9 eV as shown in Figure \ref{fig3.1}(b) and the images shown in Figures \ref{fig3.4}(d) - (e) and \ref{fig3.6}(d) - (e) are taken at 9 eV and 10 eV respectively. Like in the case of the first resonance, we are unable to see a resolved image for \ce{O-} in both \ce{H2O} and \ce{D2O} at this resonance too. The threshold for three-body breakup leading to \ce{O-} formation is 8.04 eV and earlier measurements \cite{c3rawat} show that the entire \ce{O-} formation at this resonance occurs above this energy. The \ce{O- + D + D} channel was previously observed for \ce{D2O} \cite{c3curtis}. Thus we may conclude that the \ce{O-} is formed in this case entirely by a three-body fragmentation. Assuming instantaneous and symmetric three-body fragmentation, the \ce{O-} kinetic energy would be equal to $E_{o} \cos^{2}\theta/(8+\cos^{2}\theta)$ for \ce{H2O} and $E_{o}\cos^{2}\theta/(4 + \cos^{2}\theta)$ in the case of \ce{D2O} where $E_{o}$ is the excess energy and $\theta$ is half the bond angle. The maximum KE would be $E_{\circ}/9$ and $E_{\circ}/5$ i.e. 0.11eV and 0.19 eV respectively at incident electron energy 9 eV for \ce{H2O} and \ce{D2O} respectively. For a bond angle of $105^{\circ}$, the energies work out to be 0.042 eV and 0.081 eV respectively. From the velocity images, we estimate a kinetic energy of 0.06 eV for \ce{O-} from \ce{H2O} and 0.1 eV for \ce{O-} from \ce{D2O}. The bond angles corresponding to these observed energies works out to be $86^{\circ}$ in the case of \ce{H2O} and $99^{\circ}$ in the case of \ce{D2O}. The smaller bond angles show that as the negative ion resonance dissociates, the two H (D) atoms are getting closer and that the dissociation limit may be the \ce{H2} (\ce{D2}) vibrational continuum.

Measurements by Adaniya et al. \cite{c3adaniya} show \ce{O-} from the \ce{^{2}A1} resonance exhibiting an intense and broad distribution in the forward scattering angles and a less intense and narrow distribution in the backward angles for electron energies across the resonance. Our measurements show a small unresolved blob. In the experiment by Adaniya et al. \cite{c3adaniya}, the velocity images are obtained using the Abel inversion algorithm, whereas our measurements using the slicing technique are more direct. The \ce{O-} angular distribution reported by Adaniya et al. \cite{c3adaniya} peaks at about $30^{\circ}$ and has a broad minimum about $110^{\circ}$, their theoretical prediction assuming axial recoil approximation shows the peak at close to $150^{\circ}$ and the minimum at $70^{\circ}$. Thus, in their case, the measured distribution and theoretical prediction are mirror images about the $90^{\circ}$ direction. They explain this distribution of \ce{O-} ions arising from the molecule scissoring backwards and ejecting the oxygen through the mouth of the H-O-H bond, regardless of the direction, while the hydrogen recoil in the opposite direction. Interestingly, this kind of a distribution of \ce{O-} ions with forward scattering is a seen in the images of \ce{O-} produced from the third resonance process. It may be that the observed distribution of \ce{O-} seen at energies 9.5 eV and 10 eV in their data may be due to contribution from the third resonance process. It can be seen from the excitation function curves in Figure \ref{fig3.1}(b) that these energies may have a contribution from the rising part of the third resonance structure peaking at 12 eV.

\subsection{Third resonance process at 11.8 eV}

The velocity map image of \ce{H-} (\ce{D-}) fragment ions across the third resonance centered at 11.8 eV shown in Figure \ref{fig3.3}(f), (g), (h) and Figure \ref{fig3.5}(f), (g), (h) present a very striking behaviour. These images show \ce{H-}(\ce{D-}) ions ejected mostly in the backward scattering angles. We notice three clear rings in the momentum images of \ce{H-} and \ce{D-} from \ce{H2O} and \ce{D2O} respectively. Energetically, these correspond to three different dissociation pathways of the molecular negative ion:

\begin{align*}
H_{2}O^{-*} \rightarrow & H^{-} + OH (X ^{2}\Pi) \tag{Channel 1} \\
& H^{-} + OH^{*} (A ^{2}\Sigma) \tag{Channel 2} \\
& H^{-} + H + O \tag{Channel 3}
\end{align*}

\begin{figure}[!ht]
\centering
  \subfloat[]{\includegraphics[height=4cm,width=5cm]{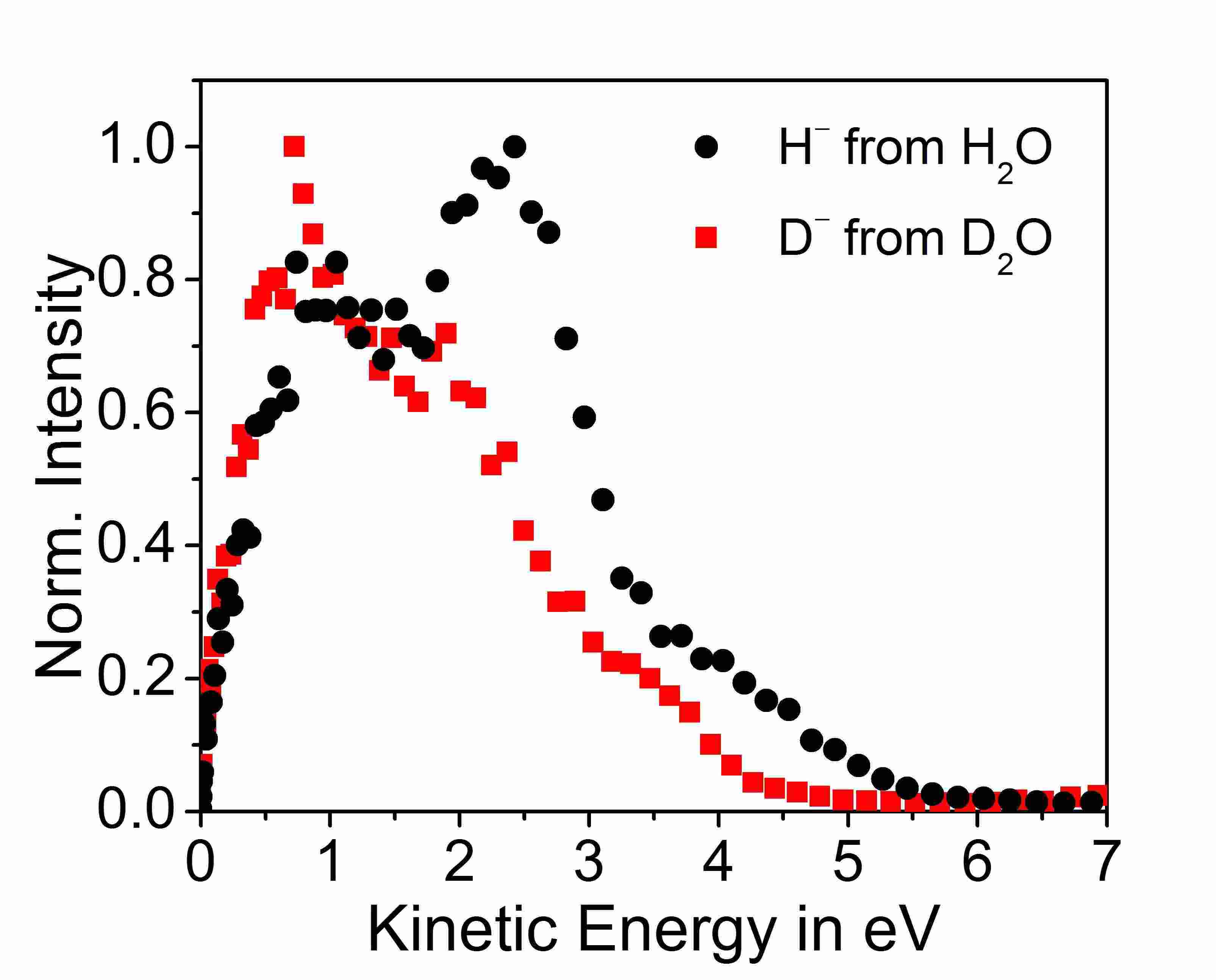}}
  \subfloat[]{\includegraphics[height=4cm,width=5cm]{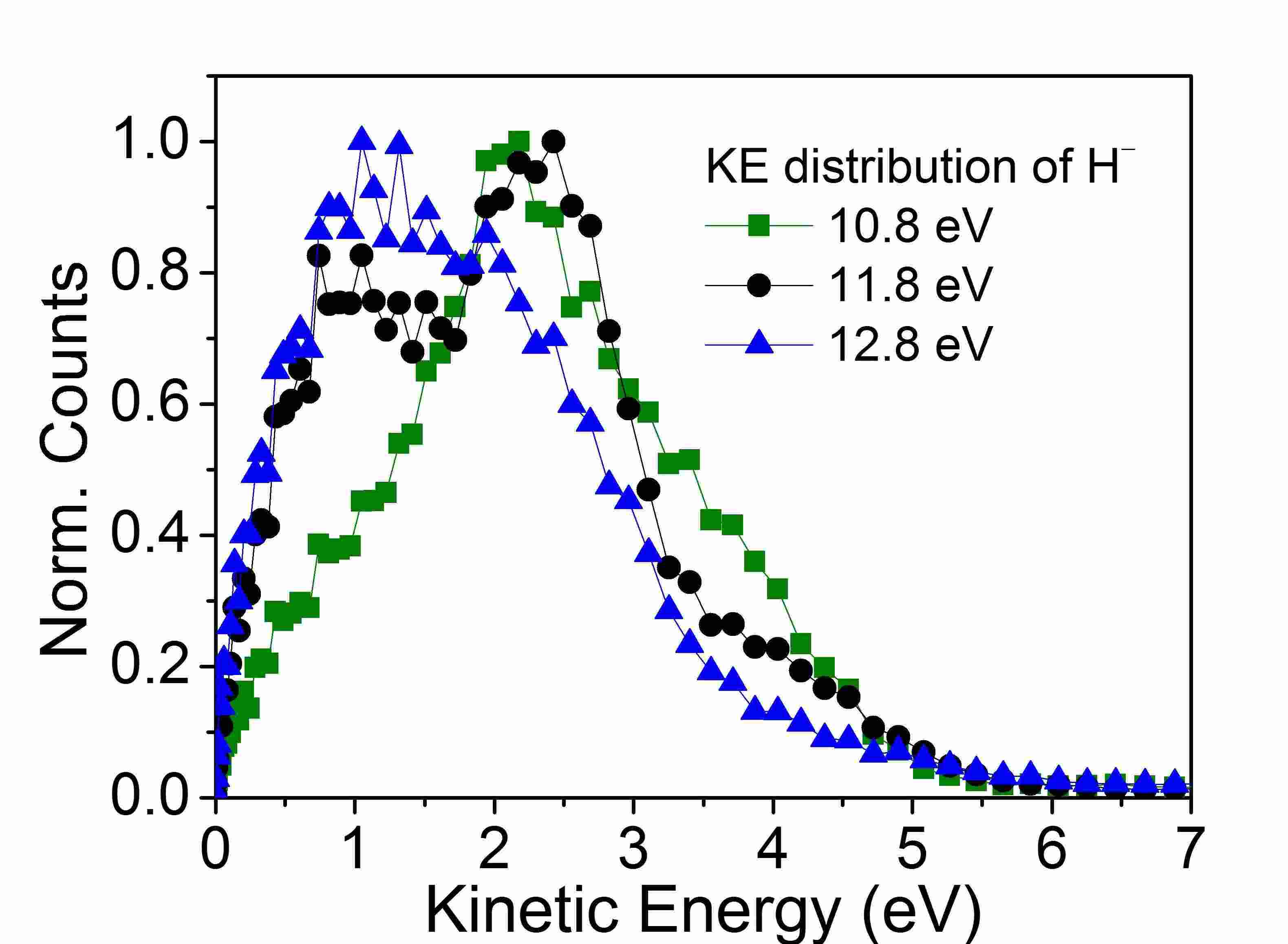}}
 \subfloat[]{\includegraphics[height=4cm,width=5cm]{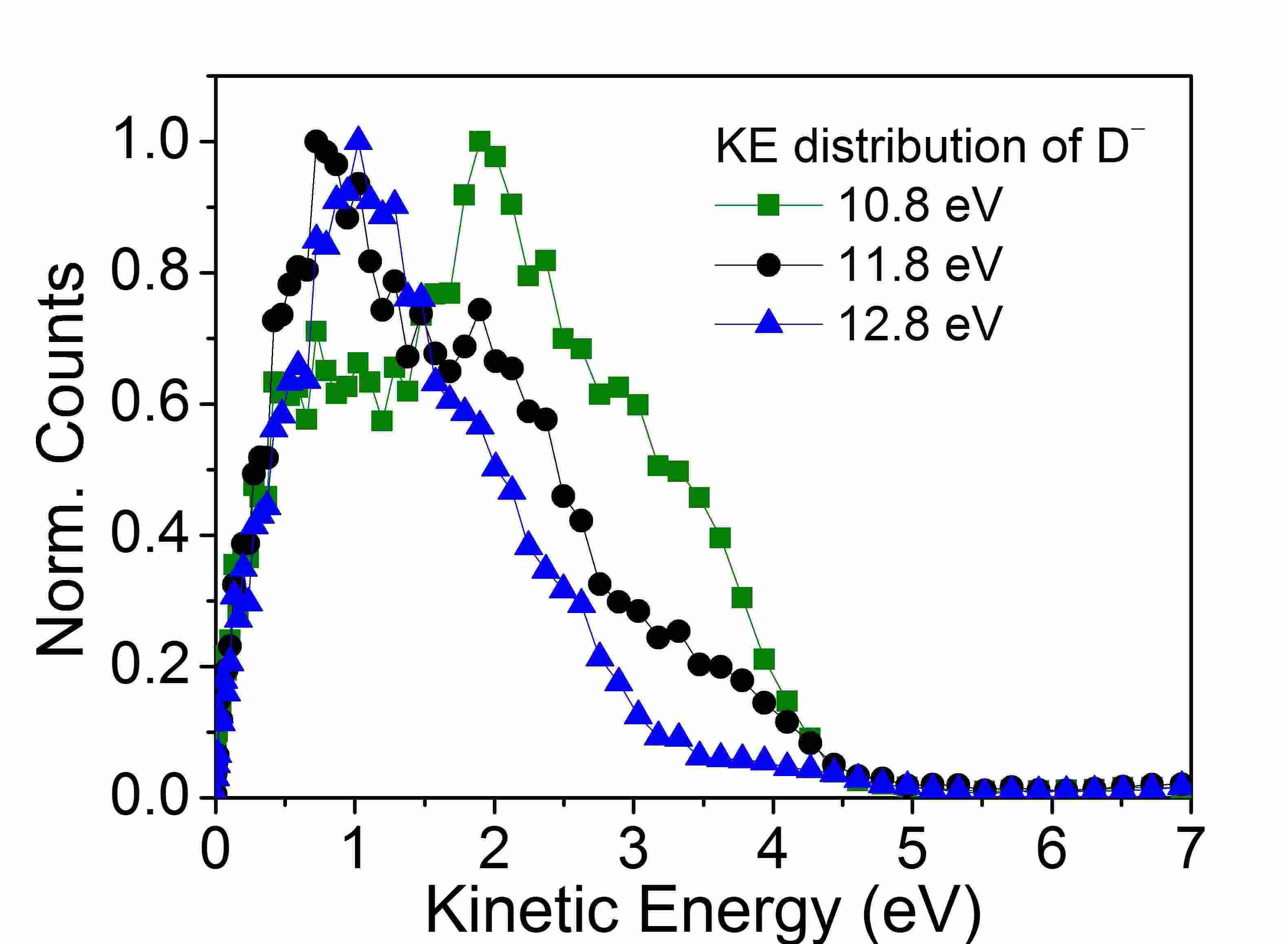}}\\ 
\subfloat[]{\includegraphics[height=4cm,width=5cm]{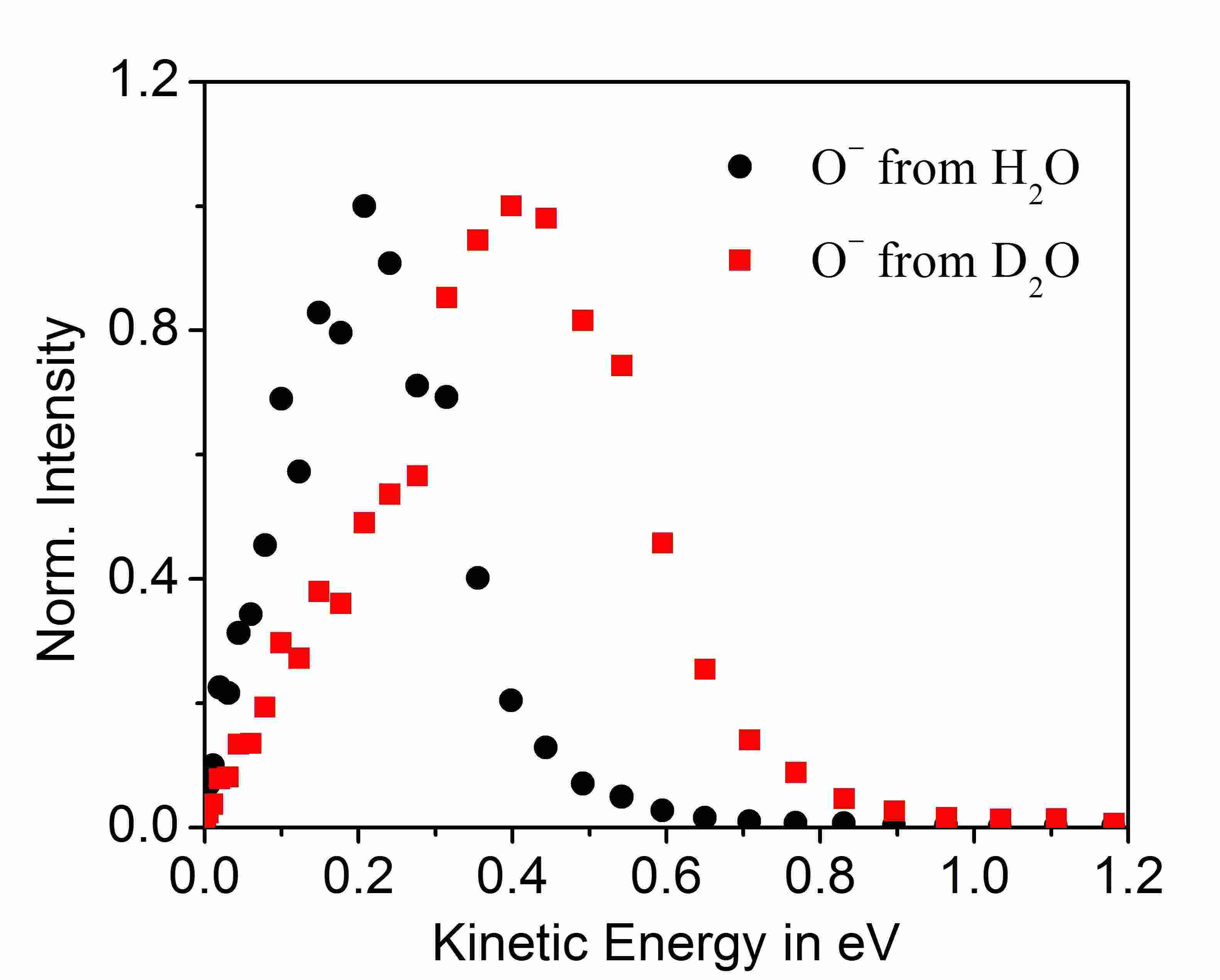}}
  \subfloat[]{\includegraphics[height=4cm,width=5cm]{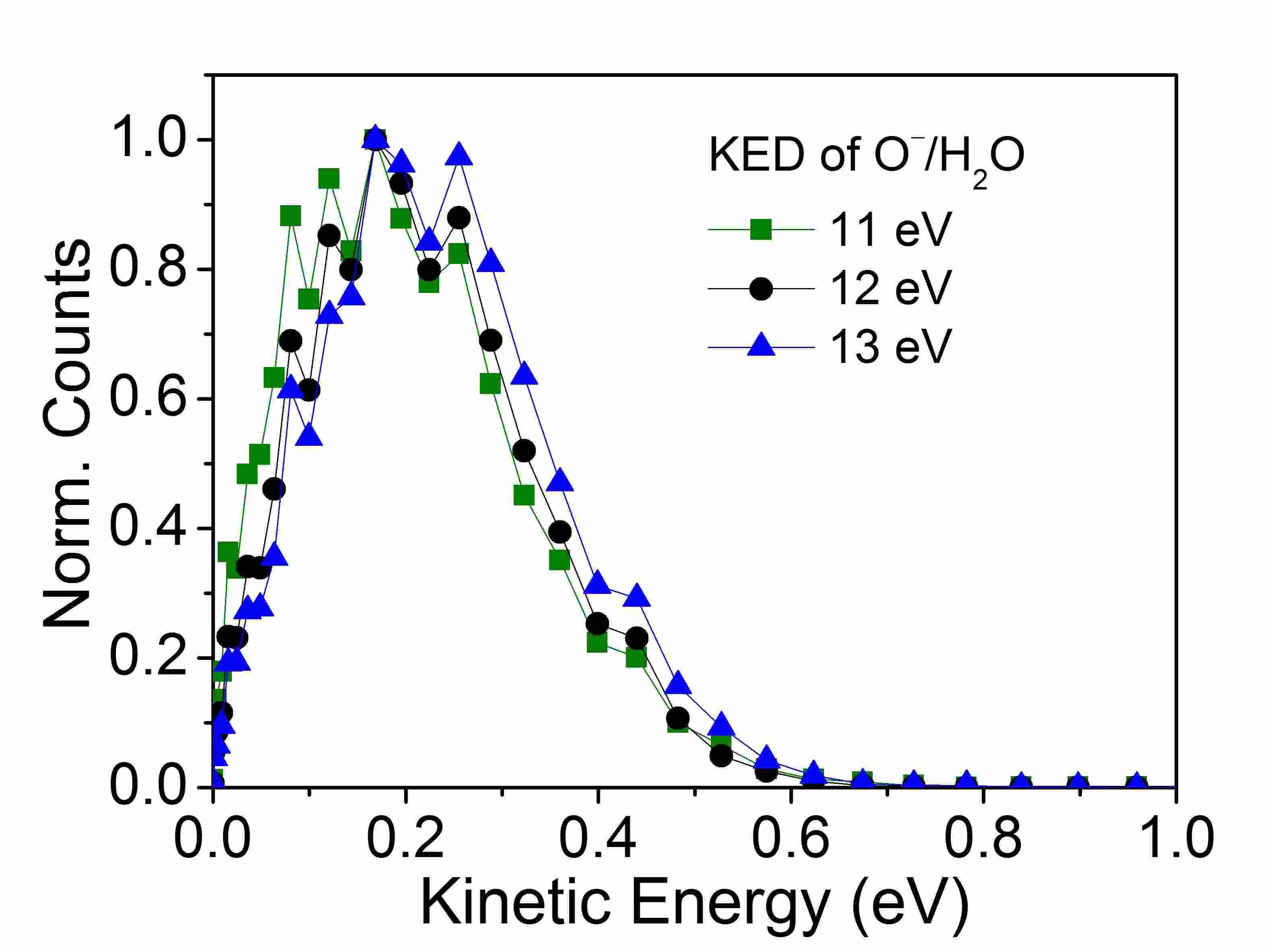}}
 \subfloat[]{\includegraphics[height=4cm,width=5cm]{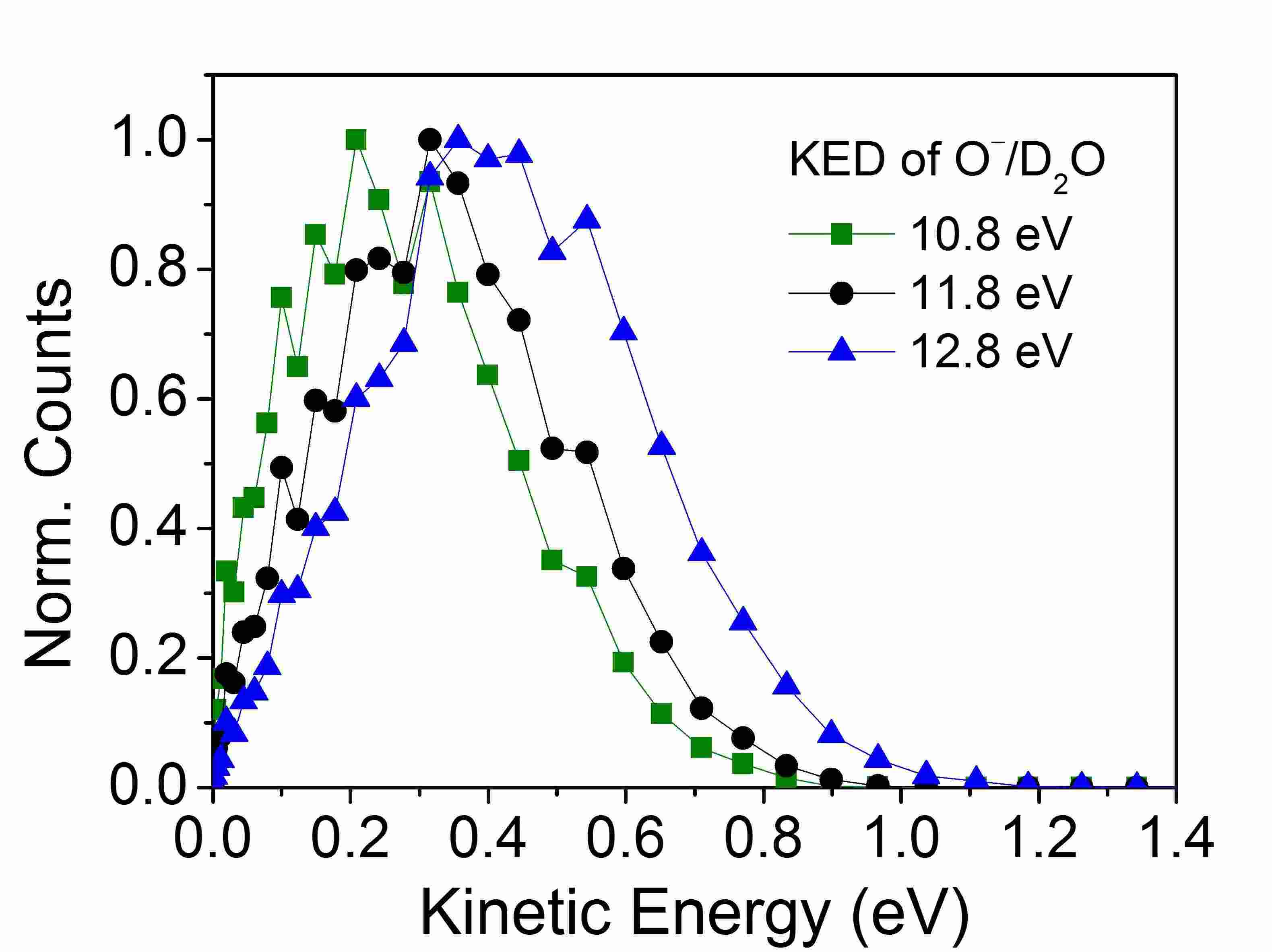}}
 \caption{(a) Kinetic energy distribution of \ce{H-} and \ce{D-} compared at 11.8 eV. (b) and (c) Variation of KE distribution of \ce{H-} and \ce{D-} ions respectively across the third resonance at electron energies 10.8 eV, 11.8 eV and 12.8 eV. (d) Kinetic energy distribution of \ce{O-}  from \ce{H2O} and \ce{D2O} at 12 eV. (e) and (f) Variation of KE distribution of \ce{O-}/\ce{H2O} and \ce{O-}/\ce{D2O} respectively at 11 eV, 12 eV and 13 eV.}
\label{fig3.13}
\end{figure}

The kinetic energy distribution of \ce{H-} (\ce{D-}) ions (Figure \ref{fig3.13}(a)) shows the three dissociation limits as three distinct structures - the first one peaking at 1 eV, second one at 2.5 eV and third one appearing as a broad structure between 3.5 and 6 eV. The outermost ring in the velocity image (Figure \ref{fig3.3}(g) and \ref{fig3.5}(g)) is faint, but its presence can be seen in the kinetic energy distribution as the broad structure from 3.5 to about 6 eV in the case of \ce{H-} from \ce{H2O} and from 3 to 4.5 eV in the case of \ce{D-} from \ce{D2O}. The \ce{H- + OH (X ^{2}$\Pi$)} channel which has a threshold of 4.35 eV will have an excess energy of 7.4 eV at 11.8 eV electron energy. The maximum KE of \ce{H-} would be 17/18 of the excess energy which is 7.0 eV in this case. The maximum KE from our measurements is 6.2 eV. This difference may be attributed to vibrational excitation of the OH fragment. In the case of \ce{D-} from \ce{D2O}, the kinetic energy distribution appears to be considerably lower, even after taking into account the difference due to higher mass. The \ce{D-} intensity drops just above 4 eV and appear to trail off to 5 eV as compared to the maximum possible energy of 6.7 eV. The lower kinetic energy distribution indicates that the \ce{OD (X ^{2}$\Pi$)} states are formed at this resonance with populations in very high levels, even higher than the corresponding case of \ce{OH (X ^{2}$\Pi$)}.

The middle ring in the VMI data gives the peak at 2.5 eV in the kinetic energy distribution and corresponds to \ce{H-} ions produced with OH in electronic excited state ($^{2}\Sigma$) with threshold energy being 8.4 eV. This distribution seems to extend up to a kinetic energy of 3.3 eV. The maximum kinetic energy of \ce{H-} in this channel will correspond to the formation of the OH ($^{2}\Sigma$) in the vibrational ground state. This energy is 17/18th of the excess energy and works out to be 3.2 eV, which is in fair agreement with what we observe. The maximum of the \ce{H-} intensity distribution for this channel observed at 2.5 eV approximately corresponds to $\nu$ = 2 vibrational excitation of the OH ($^{2}\Sigma$) radical. The kinetic energy distribution of \ce{D-} from \ce{D2O} peaks at 2 eV corresponding to the OD ($^{2}\Sigma$) channel and appears to extend approximately to 3 eV, which is close to the expected maximum energy for this channel. We note that the low kinetic energy release in the case of both \ce{H2O} and \ce{D2O} in this channel indicates formation of the hydroxyl radical in high vibrational levels. This shift to higher vibrational levels appears to be more pronounced in the case of \ce{D2O} as in Channel 1 and may be explained as due to the larger time needed for the dissociation of \ce{D2O} negative ion state.

The innermost ring in the \ce{H-} image corresponds to the kinetic energy distribution as a broad peak between 1 eV and 1.6 eV with the intensity falling off rapidly below 1 eV. For the three body breakup process (\ce{H- + H + O}; threshold 8.75 eV), the excess energy would be 3.05 eV which would be distributed as translational energy amongst the three atomic fragments. Assuming instantaneous three-body fragmentation of the resonant state under axial recoil approximation, \ce{H-} will have a kinetic energy equal to $4E_{\circ}/[8+\cos^{2}\theta]$ where $E_{\circ}$ is the excess energy and $\theta$ is half of the bond angle ($105^{\circ}$, in this case). This works out to be 1.43 eV. If we assume that the molecule undergoes bending motion during the fragmentation the value of $\theta$ will change. Assuming symmetric breaking of the two O-H bonds, depending on the value of $\theta$ the kinetic energy will be between 1.33 eV and 1.5 eV. We note that about half of the intensity distribution in the three body channel has energies below 1.3 eV. This effect seems to be more pronounced in the case of \ce{D-} from \ce{D2O}. The \ce{D-} kinetic energy for the three - body channel peaks at 0.7 eV, considerably lower than the minimum kinetic energy of 1.2 eV one expects from the instantaneous break up.

One might account for the lower than expected kinetic energies of \ce{H-} and \ce{D-} ions from the three-body break up channel as due to the formation of the O atom in the excited $^{1}D$ state, is given by:
\begin{equation}
H_{2}O^{-*} \rightarrow  H^{-} + H + O (^{1}D) \mbox{(threshold energy=10.72 eV)} \tag{Channel 4} \label{ch4}
\end{equation}

The excess energy in this case will be 1.08 eV at incident electron energy of 11.8 eV. The energy range in which the \ce{H-} ions will appear in such a channel would be between 0.48 eV and 0.54 eV and for \ce{D-} the range would be between 0.43 eV and 0.54 eV. These energy ranges are much lower than the observed distribution in both \ce{H2O} and \ce{D2O} and hence we may rule out this channel. The only way to explain the lower kinetic energy distribution from the three-body break up channel is to assume a sequential fragmentation, instead of an instantaneous three-body break up. There are two ways this sequential fragmentation might occur: 

\begin{align*}
H_{2}O^{-*} & \rightarrow  H + OH^{-*} \rightarrow H + O + H^{-} \tag{Channel 5} \label{ch5} \\
& \rightarrow  O + H_{2}^{-*} \rightarrow O + H + H^{-} \tag{Channel 6} \label{ch6}
\end{align*}

Calculations by Haxton et al. \cite{c3h5} indicate a possible decay mode for the \ce{B2} resonance through a \ce{H2^{-*} + O (^{1}D)} channel. However, this would have higher energy threshold leaving less excess energy as in the case of Channel 4. Moreover, formation of H-H bond followed by its break appears too contrived and hence we may rule out Channel 6. Though we do not have much information on excited states of the hydroxyl anion, in order to explain the results, we need to assume that the sequential break up occurs through Channel 5. Curtis and Walker \cite{c3curtis} have pointed out that the three-body break up may be happening through the sequential process as in Channel 5. They attribute the formation of \ce{O-} also through such a process in which the extra electron is bound to the O atom. As to be discussed later, the analysis of the \ce{O-} kinetic energy spectrum from \ce{D2O} also shows contribution from sequential fragmentation via \ce{OD^{-*}} state.

The kinetic energy distributions of \ce{H-} and \ce{D-} ions across the third resonance process are shown in Figure \ref{fig3.13}(b) and (c). It is seen that the peak structure between 2.5 and 3.5 eV due to the \ce{OH (^{2}$\Sigma$)} (correspondingly \ce{OD (^{2}$\Sigma$)} in \ce{D2O}) diminishes in intensity while the intensity of the peak close to 1 eV increases. This indicates that the three body fragmentation channel becomes more favoured with increase in electron energy.

The kinetic energy spectra of \ce{O-} from \ce{H2O} and \ce{D2O} respectively at 12 eV are given in Figure \ref{fig3.13}(d). As expected the kinetic energy of \ce{O-} from \ce{D2O} is about twice that of \ce{O-} from \ce{H2O} with peaks at 0.4 eV and 0.2 eV respectively. At the 12 eV resonance, the formation of \ce{O-} occurs only through a three body fragmentation process which has a threshold of 8.04 eV. Since the excess energy available for this channel is 3.96 eV, we expect the \ce{O-} kinetic energy distribution to extend up to 0.79 eV from \ce{D2O} and 0.44 eV from \ce{H2O}. The data seems to be in reasonable agreement with this. Figures \ref{fig3.13}(e) and (f) show the KE distribution of \ce{O-} across the resonance i.e at 11, 12 and 13 eV which show increase in maximum KE shifting towards higher values with increase in electron energy.  Assuming symmetric fragmentation of the resonance leads to the formation of \ce{O-}, its kinetic energy would be equal to $E_{\circ}\cos^{2}\theta/(4 + \cos^{2}\theta)$ in the case of \ce{D2O} and $E_{\circ}\cos^{2}\theta/(8+\cos^{2}\theta)$ for \ce{H2O} where $E_{\circ}$ is the excess energy and $\theta$ is half the bond angle. For a bond angle of $105^{\circ}$, the kinetic energy works out to be 0.34 eV for \ce{O-}(\ce{D2O}) and 0.18 eV for \ce{O-}(\ce{H2O}). The peaks we see in the kinetic energy distributions are slightly larger and correspond to bond angles of $96^{\circ}$ and $99^{\circ}$ for \ce{D2O} and \ce{H2O} respectively. This indicates that as the two O-H bonds are breaking the two H atoms tend to get closer. One interesting thing we note in the kinetic energy distribution of \ce{O-} from \ce{D2O} (Figure \ref{fig3.13}(d)) is the discernible drop at 0.35 eV. In terms of bond angle, this corresponds to $105^{\circ}$. We explain this as a signature of the sequential fragmentation process seen in the kinetic energy spectra of \ce{H-} and \ce{D-}. As the \ce{O-} energy increases beyond 0.35 eV the bond angle decreases indicating that the potential energy surface of the resonance has its steepest descent towards larger O-D distances and shorter D-D distances. This would mean that $105^{\circ}$ would be a cut off angle beyond which no instantaneous three body fragmentation could occur as the system cannot climb up the surface. Hence the \ce{O-} intensity seen at energies below 0.35 eV has to be explained through a sequential process similar to what is seen in the case of \ce{H-} and \ce{D-} kinetic energy spectra. In this channel, the intermediate state is \ce{OD^{-*}} and this may decay through the formation of either \ce{D-} or \ce{O-}. The kinetic energy of the fragments coming through the sequential process would be smaller since the first D would have taken away a good fraction of the available excess energy. Hence the sequential dissociation may explain the observed \ce{O-} distribution below 0.35 eV. We note that the signature of the sequential process to be present in the \ce{O-} kinetic energy distribution from \ce{H2O} as well. 

The absolute cross sections from \ce{H2O} and \ce{D2O} show isotope effects in \ce{H-}/\ce{D-} and \ce{O-} channels \cite{c3rawat}. It is more pronounced in the \ce{O-} channel and present at all three resonances. In the \ce{H-}/\ce{D-} channel it is seen only at the third resonance. The cross sections for \ce{D-} formation is about 75\% of that for \ce{H-} formation in this energy range \cite{c3rawat}. We have made an attempt to quantify the isotope dependence in terms of the two-body to three body fragmentation channels. This is shown in Table \ref{tab3.5} where we give the ratio of the two-body to three body fragmentation cross sections in the hydride ion channel at different electron energies across the third resonance. We note from Figure \ref{fig3.13}(b) and (c) that at lower energy the two body channel is higher and decreases with respect to the three body channel as the energy is increased. We also note that the three body channel is relatively stronger in the case of \ce{D2O} as compared to that of \ce{H2O} indicating that isotope effect is more pronounced in the two body fragmentation channel.

\begin{table}
\caption{The ration of two-body to three-body fragmentations in \ce{H2O} and \ce{D2O} in the hydride ion channel at the third resonance}
\begin{center}
\begin{tabular}{ccc}
\hline
\\
Electron energy (eV) & \ce{H-} (\ce{H2O}) & \ce{D-} (\ce{D2O}) \\
\\
\hline
\\
10.8 & 1.97 & 1.39 \\
\\
11.8 & 1.00 & 0.84 \\
\\
12.8 & 0.68 & 0.71 \\
\\
\hline
\end{tabular}
\end{center}
\label{tab3.5}
\end{table}

\begin{figure}[!htbp]
\centering
\subfloat[]{\includegraphics[width=0.35\columnwidth]{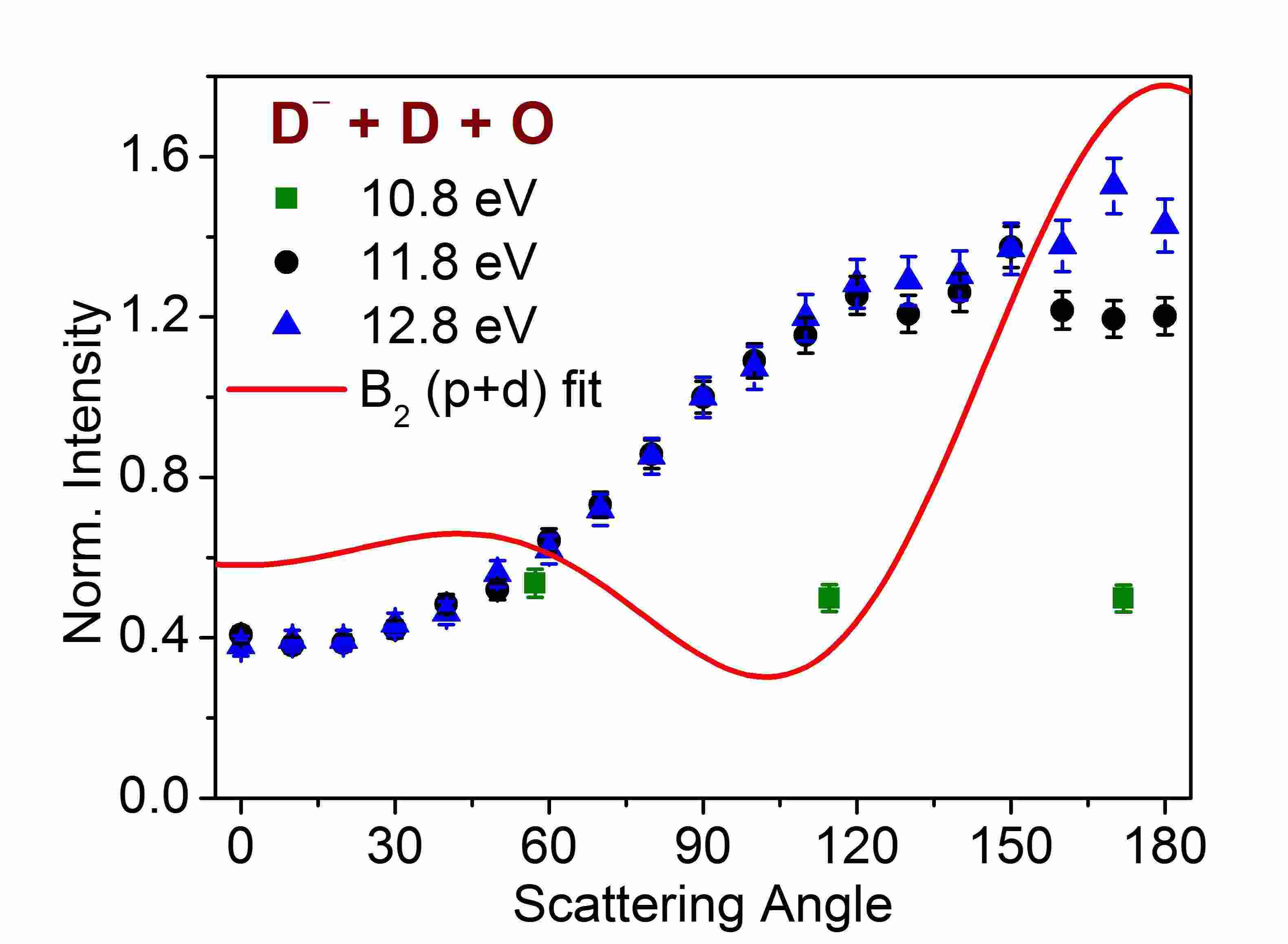}} \hspace{1cm}
\subfloat[]{\includegraphics[width=0.35\columnwidth]{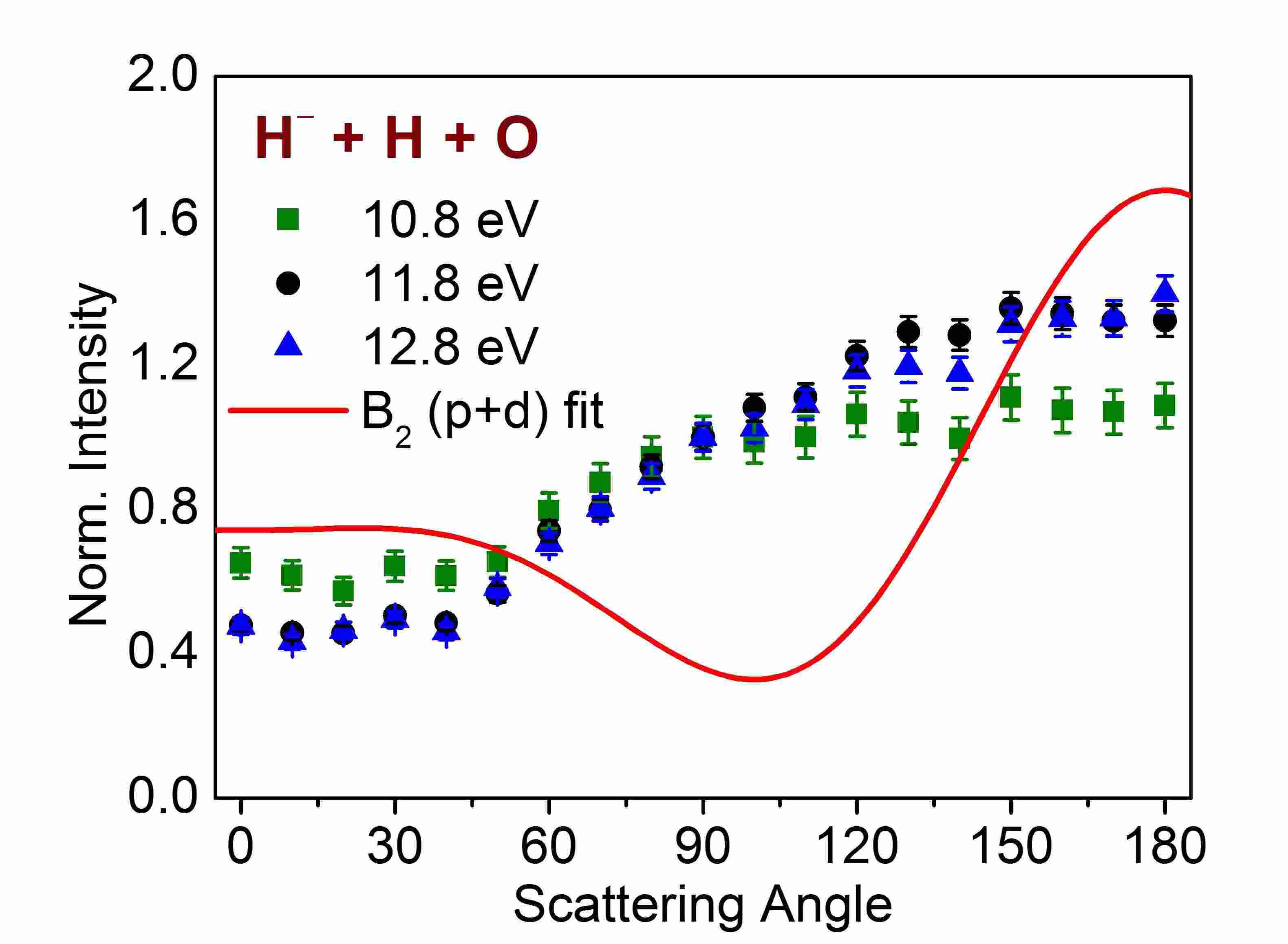}}\\
\subfloat[]{\includegraphics[width=0.35\columnwidth]{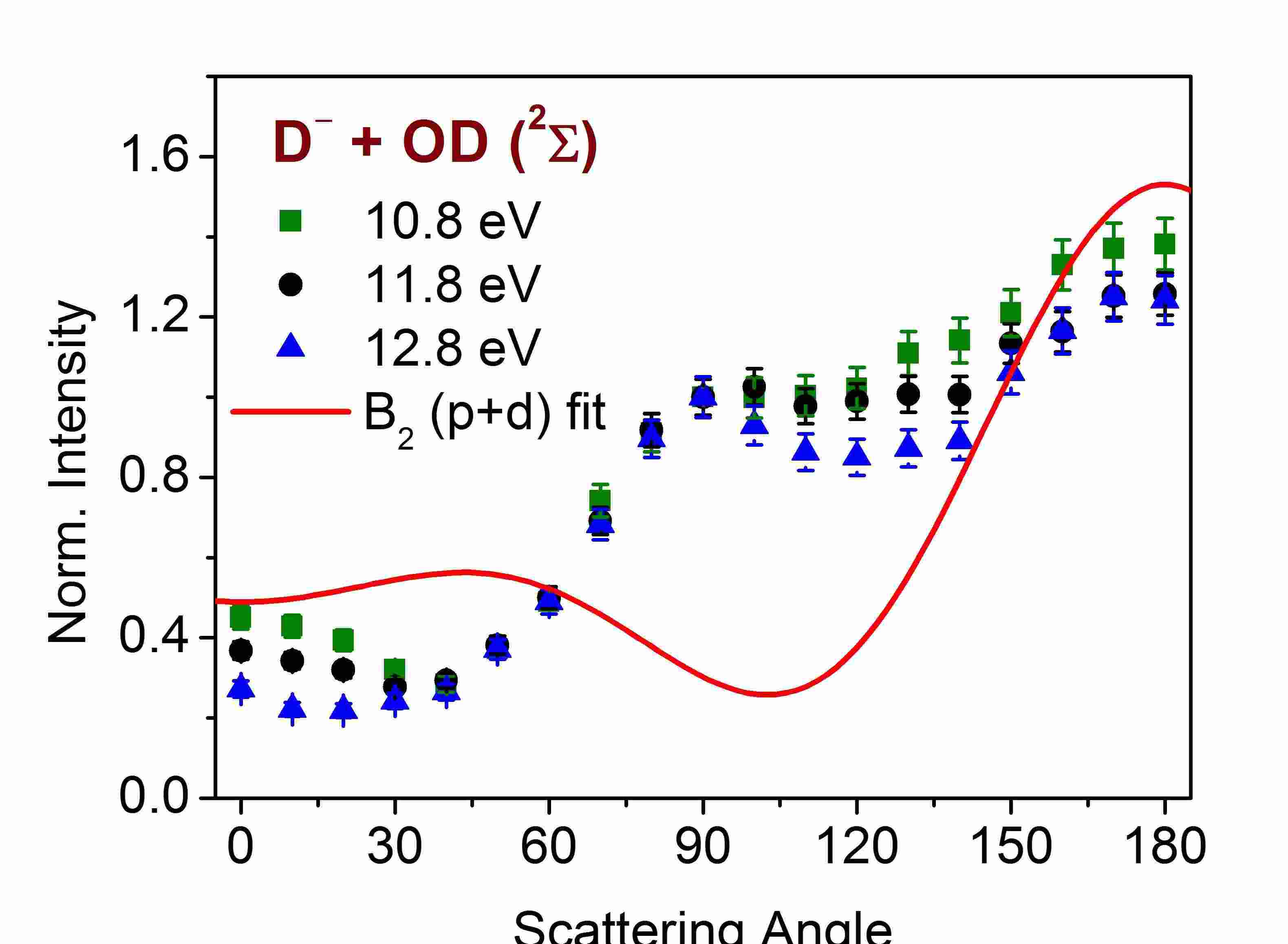}} \hspace{1cm}
\subfloat[]{\includegraphics[width=0.35\columnwidth]{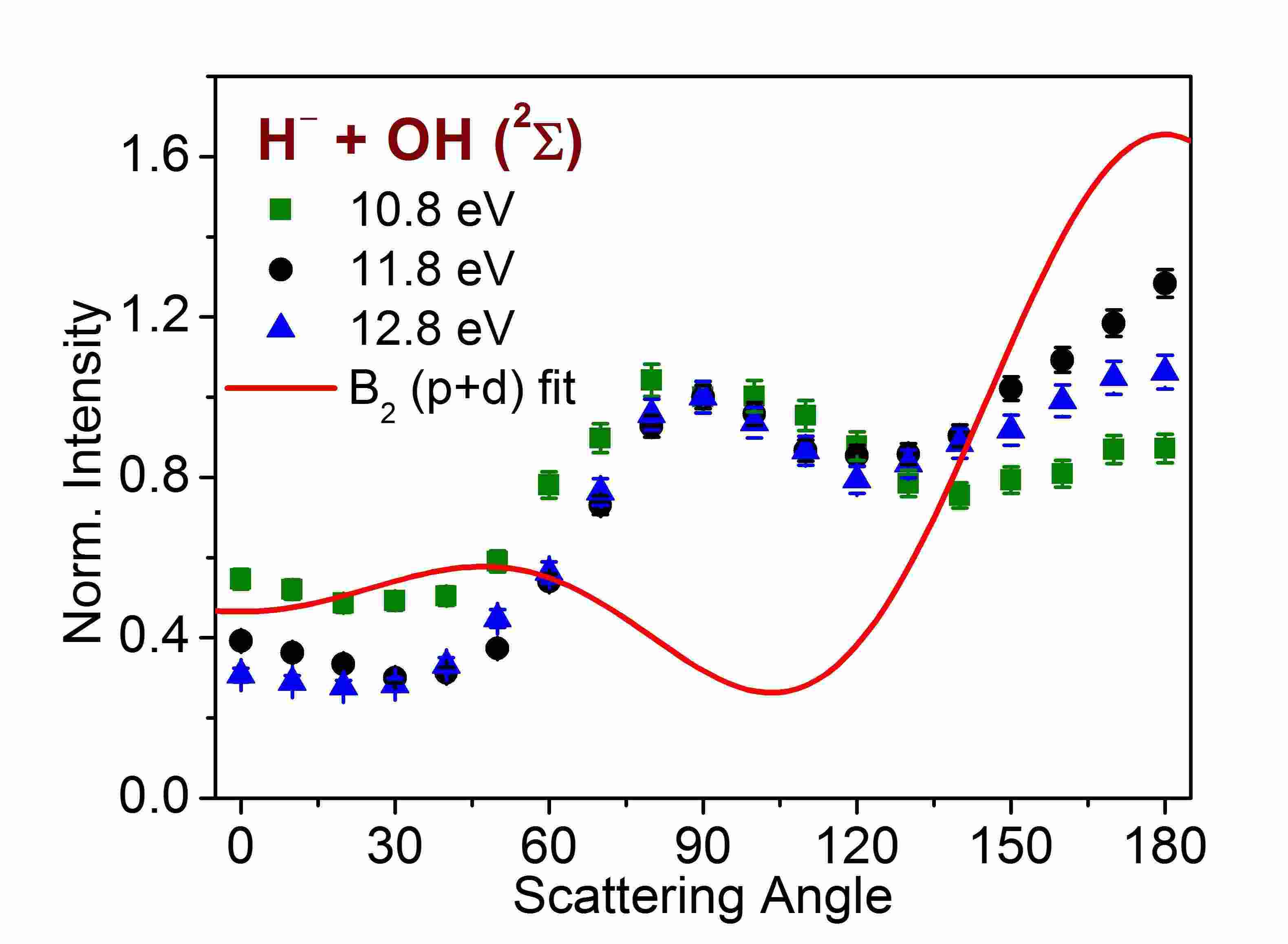}}\\
\subfloat[]{\includegraphics[width=0.35\columnwidth]{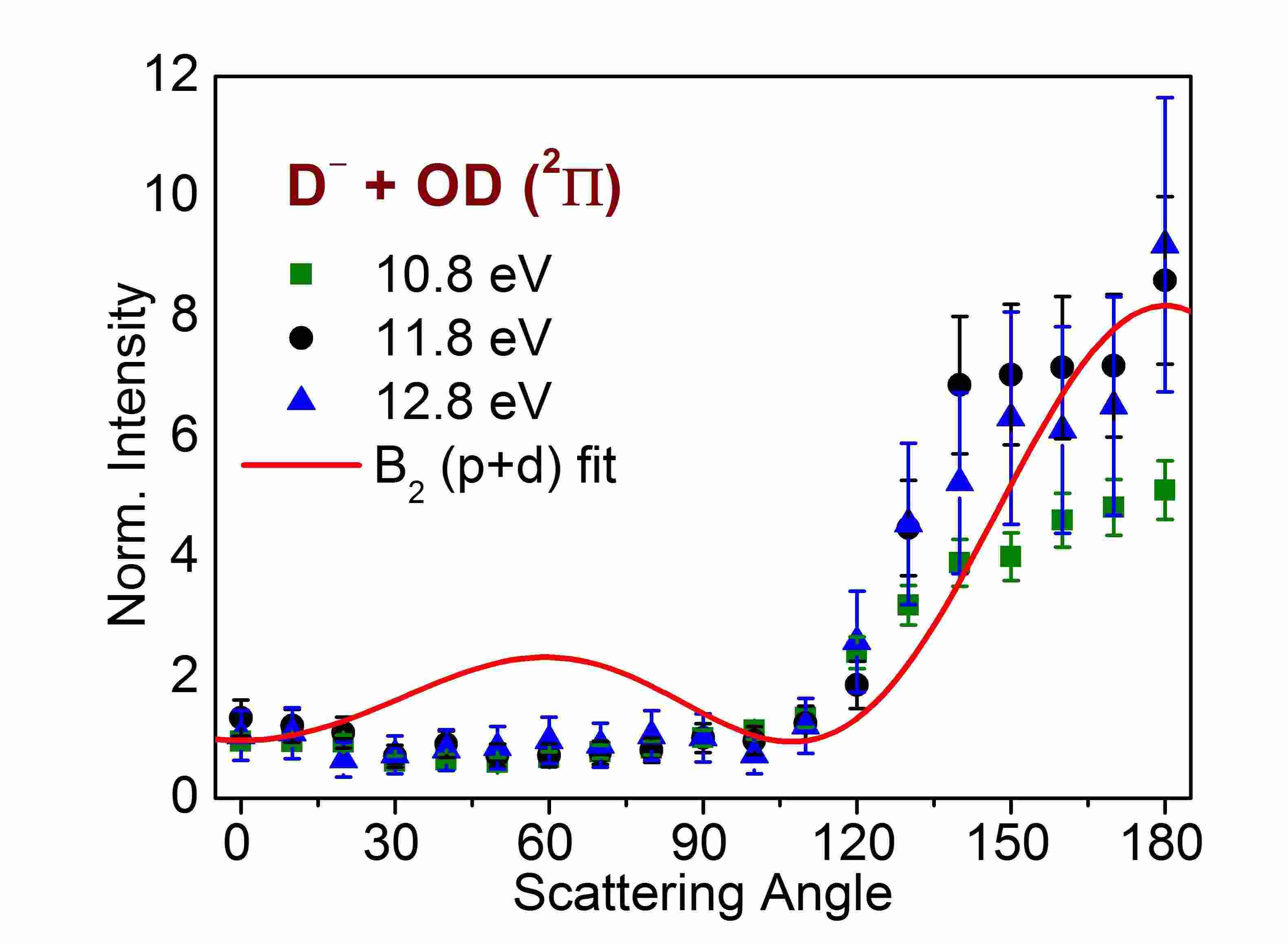}}\hspace{1cm}
\subfloat[]{\includegraphics[width=0.35\columnwidth]{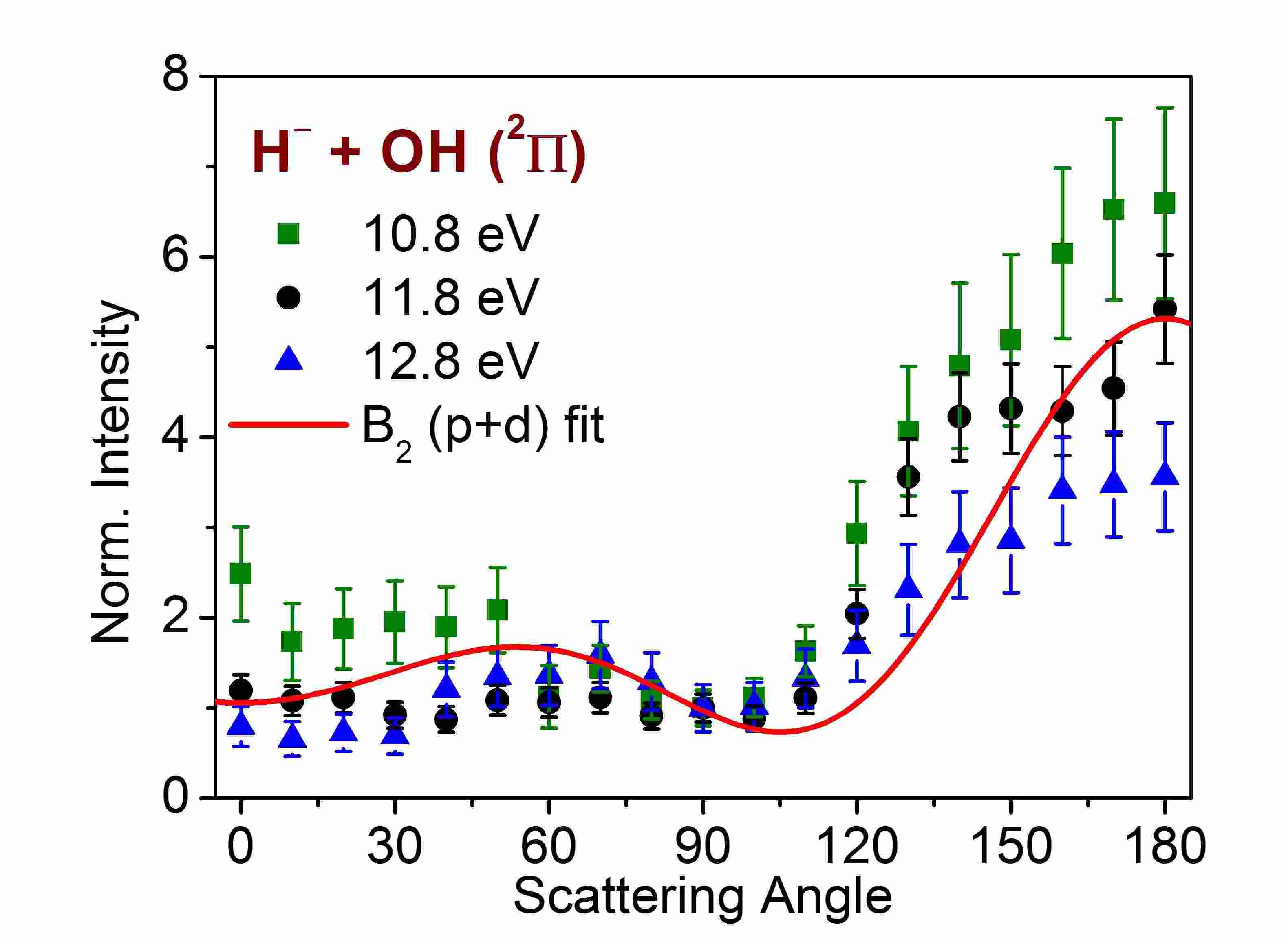}}\\
\subfloat[]{\includegraphics[width=0.35\columnwidth]{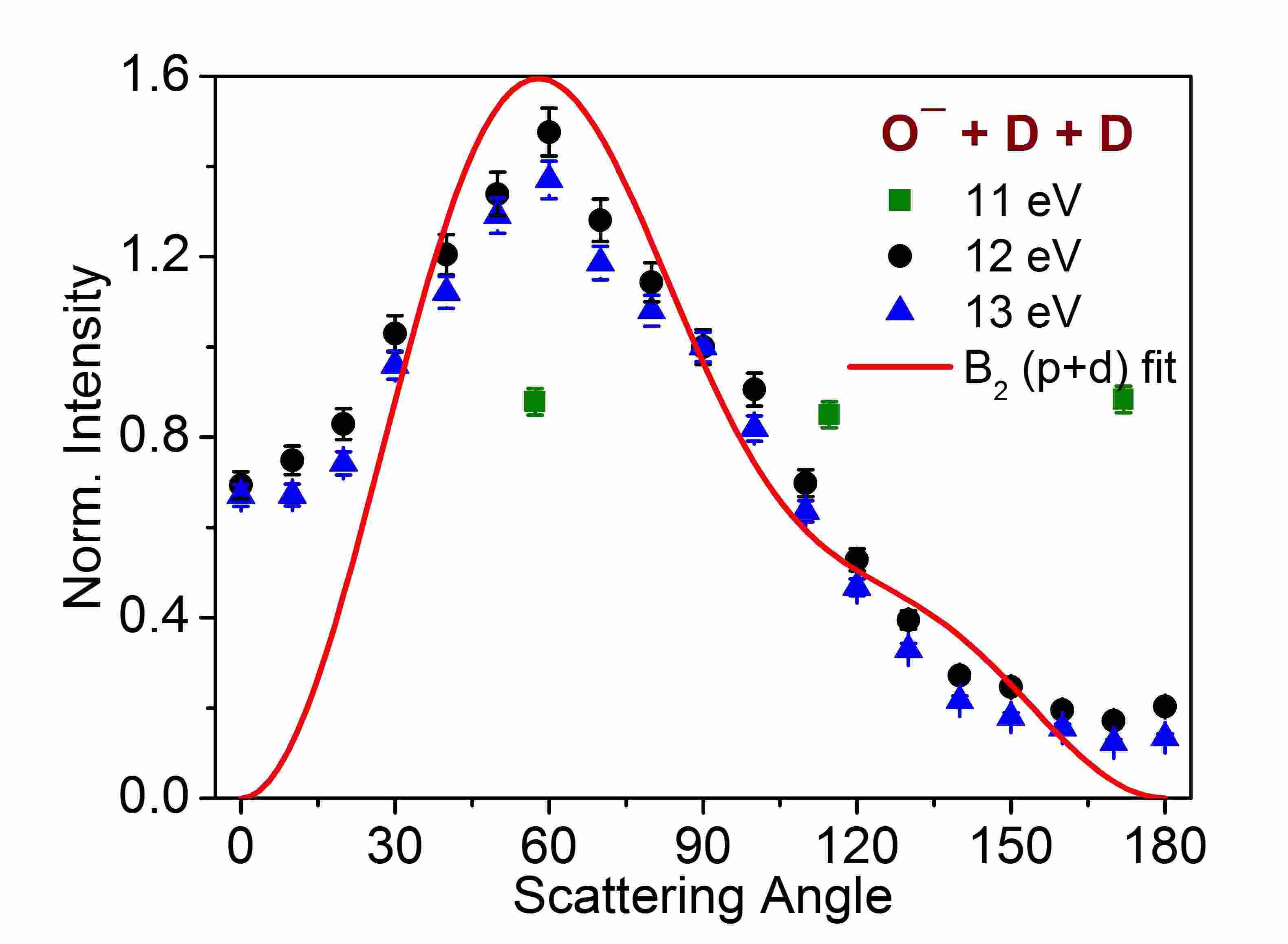}}\hspace{1cm}
\subfloat[]{\includegraphics[width=0.35\columnwidth]{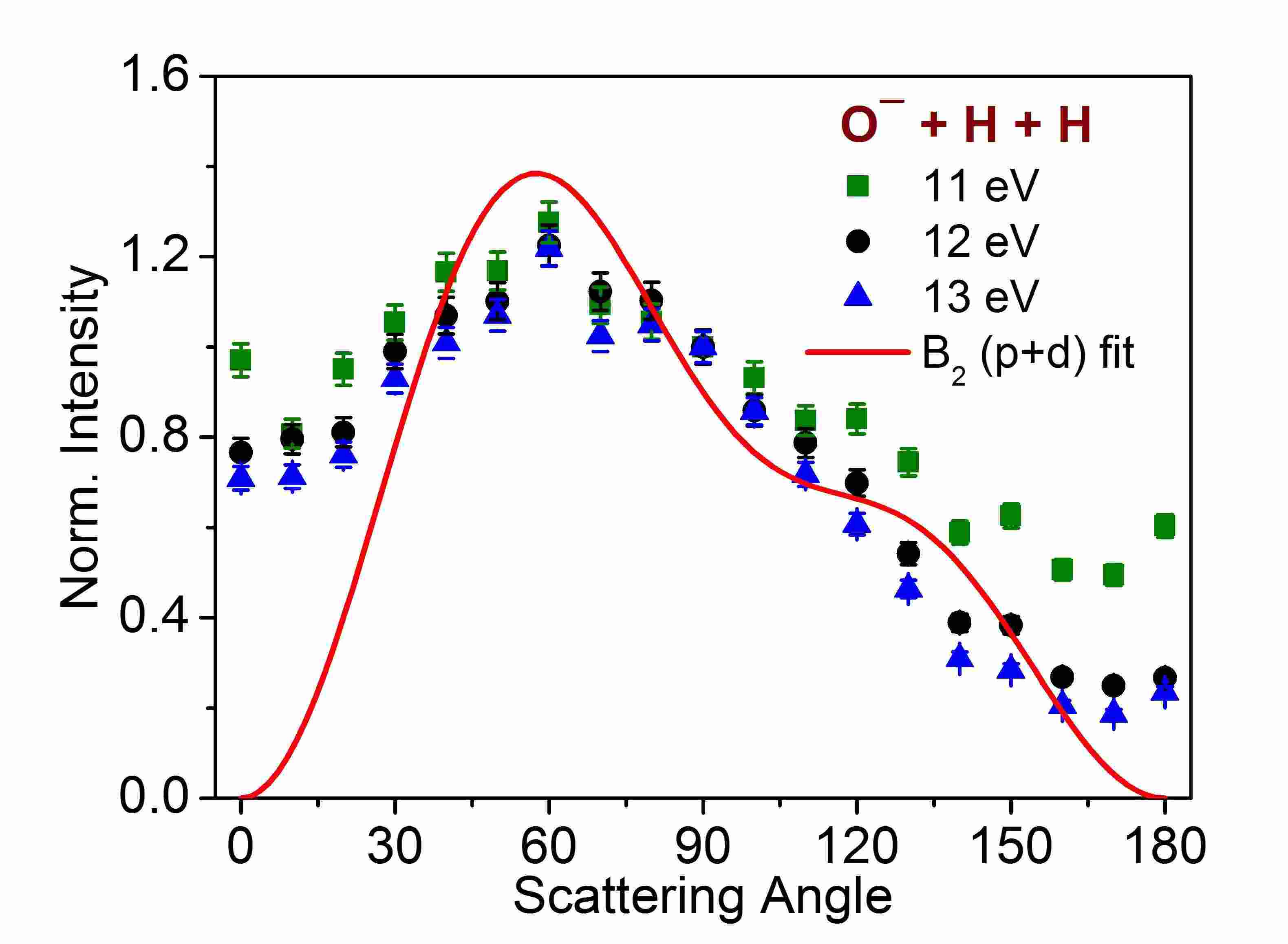}}
 \caption{Angular distribution of ions across the third resonance at three electron energies . (a) \ce{D-} from \ce{D- + D + O} channel (b) \ce{H-} from \ce{H- + H + O} channel (c) \ce{D-}  from \ce{D- + OD(^{2}$\Sigma$)} channel (d) \ce{H-} from \ce{H- + OH(^{2}$\Sigma$)} channel (e) \ce{D-} from the \ce{D- + OD(X ^{2}$\Pi$)} channel (f) \ce{H-} from the \ce{H- + OH (X ^{2}$\Pi$)} channel (g) \ce{O-} from \ce{O- + D + D} channel and (h) \ce{O-} from \ce{O- + H + H} channel. The red curve in each plot is the fit obtained for the 11.8 eV data using \ce{B2} symmetry functions.}
\label{fig3.14}
\end{figure}

In addition to the appearance of three clear rings, the most striking aspect of the momentum distribution of the ions across the 11.8 eV resonance is the strong forward - backward asymmetry in both \ce{H-}/\ce{D-} and \ce{O-} ions. The angular distribution as seen from the images appears to be very unique with the \ce{H-} being ejected predominantly in the backward hemisphere. In contrast the \ce{O-} is ejected entirely in the forward hemisphere. The previous measurements by Belic et al. \cite{c3belic} in which they identified only one kinetic energy component for \ce{H-} at this resonance appear to peak at $90^{\circ}$ as well as toward $0^{\circ}$. They did not report any angular distribution data on \ce{O-}. What we observe is considerably different from that reported by them. The results of Belic et al. \cite{c3belic} indicated the resonance to be a \ce{B2} state. We calculated individual transition amplitudes for $p$ and $d$ waves for a \ce{B2} resonance using the method described in Section 3.3 and calculated the angular distribution for \ce{H-} using these partial waves in an effort to fit the experimental results. We calculate the transition amplitude functions for \ce{A1}(neutral state) to \ce{B2}(negative ion state) for the lowest allowed $p$ and $d$ partial waves. Now combining the two individual functions along with the interference term, the general expression for scattering intensity $I_{p+d}(\theta)$ from a \ce{B2} state due to $p$ and $d$ wave capture is given by $2a^{2}(2\sin^{2}\beta \cos^{2}\theta + \cos^{2}\beta \sin^{2}\theta) + \frac{3}{2} b^{2} (\frac{1}{4}\sin^{2}2\beta \sin^{4}\theta + \cos^{2}2\beta \sin^{2}2\theta) + 2 a b \sqrt{3} (\cos\beta \cos2\beta \sin\theta \sin2\theta - \sin\beta \sin2\beta cos\theta (3\cos^2{\theta}-1)) \cos\delta$ where $a$, $b$ and $\delta$ are the fitting parameters. This equation is fit to the angular distribution data and the goodness of fit determines the efficacy of the axial recoil approximation in descibing the dissociation process. The angular distribution for the three dissociation channels (both \ce{H-} and \ce{D-} ions) at 10.8eV, 11.8 eV and 12.8 eV are shown in Figure \ref{fig3.14}. The red curve in each plot depicts the best fit for the angular distribution data at 11.8 eV. The best fits at 10.8 eV and 12.8 eV are very similar. The fits with $p$ and $d$ partial waves reproduce the forward-backward asymmetry in all three fragmentation channels. However, it is seen that except for the \ce{OH (^{2}$\Pi$)} channel, we are unable to get a reasonably good fit. It is likely that all three channels may have effects due to deviation from axial recoil approximation. However, it is difficult to visualize how this deviation from axial recoil approximation will give rise to the strong forward backward anisotropy that we observe. We believe that the sequential fragmentation seen in the three-body break up channel may have smoothed the forward-backward anisotropy as compared to the other two channels. The angular distribution for \ce{O-} is shown in Figure \ref{fig3.14}(d) along with a possible fit for a \ce{B2} resonance taking into account contribution from $p$ and $d$ partial waves. Assuming axial recoil, the \ce{O-} ejects along the \ce{C2} axis (i.e. $\beta$=0) and not along the O-H bonds. Thus, the functional form of the fitting expression is $2a^{2}\sin^{2}\theta + \frac{3}{2} b^{2}\sin^{2}2\theta + 2 a b \sqrt{3} \sin\theta \sin2\theta \cos\delta$. In this case also it is difficult to say that the fit is consistent with the experimental results, though we are able to reproduce the peak at about $60^{\circ}$. We find the ratio of the amplitudes of the $p$ to $d$ waves to close to 5 (inner ring), 4 (middle ring) and 2.5 (outer ring OH) for the three \ce{H-} channels with a phase difference close to $\pi$ radians. The ratio of $p$ to $d$ wave amplitudes for the \ce{O-} channel is close to 1.3 with a phase difference of about 1 radian.

Since we are accessing the same resonance which is decaying through three different channels, the angular distributions should correspond to the same state with identical partial waves. In order to fit the forward-backward asymmetry, we have a phase difference of almost $\pi$ radians in the two partial waves for the \ce{H-} channels, while
it is small for the \ce{O-} channel. The phase shifts among the partial waves are expected to be produced by the distortion of the incoming plane wave due to scattering by the molecule as a whole. One may argue that appropriately phase-shifted partial waves take part in the capture process leading to the observed angular distributions. This raises an interesting possibility that one set of partial waves with appropriate phase relation get captured to give rise to the \ce{H-} ions, while another set with a different set of phases get captured at the same resonance, but leading to the formation of \ce{O-} ions. It is not clear if the complex dynamics of the nuclear and electronic motion leading
to DEA could retain such fine details of the phase difference between partial waves in determining the final dissociation channel.

It appears that the most probable way we get different angular distributions are due to deviations from the axial recoil approximation in the three different channels. The angular distribution for the \ce{OH (X ^{2}$\Pi$)} channel peaking at $180^{\circ}$ with respect to the electron beam indicates that the electron is approaching along the H-O bond and that mostly this particular O-H bond is being broken. This is partly consistent with the fact that the probability for electron capture into the (\ce{1b2^{-1}, 4a1^{2}}) \ce{B2} resonance will be maximum when the electron approach the molecule along the O-H bond from the hydrogen side. In this orientation, we expect to see \ce{H-} to be ejected at $75^{\circ}$ as well since either of the O-H bonds could break following the electron capture. That we do not see a peak at $75^{\circ}$ is intriguing.  The only possible way this could happen is a strong asymmetric stretching introduced in the O-H bond along which the electron is approaching the molecule. This asymmetric stretch, caused during the attachment process, makes the resonance dissociate along that particular bond, yielding \ce{H-}.

The behaviour of the angular distributions in the two inner rings may be explained in terms of the structural changes that the molecular negative ion is undergoing after electron capture. For the outermost ring, the \ce{H-} has the largest possible kinetic energy and we expect the dissociation to be the fastest, despite some of the energy going into vibrational excitation of the OH fragment as seen from the kinetic energy spectrum. Thus this channel will mostly follow the axial recoil approximation. As the three channels are assumed to be produced from the same resonant state, this distribution should also represent the anisotropy in the overall electron capture. For the middle ring corresponding to the \ce{OH (A ^{2}$\Sigma$)} channel, the \ce{H-} energy is considerably reduced to 2.5 eV due to the electronic and vibrational excitation of the OH. This reduction in energy corresponds to a slower dissociation process. In this timescale the vibrational motions of the molecule could force some of the fragmentation to take place along the O-H bond that is away from the electron beam direction. Simultaneously, the bending mode vibration brings the two H atoms closer, producing the observed secondary peak at $80^{\circ}$ and smearing out the abrupt drop seen in the angular distribution below $140^{\circ}$ in the outermost ring. In the case of the three-body fragmentation channel, the bending mode vibration appears to smoothen out the anisotropy further. The sequential fragmentation present in this channel will also smooth the angular distribution. 

Although for the formation of \ce{O-} both the OH bonds need to be broken, one can deduce the angular distribution of \ce{O-} using the axial recoil approximation if the role of vibrations and rotations are neglected. Such comparison of the angular distribution is shown in Figure \ref{fig3.14}(g) and (h). We may qualitatively explain the difference between the fit and the observed distribution in the case of \ce{O-} also as due to deviation from the axial recoil approximation, including sequential fragmentation.

\section{Summary}	
\begin{enumerate}
\item	First imaging experiment on study of DEA to Water.
\item Velocity images of \ce{H-}(\ce{D-}) and \ce{O-} ions obtained for incident electron energies 5-13 eV.
\item	First resonance centered at 6.5 eV has \ce{^{2}B1} symmetry state and agrees with previous measurements.
\item	Second resonance centered at 8.5 eV proceeds via \ce{^{2}A1} symmetry and undergoes bending mode vibrations to linear geometry causing crossover to \ce{B1} symmetry. Revealed by \ce{H-} angular distribution changing as a function of its kinetic energy (or OH vibrational state). 
\item	\ce{OH- (^{1}$\Sigma$) + H} channel having the lowest threshold of 3.27 eV and favoured by a \ce{A1} resonance not seen.
\item	Third resonance centered at 12 eV has \ce{^{2}B2} symmetry and shows strong anisotropy in the scattering of \ce{H-} and \ce{O-} ions. 
\item	\ce{H-} from the third resonance process is produced via three dissociation channels including a three body fragmentation channel and is ejected in backward angles. 
\item	\ce{O-} produced via a three body breakup is scattered in the forward direction.
\item	Instantaneous and sequential fragmentation processes inferred from the kinetic energy spectrum leading to the formation of \ce{H-} and \ce{O-}. 
\item Deviation from axial recoil leading to complex dissociation dynamics observed in second and third resonance processes.
\end{enumerate}

\appendix
\newpage
\section{Spherical harmonics tranformed by Euler Angles $(\alpha, \beta, \gamma)$}

Spherical harmonics tranformed from one frame to another frame by rotation through Euler angles $(\alpha,\beta,\gamma)$ can be related as per the expression below (Ref: W Thompson, {\em Angular Momentum - An Illustrated Guide to Rotational Symmetries for Physical Systems}, WILEY-VCH Verlag GmbH \& Co. KGaA (2004)). The term on LHS i.e  is the spherical harmonic in the new frame and the $Y_{l,m}$s on the right are in the old frame;

\begin{eqnarray}
Y_{l,m^{\prime}}(\theta,\phi)&=&\displaystyle \sum_{m=-l}^{l} Y_{l,m}(\beta,\alpha) D^{l}_{m,m^{\prime}}(\alpha,\beta,\gamma) \nonumber \\
\mbox{where} D^{l}_{m,m^{\prime}}(\alpha,\beta,\gamma)&=&e^{-im\alpha} d^{l}_{m,m^{\prime}}(\beta) e^{-im^{\prime}\gamma} \nonumber \\
d^{l}_{m,m^{\prime}}(\beta) &=& \sqrt{(l+m^{\prime})! (l-m^{\prime})! (l+m)! (l-m)!} \nonumber \\  
&& \times \displaystyle\sum_{x}{\frac{(-1)^{x}\cos^{2l+m^{\prime}-m-2x}\hba \sin^{2x+m-m^{\prime}}\hba}{(l+m^{\prime}-x)! (l+m-x)! x! (x+m-m^{\prime})!}} \nonumber \\
\nonumber \\
&&max(0,m^{\prime}-m) \leq x \leq min(l-m,l+m^{\prime}) \nonumber 
\end{eqnarray}

Following are the expressions for spherical harmonics upto $l=3$ transformed by Euler angles $(\alpha, \beta, 0)$

\begin{eqnarray}
Y_{0,0}(\theta_{e},\phi_{e})&=&Y_{0,0} 
\\
Y_{1,-1}(\theta_{e},\phi_{e})&=&e^{i\alpha} \cos^{2} \hba   Y_{1,-1} - \frac{\sin{\beta}}{\sqrt{2}}  Y_{1,0} + e^{-i\alpha} \sin^{2} \hba  Y_{1,1}
\\
Y_{1,0}(\theta_{e},\phi_{e})&=&e^{i\alpha} \frac{\sin{\beta}}{\sqrt{2}} Y_{1,-1} + \cos{\beta} Y_{1,0} - e^{-i\alpha} \frac{\sin{\beta}}{\sqrt{2}}  Y_{1,1} 
\\
Y_{1,1}(\theta_{e},\phi_{e})&=&e^{i\alpha} \sin^{2} \hba Y_{1,-1} + \frac{\sin{\beta}}{\sqrt{2}} Y_{1,0} + e^{-i\alpha} \cos^{2} \hba Y_{1,1}
\\
Y_{2,-2}(\theta_{e},\phi_{e})&=&e^{i2\alpha} \cos^{4}\hba Y_{2,-2} - e^{i\alpha} \cos^{2}\hba\sin(\beta)Y_{2,-1} + \frac{\sqrt{3}}{2\sqrt{2}} \sin^{2}(\beta) Y_{2,0} \nonumber \\
&& - e^{-i\alpha} \sin^{2}\hba \sin{\beta} Y_{2,1} + e^{-i2\alpha} \sin^{4} \hba Y_{2,2}
\end{eqnarray}
\begin{eqnarray}
Y_{2,-1}(\theta_{e},\phi_{e})&=&e^{i2\alpha} \sin{\beta} \cos^{2}\hba Y_{2,-2} + e^{i\alpha} \cos^{2}\hba (\cos^{2}\hba-3\sin^{2}\hba) Y_{2,-1} \nonumber \\
&& - \frac{\sqrt{3}}{2\sqrt{2}} \sin{2\beta} Y_{2,0} + e^{-i\alpha} \sin^{2}\hba (3\cos^{2}\hba-\sin^{2}\hba) Y_{2,1} \nonumber \\
&& -  e^{-i2\alpha} \sin{\beta} \sin^{2}\hba Y_{2,2}
\\
Y_{2,0}(\theta_{e},\phi_{e})&=&e^{i2\alpha} \frac{\sqrt{3}}{2\sqrt{2}} \sin^{2}\beta Y_{2,-2} + e^{i\alpha} \frac{\sqrt{3}}{2\sqrt{2}} \sin{2\beta} Y_{2,-1} + \frac{3 \cos^{2}{\beta}-1}{2} Y_{2,0} \nonumber \\ 
&& - e^{-i\alpha} \frac{\sqrt{3}}{2\sqrt{2}} \sin{2\beta} Y_{2,1} + e^{-i2\alpha} \frac{\sqrt{3}}{2\sqrt{2}} \sin^{2}\beta Y_{2,2}
\\
Y_{2,1}(\theta_{e},\phi_{e})&=&e^{i2\alpha} \sin{\beta} \sin^{2}\hba Y_{2,-2} + e^{i\alpha} \sin^{2}\hba (3\cos^{2}\hba-\sin^{2}\hba) Y_{2,-1} \nonumber \\
&& + \frac{\sqrt{3}}{2\sqrt{2}} \sin{2\beta} Y_{2,0} + e^{-i\alpha} \cos^{2}\hba (\cos^{2}\hba-3\sin^{2}\hba) Y_{2,1} \nonumber \\
&& -  e^{-i2\alpha} \sin{\beta} \cos^{2}\hba Y_{2,2}
\\
Y_{2,2}(\theta_{e},\phi_{e})&=&e^{i2\alpha} \sin^{4}\hba Y_{2,-2} + e^{i\alpha} \sin{\beta} \sin^{2}\hba Y_{2,-1} + \frac{\sqrt{3}}{2\sqrt{2}} \sin^{2}{\beta} Y_{2,0} \nonumber \\
&& + e^{-i\alpha} \sin{\beta} \cos^{2}\hba Y_{2,1} + e^{-i2\alpha} \cos^{4} \hba Y_{2,2}
\\
Y_{3,-3}(\theta_{e},\phi_{e})&=&e^{i3\alpha} \cos^{6}\hba Y_{3,-3} \nonumber \\
&& - e^{i2\alpha} \sqrt{6} \cos^{5}\hba \sin\hba Y_{3,-2} + e^{i\alpha} \sqrt{15} \cos^{4}\hba \sin^{2}\hba Y_{3,-1} \nonumber \\
&& - 2\sqrt{5} \cos^{3}\hba \sin^{3}\hba Y_{3,0} + e^{-i\alpha} \sqrt{15} \cos^{2}\hba \sin^{4}\hba Y_{3,1} \nonumber \\
&& - e^{-i2\alpha} \sqrt{6} \cos\hba \sin^{5}\hba Y_{3,2} + e^{-i3\alpha} \sin^{6}\hba Y_{3,3}
\\
Y_{3,-2}(\theta_{e},\phi_{e})&=&e^{i3\alpha} \sqrt{6} \cos^{5}\hba \sin\hba Y_{3,-3} + e^{i2\alpha}  (\cos^{6}\hba - 5 \cos^{4}\hba \sin^{2}\hba ) Y_{3,-2} \nonumber \\
&& + e^{i\alpha} \sqrt{10} \cos^{3}\hba  \sin\hba (2 \sin^{2}\hba - \cos^{2} \hba) Y_{3,-1} + \frac{\sqrt{30}}{4} \sin^{2}\beta \cos\beta Y_{3,0} \nonumber \\
&& + e^{-i\alpha} \sqrt{10} \cos\hba \sin^{3}\hba (\sin^{2}\hba - 2 \cos^{2}\hba) Y_{3,1} \nonumber \\
&& + e^{-i2\alpha} \sin^{4}\hba (5 \cos^{2}\hba - \sin^{2}\hba) Y_{3,2} - e^{-i3\alpha} \sqrt{6} \cos\hba \sin^{5}\hba Y_{3,3}
\\
Y_{3,-1}(\theta_{e},\phi_{e})&=&e^{i3\alpha} \sqrt{15} \cos^{4}\hba \sin^{2}\hba Y_{3,-3} \nonumber \\
&& + e^{i2\alpha} \sqrt{10}  \cos^{3}\hba \sin\hba (\cos^{2}\hba - 2 \sin^{2}\hba) Y_{3,-2} \nonumber \\
&& + e^{i\alpha} \cos^{2}\hba (\cos^{4}\hba - 8 \cos^{2}\hba \sin^{2}\hba + 6 \sin^{4}\hba) Y_{3,-1} \nonumber \\
&& - \sqrt{3} \sin\beta (\cos^{2}\beta - \frac{\sin^{2}}{4}) Y_{3,0} \nonumber \\
&& + e^{-i\alpha} \sin^{2}\hba (6 \cos^{4}\hba - 8 \sin^{2}\hba\cos^{2}\hba + \sin^{4}\hba) Y_{3,1} \nonumber \\
&& + e^{-i2\alpha} \sqrt{10} \cos\hba \sin^{3}\hba (\sin^{2}\hba - 2 \cos^{2}\hba) Y_{3,2} \nonumber \\
&& + e^{-i3\alpha} \sqrt{15} \cos^{2}\hba \sin^{4}\hba Y_{3,3}
\end{eqnarray}
\begin{eqnarray}
Y_{3,0}(\theta_{e},\phi_{e})&=&e^{i3\alpha} \frac{\sqrt{5}}{4} \sin^{3}\beta Y_{3,-3} + e^{i2\alpha} \frac{\sqrt{15}}{2\sqrt{2}} \sin^{2}\beta \cos\beta Y_{3,-2} \nonumber \\
&& + e^{i\alpha} \sqrt{3} \sin\beta (\cos^{2}\beta - \frac{\sin^{2}\beta}{4}) Y_{3,-1} + \cos\beta (1 - \frac{5}{2} \sin^{2}\beta) Y_{3,0} \nonumber \\
&& - e^{-i\alpha} \sqrt{3} \sin{\beta} (\cos^{2}\beta - \frac{\sin^{2}}{4}) Y_{3,1} + e^{-i2\alpha} \frac{\sqrt{15}}{2\sqrt{2}} \sin^{2}\beta \cos\beta Y_{3,2} \nonumber \\
&& - e^{-i3\alpha} \frac{\sqrt{5}}{4} \sin^{3}\beta Y_{3,3}
\\
Y_{3,1}(\theta_{e},\phi_{e})&=&e^{i3\alpha} \sqrt{15} \cos^{2}\hba \sin^{4}\hba Y_{3,-3} \nonumber \\
&& + e^{i2\alpha} \sqrt{10}  \sin^{3}\hba \cos\hba (2\cos^{2}\hba - \sin^{2}\hba) Y_{3,-2} \nonumber \\
&& + e^{i\alpha} \sin^{2}\hba (\sin^{4}\hba - 8 \cos^{2}\hba \sin^{2}\hba + 6 \cos^{4}\hba) Y_{3,-1} \nonumber \\
&& + \sqrt{3} \sin\beta (\cos^{2}\beta - \frac{\sin^{2}\beta}{4}) Y_{3,0}  \nonumber \\
&& + e^{-i\alpha} \cos^{2}\hba (6 \sin^{4}\hba - 8 \sin^{2}\hba \cos^{2}\hba + \cos^{4}\hba) Y_{3,1} \nonumber \\
&& - e^{-i2\alpha} \sqrt{10} \sin\hba \cos^{3}\hba (\cos^{2}\hba - 2 \sin^{2}\hba) Y_{3,2} \nonumber \\
&& + e^{-i3\alpha} \sqrt{15} \cos^{4}\hba \sin^{2}\hba Y_{3,3}
\\
Y_{3,2}(\theta_{e},\phi_{e})&=&e^{i3\alpha} \sqrt{6} \cos\hba \sin^{5}\hba Y_{3,-3} + e^{i2\alpha} \sin^{4}\hba (5\cos^{2}\hba - \sin^{2}\hba) Y_{3,-2} \nonumber \\
&& + e^{i\alpha} \sqrt{10} \sin^{3}\hba  \cos\hba (2 \cos^{2}\hba - \sin^{2}\hba) Y_{3,-1} + \frac{\sqrt{30}}{4} \sin^{2}\beta \cos\beta Y_{3,0} \nonumber \\
&& + e^{-i\alpha} \sqrt{10} \sin\hba \cos^{3}\hba (\cos^{2}\hba - 2 \sin^{2}\hba) Y_{3,1} \nonumber \\
&& + e^{-i2\alpha} \cos^{4}\hba (\cos^{2}\hba - 5 \sin^{2}\hba) Y_{3,2} \nonumber \\
&& - e^{-i3\alpha} \sqrt{6} \sin\hba \cos^{5}\hba Y_{3,3}
\\
Y_{3,3}(\theta_{e},\phi_{e})&=&e^{i3\alpha} \sin^{6}\hba Y_{3,-3} + e^{i2\alpha} \sqrt{6} \sin^{5}\hba \cos\hba Y_{3,-2} + e^{i\alpha} \sqrt{15} \cos^{2}\hba \sin^{4}\hba Y_{3,-1} \nonumber \\
&& - \frac{\sqrt{5}}{4} \sin^{3}\beta Y_{3,0} + e^{-i\alpha} \sqrt{15} \cos^{4}\hba \sin^{2}\hba Y_{3,1} \nonumber \\
&& + e^{-i2\alpha} \sqrt{6} \sin\hba \cos^{5}\hba Y_{3,2} + e^{-i3\alpha} \cos^{6}\hba Y_{3,3}
\end{eqnarray}
%\end{document}
\newpage
\section{Angular distribution curves for $C_{2v}$ point group}

Expression for angular distribution for various symmetries under $C_{2v}$ group taking lowest allowed partial waves. $\beta$ is the angle between the dissociating bond and the molecular symmetry axis i.e. $C_{2}$ axis. $\theta$ is the scattering angle with respect to the electron beam direction in lab frame. 

\section*{$A_{1}$ to $A_{1}$ transition}
\begin{eqnarray}
I_{s}(\theta)&=& 1 \\
I_{p}(\theta)&=&\sin^{2}\beta \sin^{2}\theta + 2 \cos^{2}\beta \cos^{2}\theta \\
I_{d}(\theta)&=& \frac{9}{16} (\sin^{4}\beta \sin^{4}\theta + \sin^{2}2\beta \sin^{2}2\theta)+ \frac{1}{2}(3 \cos^{2}\beta - 1)(3 \cos^{2}\theta - 1) \\
I_{s+p}(\theta)&=& a^{2} + b^{2}(\sin^{2}\beta \sin^{2}\theta + 2 \cos^{2}\beta \cos^{2}\theta) + 4ab \cos\beta \cos\theta \cos\delta \\
I_{s+p+d}(\theta)&=& a^{2} + b^{2}(\sin^{2}\beta \sin^{2}\theta + 2 \cos^{2}\beta \cos^{2}\theta) \nonumber \\
&& + c^{2} (\frac{9}{16} (\sin^{4}\beta \sin^{4}\theta + \sin^{2}2\beta \sin^{2}2\theta)+ \frac{1}{2}(3 \cos^{2}\beta - 1)(3 \cos^{2}\theta - 1)) \nonumber \\
&&+ 4 a b \cos\beta \cos\theta \cos\delta_{1} \nonumber \\
&& + 2 b c (\frac{3}{4}\sin\beta \sin2\beta \sin\theta \sin2\theta + \frac{1}{2}\cos\beta (3\cos^{2}\beta-1)\cos\theta (3\cos^{2}\theta-1)) \cos\delta_{2} \nonumber \\
&& + a c (3\cos^{2}\beta-1)(3\cos^{2}\theta-1) \cos(\delta_{1}+\delta_{2}) 
\end{eqnarray}

\section*{$A_{1}$ to $A_{2}$ transition}
\begin{eqnarray}
I_{d}(\theta)&=&\cos^{2}\beta \sin^{4}\theta + \sin^{2}\beta \sin^{2}2\theta
\end{eqnarray}

\section*{$A_{1}$ to $B_{1}$ transition}
\begin{eqnarray}
I_{p}(\theta)&=&2\sin^{2}\theta \\
I_{d}(\theta)&=&1.5(\sin^{2}\beta \sin^{4}\theta + \cos^{2}\beta \sin^{2}2\theta) \\
I_{p+d}(\theta)&=& 2a^{2} \sin^{2}\theta + 1.5b^{2} (\sin^{2}\beta \sin^{4}\theta + \cos^{2}\beta \sin^{2}2\theta) + 2ab \sqrt{3} \cos\beta \sin\theta \sin2\theta \cos\delta
\end{eqnarray}

\section*{$A_{1}$ to $B_{2}$ transition}
\begin{eqnarray}
I_{p}(\theta)&=&2(2sin^{2}\beta \cos^{2}\theta + \cos^{2}\beta \sin^{2}\theta) \\
I_{d}(\theta)&=&\frac{3}{2}(\frac{1}{4}\sin^{2}2\beta \sin^{4}\theta + \cos^{2}2\beta \sin^{2}2\theta + \frac{\sin^{2}2\beta}{2}(3cos^{2}\theta-1)^{2}) \\
I_{p+d}(\theta)&=&2a^{2}(2sin^{2}\beta \cos^{2}\theta + \cos^{2}\beta \sin^{2}\theta) \nonumber \\
&& + \frac{3}{2} b^{2} (\frac{1}{4}\sin^{2}2\beta \sin^{4}\theta + \cos^{2}2\beta \sin^{2}2\theta) \nonumber \\
&& + 2 a b \sqrt{3} (\cos\beta \cos2\beta \sin\theta \sin2\theta - \sin\beta \sin2\beta cos\theta (3\cos^2{\theta}-1)) \cos\delta
\end{eqnarray}

\end{document}